\documentclass[aps,prx,twocolumn,superscriptaddress,nobalancelastpage]{revtex4-2}

\usepackage{makecell}
\usepackage{amsmath}
\usepackage{amssymb}
\usepackage{amsthm}
\usepackage{graphicx}
\usepackage[dvipsnames]{xcolor}

\definecolor{myred}{HTML}{E1558A}
\definecolor{myblue}{HTML}{318294}
\definecolor{myskyblue}{HTML}{3FC3E0}
\usepackage[hidelinks]{hyperref}
\hypersetup{
  colorlinks   = true, 
  urlcolor     = blue, 
  linkcolor    = myred, 
  citecolor   = myblue 
}

\newcommand{\bs}[1]{\boldsymbol{#1}}

\begin{document}

\title{Geometric decomposition of information flow for overdamped Langevin systems and optimal transport in subsystems}

\author{Sosuke Ito}
\email{sosuke.ito@ubi.s.u-tokyo.ac.jp}
\affiliation{Department of Physics, The University of Tokyo, 7-3-1 Hongo, Bunkyo-ku, Tokyo 113-0033, Japan}
\affiliation{Universal Biology Institute, The University of Tokyo, 7-3-1 Hongo, Bunkyo-ku, Tokyo 113-0033, Japan}
\author{Yoh Maekawa}
\email{yoh.maekawa@ubi.s.u-tokyo.ac.jp}
\affiliation{Department of Physics, The University of Tokyo, 7-3-1 Hongo, Bunkyo-ku, Tokyo 113-0033, Japan}
\author{Ryuna Nagayama}
\affiliation{Department of Physics, The University of Tokyo, 7-3-1 Hongo, Bunkyo-ku, Tokyo 113-0033, Japan}
\author{Andreas Dechant}
\affiliation{Department of Physics $\#$1, Graduate School of Science, Kyoto University, Kyoto 606-8502, Japan}
\author{Kohei Yoshimura}
\affiliation{Universal Biology Institute, The University of Tokyo, 7-3-1 Hongo, Bunkyo-ku, Tokyo 113-0033, Japan}
\affiliation{Nonequilibrium Quantum Statistical Mechanics RIKEN Hakubi Research Team, Pioneering Research Institute (PRI), RIKEN, 2-1 Hirosawa, Wako, Saitama 351-0198, Japan}

\date{\today}

\begin{abstract}

Information flow between subsystems is a central concept in information thermodynamics, which provides the second-law-like inequalities for subsystems. 
This paper discusses the geometric decomposition of information flow, which was introduced for Markov jump systems [\href{https://doi.org/10.1103/trrq-tkw5}{Physical Review Research, {\bf 8}, 023292 (2026)}], and applies it to overdamped Langevin systems. For overdamped Langevin systems, the geometric decomposition of information flow into excess and housekeeping contributions is related to the conventional definition of the $2$-Wasserstein distance between marginal distributions in optimal transport theory. This formulation offers an optimal-transport interpretation of subsystem dynamics, and this optimal-transport formulation is simpler for overdamped Langevin systems than for general Markov jump systems. It is also possible to handle features that are specific to overdamped Langevin systems, such as representations based on the Koopman mode decomposition, as well as their relationship with the Fisher information matrix.  As with the results for Markov jump systems, we generalize the second law of information thermodynamics using housekeeping and excess information flow, leading to the concepts of excess and housekeeping demons. We also derive a thermodynamic uncertainty relation and an information-thermodynamic speed limit incorporating excess information flow. These results are illustrated for the Gaussian case, and we discuss the conditions under which the excess and housekeeping demons emerge.
\end{abstract}

\maketitle

\section{Introduction}
A central challenge in nonequilibrium physics is to identify principles that characterize and quantify departures from equilibrium. One fruitful approach is to examine the interactions between subsystems, signaling nonequilibrium behavior through the emergence of directional asymmetries. Such asymmetries can be detected using probability fluxes~\cite{schnakenberg1976, sekimoto2010stochastic,de2013non, seifert2025stochastic}, affinities~\cite{schnakenberg1976, sekimoto2010stochastic,de2013non,seifert2025stochastic}, cross-correlation functions~\cite{casimir1945, ohga2023}, or response functions~\cite{Onsager1931,Onsager1931-2, Owen2020}, and reflect a violation of the detailed balance condition. Among the various frameworks developed along these lines, information flow~\cite{parrondo2015thermodynamics} has attracted considerable attention as a quantitative measure of directed interaction between two subsystems. Broadly speaking, information flow aims to capture how the state of one subsystem influences the future evolution of another, thereby providing an information-theoretic perspective on nonequilibrium coupling between subsystems. Existing formulations of information flow have been introduced in various forms for a variety of systems~\cite{massey1990causality,Schreiber2000,Touchette2000, sagawa2008, Allahverdyan_2009, horowitz2010, sagawa2012,still2012, sagawapre2012, ito2013, Hartich_2014, horowitzesposito2014,horowitz2014second, ito2015maxwell, ito2016information, ito2016backward,rosinberg2016continuous,spinney2016,auconi2019information,Crooks_2019, wolpert2020uncertainty,yada2022}, most commonly based on temporal changes in mutual information or conditional mutual information between the two subsystems~\cite{cover1999elements}.

Particularly within the framework of information thermodynamics, the concept of information flow emerged from considerations inspired by Maxwell’s demon~\cite{parrondo2015thermodynamics,sagawa2008,sagawa2010,toyabe2010experimental}, where information processing appears to enable apparent violations of the second law of thermodynamics. In this context, information flow was introduced as an additional term that compensates for such apparent violations by accounting for the informational contribution exchanged between interacting subsystems. Corresponding expressions have been formulated for a variety of settings, including general feedback-controlled systems~\cite{sagawa2010,horowitz2010,sagawapre2012} and the dynamics of non-Markovian processes~\cite{ito2013,ito2016backward,spinney2016,Crooks_2019,wolpert2020uncertainty}. Among these formulations, a specific expression~\cite{Allahverdyan_2009,ito2013,Hartich_2014,horowitzesposito2014} defined for a restricted class of Markovian dynamics, known as bipartite systems, has attracted attention in recent years~\cite{amano2022insights,Ryota2022,Leignton2024}. Notably, the resulting information flow does not vanish even in the steady state. Since then, it has provided a theoretical explanation for Maxwell's demon that works in the steady state, which is referred to as the autonomous demon~\cite{strasberg2013}. 

On the other hand, the conventional Maxwell’s demon can also be analyzed within the framework of explicitly time-dependent, nonstationary dynamics~\cite{szilard1964decrease,sagawa2010,horowitz2010,sagawa2012}. In such situations, Maxwell's demon is usually explained using measurement and feedback processes. In this model, information acquired through measurement is used to control the system. This perspective has led to the development of a well-established thermodynamic description of Maxwell's demon, which can be applied even when the system is far from the steady state, and to non-Markovian dynamics~\cite{ito2013,ito2016backward,spinney2016,Crooks_2019,wolpert2020uncertainty}. Because both can be discussed using the same thermodynamic inequality, i.e., the second law of information thermodynamics~\cite{parrondo2015thermodynamics} with the same information flow, the conceptual distinction between the autonomous demon that works in the steady state and the conventional Maxwell’s demon that works in nonstationary dynamics has received relatively little attention.  

\begin{figure}
    \centering
    \includegraphics[width=\linewidth]{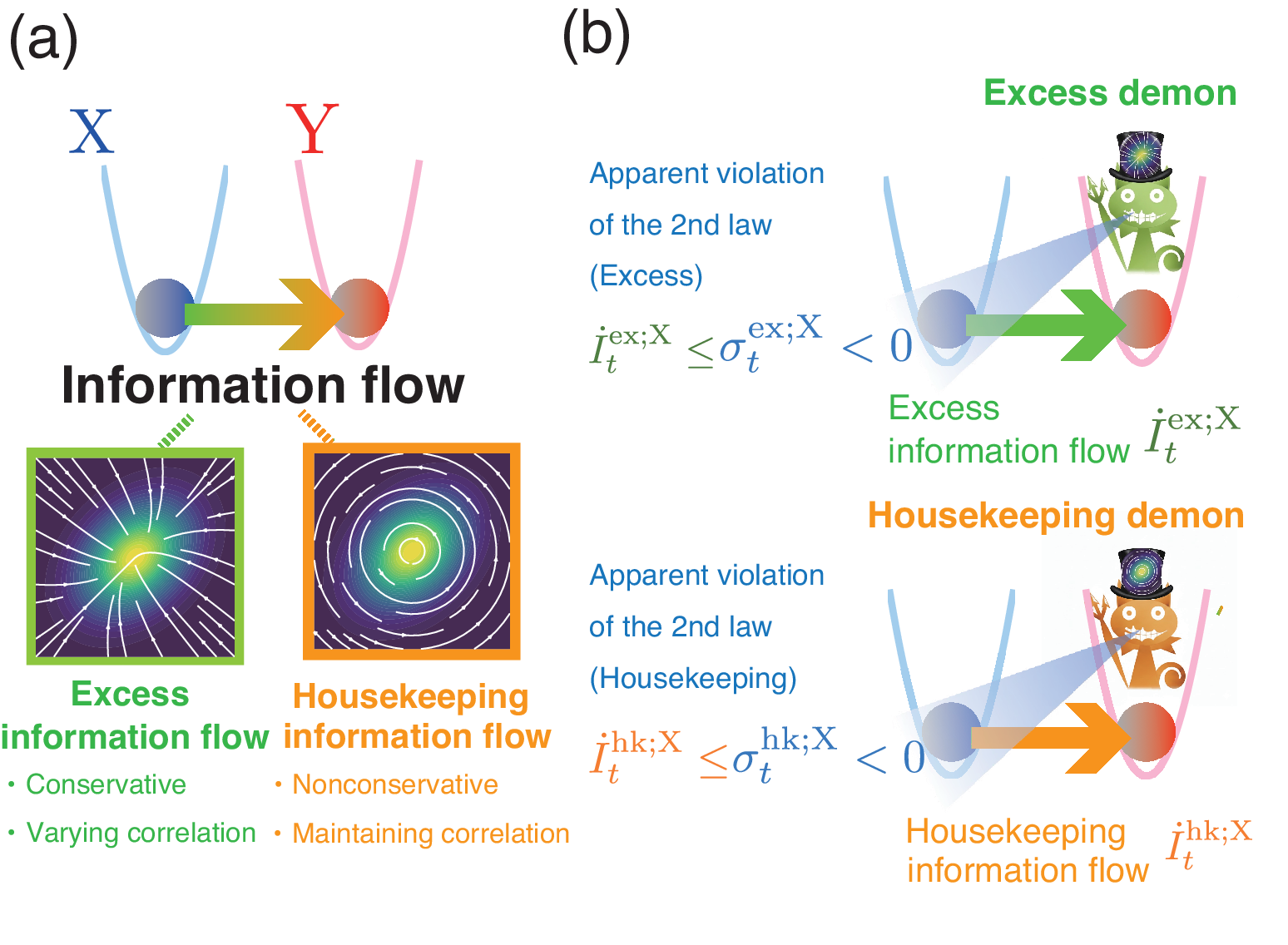}
    \caption{ (a) Schematic showing the geometric decomposition of information flow. Excess information flow represents a conservative contribution and can vary the correlation. Housekeeping information flow represents a nonconservative contribution and maintains the correlation.
    (b) Schematic showing the excess demon and the housekeeping demon. The excess demon in system $\rm Y$ uses excess information flow $\dot{I}^{\rm ex;X}_t$ to make the apparent excess entropy change rate $\sigma^{\rm ex;X}_t$ in system $\rm X$ negative. The housekeeping demon in system $\rm Y$ uses housekeeping information flow $\dot{I}^{\rm hk;X}_t$ to make the apparent housekeeping entropy change rate $\sigma^{\rm hk;X}_t$ in system $\rm X$ negative.}
    \label{exhkdemon}
\end{figure}

Some of the authors of the present paper have previously addressed this issue within the framework of bipartite discrete Markov jump systems~\cite{maekawa2025geometric}. In this setting, it was demonstrated that information flow can be divided into two distinct components: excess information flow, which is conservative and involves varying correlations between subsystems; and housekeeping information flow, which is nonconservative and maintains correlations between subsystems (see also Fig.~\ref{exhkdemon}(a)). In this paper, we investigate the overdamped Langevin representation of this decomposition. By formulating the excess and housekeeping components of information flow within continuous-state stochastic dynamics, we clarify their physical interpretation in a more accessible and transparent form based on overdamped Langevin dynamics.

Restricting our analysis to continuous-state Langevin systems clarifies the connection to optimal transport theory~\cite{villani2008optimal,ito2024geometric}, within which a natural decomposition into housekeeping and excess contributions~\cite{ito2024geometric,dechant2022geometric,dechant2022geometric2,yoshimura2023housekeeping,nagayama2025geometric} emerges. Optimal transport has recently attracted considerable attention in stochastic thermodynamics as a unifying framework for formulating thermodynamic trade-off relations~\cite{ito2024geometric}, including uncertainty relations~\cite{baratotur2015,horowitz2020thermodynamic,dechant2022geometric,dechant2022geometric2,Dechantwasser_2022,li2023wasserstein,yoshimura2023housekeeping,vu2023,kolchinsky2024generalized,Delvenne2024,nagayama2025geometric} and speed limits~\cite{aurell2012refined,chen2020, nakazato2021geometrical,yoshimura2023housekeeping,Dechantwasser_2022,vu2023,kolchinsky2024generalized,nagase2024thermodynamically,nagayama2025geometric,nagayama2025infinite,sabbagh2024wasserstein,kwon2024,oikawa2025experimentally,kamijima2025optimal}. In the context of continuous-state Langevin dynamics, the use of the $2$-Wasserstein distance~\cite{villani2008optimal} plays a central role, as it provides natural generalizations of these thermodynamic uncertainty relations~\cite{dechant2022geometric,ito2024geometric} and speed limits~\cite{aurell2012refined,chen2020,ito2013,ito2024geometric,kamijima2025optimal}. Unlike discrete-state formulations based on extensions of the Benamou–Brenier formula~\cite{maas2011gradient,yoshimura2023housekeeping}, the continuous-state approach based on the Benamou-Brenier formula~\cite{benamou2000computational} considered here is also related to the Monge–Kantorovich formulation of optimal transport~\cite{villani2008optimal}. This perspective yields a natural differential-geometric interpretation of stochastic thermodynamics~\cite{ito2024geometric, jordan1998, maas2011gradient,nakazato2021geometrical,olga2022,yoshimura2023housekeeping,chennakesavalu2023, zhong2024,nagayama2025geometric}, in which nonequilibrium processes are described as geometric structures in the space of probability distributions. Building on the decomposition of information flow into housekeeping and excess components in overdamped Langevin systems, we establish explicit connections between various results in information thermodynamics~\cite{Allahverdyan_2009, horowitz2014second}, thermodynamic trade-off relations~\cite{nakazato2021geometrical,dechant2022geometric}, and optimal-transport-based formulations of stochastic thermodynamics~\cite{ito2024geometric}. Furthermore, we generalize the results for continuous states, such as the Koopman mode decomposition~\cite{sekizawa2025koopman,sekizawa2024decomposing} and its relationship with Fisher information~\cite{matsumoto2025learning}. This unified approach clarifies their connection to housekeeping and excess information flow.

Particularly when considered from the perspective of generalizing the second law of information thermodynamics, this research provides a deeper understanding of the concepts of the housekeeping demon and the excess demon, which are new concepts introduced in Ref.~\cite{maekawa2025geometric} regarding Maxwell's demon. The housekeeping demon and the excess demon are introduced to characterize apparent violations of the generalized second law of thermodynamics via housekeeping information flow and excess information flow, respectively (see also Fig.~\ref{exhkdemon}(b)). The timing of their emergence can be discussed in detail using the Gaussian case. Specifically, in the Gaussian case, not only can each term be calculated analytically, but it also becomes possible to numerically verify when the excess demon, which can only occur during transient time evolution, and the housekeeping demon, which can only occur when a nonconservative force is present, emerge.

\begin{figure*}
    \centering
    \includegraphics[width=\linewidth]{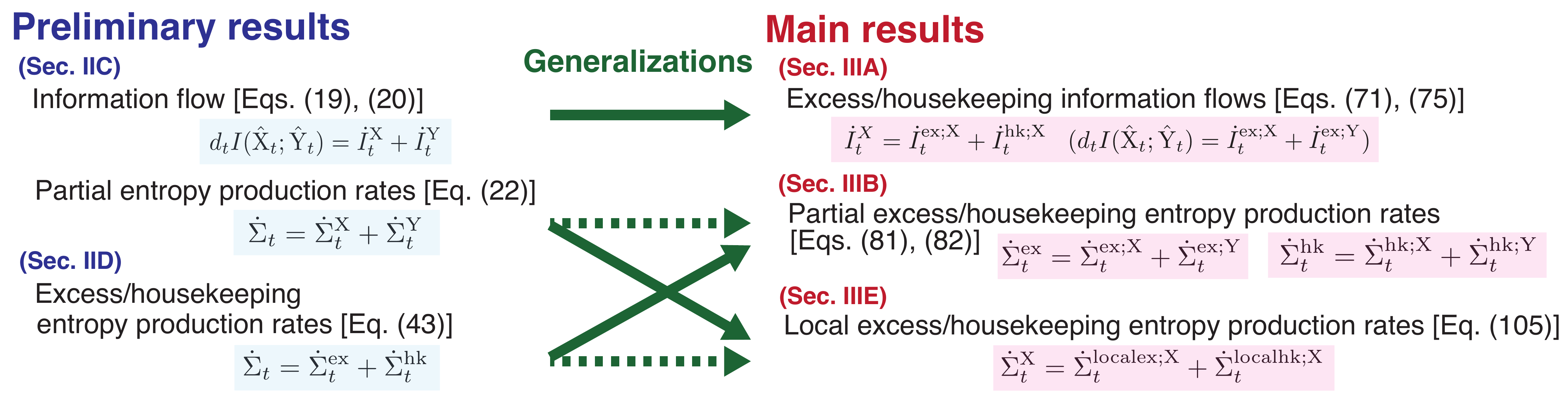}
    \caption{Connection between the review and main results sections. The main results extend the decomposition frameworks introduced in the review section: information flows $\dot{I}^{\rm X}_t$ and $\dot{I}^{\rm Y}_t$ decompose the time derivative of mutual information $d_t I( \hat{ \rm X}_t; \hat{\rm Y}_t)$, while partial entropy production rates $\dot{\Sigma}^{\rm X}_t$ and $\dot{\Sigma}^{\rm Y}_t$, and excess/housekeeping entropy production rates $\dot{\Sigma}^{\rm ex}_t$ and $\dot{\Sigma}^{\rm hk}_t$ provide complementary decompositions of the entropy production rate $\dot{\Sigma}_t$. By combining the different types of decomposition introduced in the review section, the main results section introduces excess/housekeeping information flows $\dot{I}^{\rm ex;X}_t$ and $\dot{I}^{\rm hk;X}_t$, partial excess/housekeeping entropy production rates $\dot{\Sigma}^{\rm ex;X}_t$ and $\dot{\Sigma}^{\rm hk;X}_t$, and local excess/housekeeping entropy production rates $\dot{\Sigma}^{\rm localex;X}_t$ and $\dot{\Sigma}^{\rm localhk;X}_t$.  This generalizes various properties discussed in the review section, such as the second law of information thermodynamics, variational formulas, optimal transport, thermodynamic uncertainty relations, information-thermodynamic speed limits and the Koopman mode decomposition.}
    \label{concept}
\end{figure*}
This paper is organized as follows (see also Fig.~\ref{concept}). In Sec.~\ref{sec2}, we review the stochastic-thermodynamic framework for bipartite overdamped Langevin systems (Sec.~\ref{sec2a}), and introduce the entropy production rate and related quantities (Sec.~\ref{sec2b}). We then introduce information flow in information thermodynamics, and discuss mathematical properties of information flow. We also introduce the partial entropy production rate, which provides the second law of information thermodynamics (Sec.~\ref{sec2c}). We next introduce a geometric housekeeping–excess decomposition of the entropy production rate, which is directly related to the decomposition of information flow in Sec.~\ref{subsec:geometricdecomposition}.
We also discuss how the geometric decomposition relates to variational formulas, optimal transport, thermodynamic uncertainty relations, information-thermodynamic speed limits, and the Koopman mode decomposition. In Sec.~\ref{sec3}, we present our main results. We formulate the geometric decomposition of information flow into excess and housekeeping contributions (Sec.~\ref{sec3a}) and derive the corresponding generalized second law of information thermodynamics for subsystems by introducing partial excess/housekeeping entropy production rates (Sec.~\ref{sec3b}). We further develop a Koopman mode decomposition of the partial housekeeping entropy production rate (Sec.~\ref{sec3c}), and establish thermodynamic uncertainty relations for the partial excess entropy production rate (Sec.~\ref{sec3d}). We introduce local excess/housekeeping entropy production rates  (Sec.~\ref{sec3e}) together with variational characterizations (Sec.~\ref{sec3f}). Building on these, we connect the excess contribution to an optimal transport problem in the subsystems (Sec.~\ref{sec3g}), and derive information-thermodynamic speed limits for subsystems based on the generalized $2$-Wasserstein distance between the marginal distributions (Sec.~\ref{sec3h}). We also relate excess/housekeeping information flows to conditional Fisher information (Sec.~\ref{sec3i}). In Sec.~\ref{sec4}, we illustrate the theory in an analytically tractable Gaussian case (Sec.~\ref{sec4a}) and with numerical examples (Sec.~\ref{sec4b}), thereby clarifying when the excess demon and the housekeeping demon emerge. We conclude with a discussion and outlook in Sec.~\ref{sec5}.

\section{Review on stochastic thermodynamics}
\label{sec2}
\subsection{Setup}
\label{sec2a}
We first discuss several results in stochastic thermodynamics. 
We now consider the $d$-dimensional overdamped Langevin equation 
\begin{align}
   \dot{\boldsymbol{z}}(t) = \mathsf{\mu} \boldsymbol{F}_t (\boldsymbol{z}(t)) + \sqrt{2 T} \mathsf{B}\boldsymbol{\xi}_t,
   \label{Langevineq}
\end{align}
where $\boldsymbol{z}(t) (\in \mathbb{R}^{d})$ is the state of the system, $\dot{\boldsymbol{z}}(t) ( \in \mathbb{R}^{d})$ is the time derivative of the state, $\boldsymbol{F}_t (\boldsymbol{z}(t)) (\in \mathbb{R}^{d})$ is the force at time $t$, $T (>0)$ is the temperature of the heat bath, and $\mathsf{\mu} (\in \mathbb{R}^{d \times d})$ is the mobility matrix that is assumed to be positive definite. 
The white Gaussian noise $\boldsymbol{\xi}_t (\in \mathbb{R}^{d})$ satisfies $\mathbb{E} [\boldsymbol{\xi}_t ]= \boldsymbol{0}$, $\mathbb{E}[ \boldsymbol{\xi}_t \boldsymbol{\xi}_{t'}^{\top} ]= \delta(t-t') \mathsf{I}$, where $\mathbb{E}[\dots]$ stands for the expected value, $\delta(t-t')$ stands for the Dirac delta function, and 
$\mathsf{I}$ is the identity matrix. The matrix $\mathsf{B}(\in \mathbb{R}^{d \times d})$, which does not have to be symmetric, satisfies $\mathsf{\mu}=\mathsf{B} \mathsf{B}^{\top}$.
Here, the Boltzmann constant has been set to unity, i.e., $k_{\rm B}=1$. 

We consider the probability distribution $p_t(\boldsymbol{z})$. Its time evolution corresponding to Eq.~\eqref{Langevineq} can be described by 
the Fokker-Planck equation, which is given by the following continuity equation
\begin{align}
\partial_t p_t(\boldsymbol{z}) &= - \nabla \cdot \boldsymbol{j}_t (\boldsymbol{z}),
\label{continuityeq}
\end{align}
where $\partial_t := \partial/(\partial t)$ stands for the partial derivative operator, $\nabla$ is the del operator, and $\boldsymbol{j}_t (\boldsymbol{z})$ is the flux. Here, the flux is given by
\begin{align}
\boldsymbol{j}_t (\boldsymbol{z})&= p_t(\boldsymbol{z})\mathsf{D} \boldsymbol{f}_t (\boldsymbol{z}) , \label{onsager}\\
\boldsymbol{f}_t (\boldsymbol{z}) &:=  \frac{\boldsymbol{F}_t (\boldsymbol{z})}{T}-   \nabla \ln p_t (\boldsymbol{z}),
\end{align}
where $\mathsf{D}  :=T \mu$ is the diffusion matrix and $\boldsymbol{f}_t(\boldsymbol{z})$ is the thermodynamic force. Because Eq.~\eqref{onsager} is a linear relationship between the flux and the thermodynamic force, the
quantity $p_t(\boldsymbol{z})\mathsf{D}$ can be regarded as the Onsager matrix. We also introduce the velocity field $\boldsymbol{\nu}_t (\boldsymbol{z}) = \mathsf{D} \boldsymbol{f}_t(\boldsymbol{z})$ to rewrite the continuity equation as $\partial_t p_t(\boldsymbol{z}) = - \nabla \cdot (\boldsymbol{\nu}_t (\boldsymbol{z})  p_t(\boldsymbol{z}))$.

To discuss information thermodynamics, we assume that the total system is well decomposed into two subsystems $\mathrm{X}$ and $\mathrm{Y}$. Similarly, the state $\boldsymbol{z}$ is also decomposed as $\boldsymbol{z} =(\boldsymbol{x}, \boldsymbol{y})$, where $\boldsymbol{x} (\in \mathbb{R}^{d^{\rm X}})$ and $\boldsymbol{y} (\in \mathbb{R}^{d^{\rm Y}})$ are the states of $\rm X$ and $\rm Y$, respectively, so that $d=d^{\rm X}+d^{\rm Y}$ is satisfied.

The fact that two subsystems can be well decomposed corresponds to the total system being bipartite. To use vector and matrix notation, let $\boldsymbol{x}$ and $\boldsymbol{y}$ denote column vectors, and let $\boldsymbol{z} \in \mathbb{R}^d$ denote the column vector formed by arranging $\boldsymbol{x}$ and $\boldsymbol{y}$ in that order. The bipartite property is given by the block diagonal condition  
\begin{align}
   \mathsf{B} = \begin{pmatrix}
   \mathsf{B}^{\mathrm X}& \mathsf{O} \\
   \mathsf{O} & \mathsf{B}^{\mathrm Y}
\end{pmatrix},
\end{align}
and thus
\begin{align}
   \mathsf{\mu} = \begin{pmatrix}
   \mathsf{\mu}^{\mathrm X}& \mathsf{O} \\
   \mathsf{O} & \mathsf{\mu}^{\mathrm Y}
\end{pmatrix},
\label{blockdiag}
\end{align}
where $\mathsf{O}$ is the zero matrix of the appropriate size, $\mathsf{B}^{\mathrm X} (\in \mathbb{R}^{d^{\rm X} \times d^{\rm X} })$ and $\mathsf{B}^{\mathrm Y} (\in \mathbb{R}^{d^{\rm Y} \times d^{\rm Y} })$ are the square matrices, and $\mathsf{\mu}^{\mathrm X} := \mathsf{B}^{\mathrm X} (\mathsf{B}^{\mathrm X})^{\top} (\in \mathbb{R}^{d^{\rm X} \times d^{\rm X} })$ and $\mathsf{\mu}^{\mathrm Y} := \mathsf{B}^{\mathrm Y} (\mathsf{B}^{\mathrm Y})^{\top} (\in \mathbb{R}^{d^{\rm Y} \times d^{\rm Y} })$, which are the mobility matrices for $\rm X$ and $\rm Y$, respectively. In this paper, we assume this condition [Eq.~\eqref{blockdiag}], which corresponds to the noise acting on system $\rm X$ and the noise acting on system $\rm Y$ being uncorrelated in the Langevin equation [Eq.~\eqref{Langevineq}].

Under this condition, the Langevin equations of two subsystems ${\rm X}$ and ${\rm Y}$ are given by 
\begin{align}
    \dot{\boldsymbol{x}}(t) &= \mathsf{\mu}^{\mathrm X} \boldsymbol{F}^{\mathrm X}_t (\boldsymbol{x}(t), \boldsymbol{y}(t)) + \sqrt{2 T }\mathsf{B}^{\mathrm X} \boldsymbol{\xi}^{\mathrm X}_t , \label{LangevinX}\\
    \dot{\boldsymbol{y}}(t) &= \mathsf{\mu}^{\mathrm Y} \boldsymbol{F}^{\mathrm Y}_t (\boldsymbol{x}(t), \boldsymbol{y}(t)) + \sqrt{2 T} \mathsf{B}^{\mathrm Y}\boldsymbol{\xi}^{\mathrm Y}_t, \label{LangevinY}
\end{align}
where $\boldsymbol{F}^{\mathrm X}_t  (\boldsymbol{x}, \boldsymbol{y}) (\in \mathbb{R}^{d^{\rm X}})$, $\boldsymbol{F}^{\mathrm Y}_t  (\boldsymbol{x}, \boldsymbol{y}) (\in \mathbb{R}^{d^{\rm Y}})$, $\boldsymbol{\xi}^{\mathrm X}_t (\in \mathbb{R}^{d^{\rm X}})$ and $\boldsymbol{\xi}^{\mathrm Y}_t (\in \mathbb{R}^{d^{\rm Y}})$ are given by
\begin{align}
\boldsymbol{F}_t  (\boldsymbol{z}) = \begin{pmatrix}
   \boldsymbol{F}^{\mathrm X}_t  (\boldsymbol{x}, \boldsymbol{y}) \\
   \boldsymbol{F}^{\mathrm Y}_t (\boldsymbol{x}, \boldsymbol{y})
\end{pmatrix}, \: 
\boldsymbol{\xi}_t = \begin{pmatrix}
   \boldsymbol{\xi}^{\mathrm X}_t   \\
   \boldsymbol{\xi}^{\mathrm Y}_t 
\end{pmatrix}.
\end{align}
Here, $\boldsymbol{z}$ is set to $(\boldsymbol{x}, \boldsymbol{y})$ to clearly show that the argument depends on the states of the two subsystems. The white Gaussian noises $\boldsymbol{\xi}^{\rm X}_t$ and $\boldsymbol{\xi}^{\rm Y}_t$ satisfy $\mathbb{E}[ \boldsymbol{\xi}^{\rm X}_t (\boldsymbol{\xi}^{\rm X}_{t'})^{\top} ]= \delta(t-t') \mathsf{I}$, $\mathbb{E}[ \boldsymbol{\xi}^{\rm Y}_t (\boldsymbol{\xi}^{\rm Y}_{t'})^{\top} ]= \delta(t-t') \mathsf{I}$, and $\mathbb{E}[ \boldsymbol{\xi}^{\rm X}_t (\boldsymbol{\xi}^{\rm Y}_{t'})^{\top} ]= \mathsf{O}$.

As an expression focusing on the decomposition of the system, the Fokker-Planck equation [Eq.~\eqref{continuityeq}] is also rewritten as $\partial_t p_t(\boldsymbol{x}, \boldsymbol{y}) = - \nabla_{\boldsymbol{x}} \cdot \boldsymbol{j}_t^{\rm X} (\boldsymbol{x}, \boldsymbol{y}) -\nabla_{\boldsymbol{y}} \cdot \boldsymbol{j}_t^{\rm Y} (\boldsymbol{x}, \boldsymbol{y})$, where $\nabla_{\boldsymbol{x}}$ ($\nabla_{\boldsymbol{y}}$) is the del operator for system $\rm X$ ($\rm Y$) and $\boldsymbol{j}_t^{\rm X} (\boldsymbol{x}, \boldsymbol{y})$ ($\boldsymbol{j}_t^{\rm Y} (\boldsymbol{x}, \boldsymbol{y})$) is the flux for system $\rm X$ ($\rm Y$). Here, $\nabla_{\boldsymbol{x}}$ ($\nabla_{\boldsymbol{y}}$) acts only on $\boldsymbol{x}$ ($\boldsymbol{y}$), while $\nabla$ acts on $\boldsymbol{z}=(\boldsymbol{x},\boldsymbol{y})$.
The fluxes are given by $\boldsymbol{j}^{\rm X}_t (\boldsymbol{x},\boldsymbol{y}) := p_t(\boldsymbol{x},\boldsymbol{y}) \mathsf{D}^{\rm X}\boldsymbol{f}^{\rm X}_t(\boldsymbol{x},\boldsymbol{y})$ and $\boldsymbol{j}^{\rm Y}_t (\boldsymbol{x},\boldsymbol{y}) := p_t(\boldsymbol{x},\boldsymbol{y}) \mathsf{D}^{\rm Y}\boldsymbol{f}^{\rm Y}_t(\boldsymbol{x},\boldsymbol{y})$, where $\mathsf{D}^{\rm X}=T\mu^{\rm X}$, $\mathsf{D}^{\rm Y}=T\mu^{\rm Y}$, and the thermodynamic forces for $\rm X$ and $\rm Y$ are defined as
\begin{align}
\boldsymbol{f}^{\rm X}_t (\boldsymbol{x}, \boldsymbol{y}) &:= \frac{\boldsymbol{F}^{\rm X}_t (\boldsymbol{x}, \boldsymbol{y})}{T} -   \nabla_{\boldsymbol{x}}\ln p_t (\boldsymbol{x}, \boldsymbol{y}), \nonumber\\
\boldsymbol{f}^{\rm Y}_t (\boldsymbol{x}, \boldsymbol{y}) &:= \frac{\boldsymbol{F}^{\rm Y}_t (\boldsymbol{x}, \boldsymbol{y})}{T} -  \nabla_{\boldsymbol{y}} \ln p_t (\boldsymbol{x}, \boldsymbol{y}).
\label{thermodynamicforce}
\end{align}
Furthermore, if we define the velocity fields for $\rm X$ and $\rm Y$ as $\boldsymbol{\nu}^{\rm X}_t (\boldsymbol{x}, \boldsymbol{y}) := \mathsf{D}^{\rm X} \boldsymbol{f}^{\rm X}_t (\boldsymbol{x}, \boldsymbol{y})$ and $\boldsymbol{\nu}^{\rm Y}_t (\boldsymbol{x}, \boldsymbol{y}) := \mathsf{D}^{\rm Y} \boldsymbol{f}^{\rm Y}_t (\boldsymbol{x}, \boldsymbol{y})$, respectively, the Fokker-Planck equation [Eq.~\eqref{continuityeq}] can also be expressed as 
$\partial_t p_t(\boldsymbol{x}, \boldsymbol{y}) = -\nabla_{\boldsymbol{x}} \cdot (\boldsymbol{\nu}^{\rm X}_t (\boldsymbol{x}, \boldsymbol{y}) p_t(\boldsymbol{x}, \boldsymbol{y})) -\nabla_{\boldsymbol{y}} \cdot ( \boldsymbol{\nu}^{\rm Y}_t (\boldsymbol{x}, \boldsymbol{y})p_t(\boldsymbol{x}, \boldsymbol{y}))$.

\subsection{Entropy production rate}
\label{sec2b}
Before discussing information thermodynamics, let us first discuss the entropy production rate for the total system~\cite{de2013non,seifert2025stochastic}. We define the entropy production rate as
\begin{align}
\dot{\Sigma}_t &:= \int d\boldsymbol{z} \boldsymbol{f}_t (\boldsymbol{z}) \cdot \boldsymbol{j}_t (\boldsymbol{z}).
\end{align}
If we define the inner product between $\boldsymbol{a}(\boldsymbol{z})$ and $\boldsymbol{b}(\boldsymbol{z})$ with the metric $\mathsf{C}(\boldsymbol{z})$ as 
$\langle \boldsymbol{a}, \boldsymbol{b} \rangle_{\mathsf{C}} := \int d\boldsymbol{z} [\boldsymbol{a} (\boldsymbol{z}) ]^{\top} \mathsf{C} ( \boldsymbol{z}) \boldsymbol{b}(\boldsymbol{z})$, the entropy production rate is rewritten as
\begin{align}
\dot{\Sigma}_t &= \langle \boldsymbol{f}_t, \boldsymbol{f}_t\rangle_{p_t \mathsf{D} }.
\end{align}
Because the entropy production rate is the squared norm of the thermodynamic force when the Onsager matrix is used as the metric, the entropy production rate should be nonnegative, i.e., $\dot{\Sigma}_t= \langle \boldsymbol{f}_t, \boldsymbol{f}_t\rangle_{p_t \mathsf{D} } \geq 0$. This nonnegativity of the entropy production rate implies the second law of thermodynamics. 

The entropy production rate is given by the sum of  the entropy change rate of the total system $d_t S^{\rm sys}_t$ and the entropy change rate of the bath  $\dot{S}^{\rm bath}_t$ in contact with the total system, i.e., $\dot{\Sigma}_t = d_t S^{\rm sys}_t+\dot{S}^{\rm bath}_t $. Here, $d_t:=d/dt$ stands for the time derivative.  Thus, the second law of thermodynamics can be rewritten as
\begin{align}
\dot{\Sigma}_t=  d_t S^{\rm sys}_t+\dot{S}^{\rm bath}_t  \geq 0.
\label{secondlaw}
\end{align}
The entropy change rate of the total system is defined as $d_t S^{\rm sys}_t := \langle \boldsymbol{f}_t, -\nabla\ln p_t  \rangle_{p_t\mathsf{D}}$. This entropy $S^{\rm sys}_t$ can be regarded as the differential entropy $H_t := - \int d\boldsymbol{z} p_t(\boldsymbol{z})\ln p_t(\boldsymbol{z})$ as follows:
\begin{align}
d_t S^{\rm sys}_t &= \int d\boldsymbol{z} \nabla
 \cdot (\mathsf{D} p_t(\boldsymbol{z}) \boldsymbol{f}_t(\boldsymbol{z}))\ln p_t(\boldsymbol{z}) \nonumber\\
 &= -\int d\boldsymbol{z} [\partial_t p_t(\boldsymbol{z})] \ln p_t(\boldsymbol{z}) = d_t H_t.
\end{align}
Here, we used $\int d\boldsymbol{z} \partial_t p_t(\boldsymbol{z})=0$ and performed integration by parts, assuming boundary terms vanish. Throughout this paper, we make the assumption that the decay of $p_t(\boldsymbol{z})$ is sufficiently fast so that all boundary terms vanish. The entropy change rate of the bath $\dot{S}^{\rm bath}_t$ is defined as $\dot{S}^{\rm bath}_t:=(1/T)\int d\boldsymbol{z} \boldsymbol{j}_t(\boldsymbol{z} ) \cdot \boldsymbol{F}_t (\boldsymbol{z} )= \langle \boldsymbol{f}_t,\boldsymbol{F}_t/T \rangle_{p_t\mathsf{D}}$. Here, $\langle \boldsymbol{f}_t,\boldsymbol{F}_t \rangle_{p_t \mathsf{D} }=\int d\boldsymbol{z} \boldsymbol{j}_t(\boldsymbol{z} ) \cdot \boldsymbol{F}_t (\boldsymbol{z})$ can be regarded as heat flux from the total system. In particular, when $\boldsymbol{F}_t (\boldsymbol{z})$ is a conservative force and is expressed as $\boldsymbol{F}_t (\boldsymbol{z})=-\nabla U_t(\boldsymbol{z})$ in terms of the potential energy $U_t(\boldsymbol{z})$, heat flux is given by $\dot{Q}_t := -\langle \boldsymbol{f}_t,-\nabla U_t\rangle_{p_t \mathsf{D} } =\int d\boldsymbol{z}  U_t(\boldsymbol{z}) [\partial_t p_t (\boldsymbol{z})]$ and the following first law of thermodynamics~\cite{sekimoto2010stochastic} 
\begin{align}
 d_t \mathbb{E}_{p_t}[U_t]&:= \dot{W}_t +\dot{Q}_t,
\end{align}
holds, where $\mathbb{E}_{p_t}[U_t]:= \int d\boldsymbol{z} U_t(\boldsymbol{z}) p_t (\boldsymbol{z})$ is the expected value of the potential, and $\dot{W}_t:=\int d\boldsymbol{z} [\partial_t U_t(\boldsymbol{z})] p_t (\boldsymbol{z})$ is the work done on the total system. Therefore, if the force is given by the potential force $\boldsymbol{F}_t (\boldsymbol{z})=-\nabla U_t(\boldsymbol{z})$, the entropy production rate can be rewritten as
\begin{align}
\dot{\Sigma}_t=  d_t S^{\rm sys}_t - \frac{\dot{Q}_t}{T} \: (\geq 0).
\label{clausiusheat}
\end{align}
This expression corresponds to the Clausius heat
theorem in classical thermodynamics.

In the steady state $p_t(\boldsymbol{z})=p^{\rm st} (\boldsymbol{z})$ that satisfies $\left. \partial_t p_t(\boldsymbol{z}) \right|_{p_t=p^{\rm st}}=\left. - \nabla \cdot \boldsymbol{j}_t (\boldsymbol{z})\right|_{p_t=p^{\rm st}} =0$, the entropy change rate of the system becomes zero, i.e., $\left. d_t S^{\rm sys}_t\right|_{p_t=p^{\rm st}} =0$. This means that the steady-state entropy production rate $\dot{\Sigma}_t^{\rm st}:=\left. \dot{\Sigma}_t \right|_{p_t=p^{\rm st}}$ is only given by the entropy change rate of the bath $\dot{\Sigma}_t^{\rm st}= \left. \dot{S}^{\rm bath}_t \right|_{p_t=p^{\rm st}}$. If the force is given by a potential force $\boldsymbol{F}_t (\boldsymbol{z})=-\nabla U_t(\boldsymbol{z})$, the system is conservative. For the conservative system, the steady-state entropy production rate becomes zero ($\dot{\Sigma}_t^{\rm st}=0$), and this steady state can be regarded as the equilibrium state.
\subsection{Information thermodynamics}
\label{sec2c}
In information thermodynamics~\cite{parrondo2015thermodynamics}, stochastic thermodynamics is considered for each subsystem. In this case, the apparent difference from stochastic thermodynamics for the total system can be captured by the change in mutual information, that is, by the flow of information. 

Here, we introduce the concept of information flow. We first consider the mutual information between system $\rm X$ and system $\rm Y$, 
\begin{align}
I(\hat{\rm X}_t; \hat{\rm Y}_t):= \int d\boldsymbol{x} \int d\boldsymbol{y} p_t(\boldsymbol{x},\boldsymbol{y})i_t(\boldsymbol{x},\boldsymbol{y}),
\end{align}
where the stochastic mutual information $i_t(\boldsymbol{x},\boldsymbol{y})$ is defined as
\begin{align}
i_t(\boldsymbol{x},\boldsymbol{y}):=\ln \frac{p_t(\boldsymbol{x},\boldsymbol{y})} {p^{\rm X}_t(\boldsymbol{x})p^{\rm Y}_t(\boldsymbol{y})}.
\end{align}
Here, $p^{\rm X}_t(\boldsymbol{x}):= \int d\boldsymbol{y}p_t (\boldsymbol{x},\boldsymbol{y})$ and $p^{\rm Y}_t(\boldsymbol{y}):= \int d\boldsymbol{x} p_t(\boldsymbol{x},\boldsymbol{y})$ are marginal distributions. The symbol $\hat{\rm X}_t$ ($\hat{\rm Y}_t$) stands for the random variable for the state of the system $\rm X$ ($\rm Y$) at time $t$. The time derivative of mutual information can be decomposed into two information flows~\cite{Allahverdyan_2009,horowitz2014second,nakazato2021geometrical} defined as
\begin{align}
\dot{I}^{\rm X}_t :=& \int d\boldsymbol{x} \int d\boldsymbol{y} [-\nabla_{\boldsymbol{x}} \cdot \boldsymbol{j}^{\rm X}_t (\boldsymbol{x}, \boldsymbol{y})] i_t(\boldsymbol{x},\boldsymbol{y}), \nonumber\\
\dot{I}^{\rm Y}_t :=& \int d\boldsymbol{x} \int d\boldsymbol{y} [-\nabla_{\boldsymbol{y}} \cdot \boldsymbol{j}^{\rm Y}_t (\boldsymbol{x}, \boldsymbol{y})] i_t(\boldsymbol{x},\boldsymbol{y}).
\label{informationflow}
\end{align}
Because $\partial_t p_t(\boldsymbol{x}, \boldsymbol{y}) = - \nabla_{\boldsymbol{x}} \cdot \boldsymbol{j}_t^{\rm X} (\boldsymbol{x}, \boldsymbol{y}) -\nabla_{\boldsymbol{y}} \cdot \boldsymbol{j}_t^{\rm Y} (\boldsymbol{x}, \boldsymbol{y})$ and $\int d\boldsymbol{x} \int d\boldsymbol{y}\partial_t p_t(\boldsymbol{x},\boldsymbol{y} )=\int d\boldsymbol{x} \partial_t p^{\rm X}_t(\boldsymbol{x})=\int d\boldsymbol{y} \partial_t p^{\rm Y}_t(\boldsymbol{y})=0$, we obtain the decomposition
\begin{align}
d_t I(\hat{\rm X}_t; \hat{\rm Y}_t)= \dot{I}^{\rm X}_t + \dot{I}^{\rm Y}_t.
\label{decompositioninfo}
\end{align}
These information flows can also be regarded as partial derivatives of mutual information, i.e., $\dot{I}^{\rm X}_t = \left. \partial_s I(\hat{\rm X}_s; \hat{\rm Y}_t) \right|_{s=t}$ and $\dot{I}^{\rm Y}_t = \left. \partial_s I(\hat{\rm X}_t; \hat{\rm Y}_s) \right|_{s=t}$. These information flows appear in the second law of information thermodynamics, which corresponds to the second law of thermodynamics [Eq.~\eqref{secondlaw}]. 

We note that this information flow can be expressed in the same way as the representation in Ref.~\cite{ito2013}, which includes  the transfer entropy~\cite{Schreiber2000}. This representation can also be rewritten as a contribution involving simple simultaneous mutual information, the transfer entropy, and the backward transfer entropy (see Refs.~\cite{ito2015maxwell,ito2016backward}).

Interestingly, information flow does not vanish even in the steady state. On the other hand, $\left. d_t I(\hat{\rm X}_t; \hat{\rm Y}_t) \right|_{p_t=p^{\rm st}}$ becomes zero in the steady state.
From Eq.~\eqref{decompositioninfo}, we obtain an antisymmetric relation in the steady state as follows:
\begin{align}
\left. \dot{I}^{\rm X}_t \right|_{p_t=p^{\rm st}}=- \left.\dot{I}^{\rm Y}_t \right|_{p_t=p^{\rm st}}.
\label{antisymmetric}
\end{align}

To discuss the role of information flow, we start with the definition of the partial entropy production rate. Since the system is bipartite and $\mathsf{D}=T\mu$ is a block diagonal matrix, the total entropy production rate $\dot{\Sigma}_t$ can be decomposed into the partial entropy production rates of the subsystems $\dot{\Sigma}_t^{\rm X} :=\langle \boldsymbol{f}^{\rm X}_t, \boldsymbol{f}^{\rm X}_t\rangle_{p_t \mathsf{D}^{\rm X}}$ and $\dot{\Sigma}_t^{\rm Y} :=\langle \boldsymbol{f}^{\rm Y}_t, \boldsymbol{f}^{\rm Y}_t\rangle_{p_t \mathsf{D}^{\rm Y}}$ as follows,
\begin{align}
 \dot{\Sigma}_t= \dot{\Sigma}_t^{\rm X}+\dot{\Sigma}_t^{\rm Y}.
\end{align}
Here, both partial entropy production rates are nonnegative, i.e., $\dot{\Sigma}_t^{\rm Y} \geq 0$ and $\dot{\Sigma}_t^{\rm X} \geq 0$. Their nonnegativity corresponds to the second law of information thermodynamics, which explains the role of information flow.

To discuss the second law of information thermodynamics, we consider the decomposition of the partial entropy production rate as follows,
\begin{align}
 \dot{\Sigma}_t^{\rm X}= d_t S^{\rm sys;X}_t + \dot{S}^{\rm bath;X}_t - \dot{I}^{\rm X}_t,
 \label{decomppartialepr}
\end{align}
where $d_t S^{\rm sys;X}_t$ is the entropy change rate of system $\rm X$ and $\dot{S}^{\rm bath;X}_t$ is the entropy change rate of the bath in contact with system $\rm X$. We present the results for subsystem $\rm X$; those for $\rm Y$ follow analogously. The entropy change rate of system $\rm X$, $d_t S^{\rm sys;X}_t$, is defined as $d_t S^{\rm sys;X}_t:=d_t H_t^{\rm X}$, where $H_t^{\rm X}= -\int d\boldsymbol{x} p^{\rm X}_t(\boldsymbol{x}) \ln p^{\rm X}_t(\boldsymbol{x})$ is the differential entropy for system $\rm X$. The entropy change rate of the bath $\dot{S}^{\rm bath;X}_t$ is defined as $\dot{S}^{\rm bath;X}_t= (1/T) \int d\boldsymbol{x}\int d\boldsymbol{y} \boldsymbol{j}^{\rm X}_t (\boldsymbol{x},\boldsymbol{y}) \cdot \boldsymbol{F}^{\rm X}_t (\boldsymbol{x},\boldsymbol{y}) = \langle \boldsymbol{f}^{\rm X}_t , \boldsymbol{F}^{\rm X}_t/T \rangle_{ p_t\mathsf{D}^{\rm X} }$. Here, the
term $- \dot{Q}^{\rm X}_t:=\langle \boldsymbol{f}^{\rm X}_t , \boldsymbol{F}^{\rm X}_t\rangle_{ p_t\mathsf{D}^{\rm X}}$ can be regarded as heat flux from system $\rm X$. Similarly, if heat flux from system {\rm Y} is defined as $- \dot{Q}^{\rm Y}_t:=\langle \boldsymbol{f}^{\rm Y}_t , \boldsymbol{F}^{\rm Y}_t\rangle_{ p_t\mathsf{D}^{\rm Y}}$, then additivity $\dot{Q}_t =\dot{Q}^{\rm X}_t+\dot{Q}^{\rm Y}_t$ holds. Therefore, 
\begin{align}
\dot{S}^{\rm bath}_t =\dot{S}^{\rm bath;X}_t+\dot{S}^{\rm bath;Y}_t
\end{align}
also holds, where $\dot{S}^{\rm bath;Y}_t:= -\dot{Q}^{\rm Y}_t/T$. On the other hand, regarding the entropy change rate of the system, additivity holds when information flows are included as follows
\begin{align}
d_t S^{\rm sys}_t=d_t S^{\rm sys;X}_t  - \dot{I}^{\rm X}_t +d_t S^{\rm sys;Y}_t  - \dot{I}^{\rm Y}_t,
\label{mutualinfo}
\end{align}
where $d_t S^{\rm sys;Y}_t= d_t H^{\rm Y}_t$ and the differential entropy $H_t^{\rm Y}$ is defined as $H_t^{\rm Y}= -\int d\boldsymbol{y} p^{\rm Y}_t(\boldsymbol{y}) \ln p^{\rm Y}_t(\boldsymbol{y})$. To derive Eq.~\eqref{mutualinfo}, we use $\dot{I}^{\rm X}_t + \dot{I}^{\rm Y}_t = d_t I(\hat{\rm X}_t; \hat{\rm Y}_t) = d_t H_t^{\rm X}+d_t H_t^{\rm Y} -d_t H_t$.

To confirm Eq.~\eqref{decomppartialepr}, we rewrite information flow [Eq.~\eqref{informationflow}] as $\dot{I}^{\rm X}_t= \langle \boldsymbol{f}^{\rm X}_t,\nabla_{\boldsymbol{x}} i_t \rangle_{ p_t \mathsf{D}^{\rm X}}$, where we performed integration by parts. We also rewrite $d_t S^{\rm sys;X}_t$ as 
\begin{align}
&d_t S^{\rm sys;X}_t \nonumber\\
&= -\int d\boldsymbol{x} \int d\boldsymbol{y} [\partial_t p_t (\boldsymbol{x},\boldsymbol{y})] \ln  p^{\rm X}_t (\boldsymbol{x}) \nonumber\\
&=  \int d\boldsymbol{x} \int d\boldsymbol{y} \nabla_{\boldsymbol{x}} \cdot(p_t(\boldsymbol{x},\boldsymbol{y}) \mathsf{D}^{\rm X} \boldsymbol{f}_t^{\rm X}(\boldsymbol{x},\boldsymbol{y}) ) \ln  p^{\rm X}_t (\boldsymbol{x}) \nonumber\\
&= \langle \boldsymbol{f}_t^{\rm X}, -\nabla_{\boldsymbol{x}}\ln  p^{\rm X}_t \rangle_{p_t \mathsf{D}^{\rm X}},
\end{align}
where we used $\int d\boldsymbol{y} p_t(\boldsymbol{x}, \boldsymbol{y})=p^{\rm X}_t(\boldsymbol{x})$, $\int d\boldsymbol{x} \partial_t p_t^{\rm X}(\boldsymbol{x})=0$ and performed integration by parts.
Using these expressions, Eq.~\eqref{thermodynamicforce} and $\nabla_{\boldsymbol{x}} (i_t(\boldsymbol{x},\boldsymbol{y}) + \ln p^{\rm X}_t(\boldsymbol{x})) = \nabla_{\boldsymbol{x}} \ln p_t (\boldsymbol{x},\boldsymbol{y})$, the partial entropy production rate is calculated as
\begin{align}
\dot{\Sigma}_t^{\rm X} &= \langle \boldsymbol{f}^{\rm X}_t, \boldsymbol{f}^{\rm X}_t\rangle_{p_t \mathsf{D}^{\rm X}} \nonumber\\
&=\langle \boldsymbol{f}^{\rm X}_t, \boldsymbol{F}^{\rm X}_t/T\rangle_{p_t \mathsf{D}^{\rm X}} - \langle \boldsymbol{f}^{\rm X}_t, \nabla_{\boldsymbol{x}} (i_t +\ln p^{\rm X}_t)\rangle_{p_t \mathsf{D}^{\rm X}} \nonumber\\
&=  \dot{S}^{\rm bath;X}_t - \dot{I}^{\rm X}_t +d_t S^{\rm sys;X}_t,
\end{align}
which is equivalent to Eq.~\eqref{decomppartialepr}. 

The nonnegativity of the partial entropy production rates is regarded as the second law of information thermodynamics. Similarly to the entropy production rate for the total system, we introduce the entropy change rate of the subsystem and the bath as follows:
\begin{align}
 \sigma_t^{\rm X} &= d_t S^{\rm sys;X}_t + \dot{S}^{\rm bath;X}_t, \nonumber \\
  \sigma_t^{\rm Y}&= d_t S^{\rm sys;Y}_t + \dot{S}^{\rm bath;Y}_t.
  \label{apparentep}
\end{align}
Since Eq.~\eqref{decomppartialepr} and the corresponding result for system {\rm Y} hold, the nonnegativity of the partial entropy production rates $\dot{\Sigma}_t^{\rm X} ( \geq 0)$ and $\dot{\Sigma}_t^{\rm Y} ( \geq 0)$ leads to the inequalities
\begin{align}
  \sigma_t^{\rm X} &\geq \dot{I}^{\rm X}_t, \nonumber \\
  \sigma_t^{\rm Y} &\geq \dot{I}^{\rm Y}_t,
\label{2ndlawinfothermo}
\end{align}
which are called the second law of information thermodynamics~\cite{parrondo2015thermodynamics}. While the total entropy production rate and the partial entropy production rates should be nonnegative, the changes in entropy of the subsystems $\sigma_t^{\rm X}$ and $\sigma_t^{\rm Y}$ can be negative. This negativity is regarded as the apparent violation of the second law of thermodynamics, and this effect is historically well discussed in terms of Maxwell's demon. According to the second law of information thermodynamics [Eq.~\eqref{2ndlawinfothermo}], this apparent violation must be compensated for by a negative value of the information flow, which is considered a partial change in the correlation between the two subsystems.

In the steady state, $\left. \sigma_t^{\rm X} \right|_{p_t=p^{\rm st}} = \left. \dot{S}^{\rm bath;X}_t \right|_{p_t=p^{\rm st}}$ holds, and this contribution itself has the same form as the steady-state entropy production rate $\dot{\Sigma}_t^{\rm st}= \left. \dot{S}^{\rm bath}_t \right|_{p_t=p^{\rm st}}$ in the total system. On the other hand, the second law of thermodynamics for the steady state still apparently seems to be violated unless the contribution of information flow $\left. \dot{I}^{\rm X}_t \right|_{p_t=p^{\rm st}}$ is taken into account, i.e., $\left. \dot{S}^{\rm bath;X}_t \right|_{p_t=p^{\rm st}} \geq \left. \dot{I}^{\rm X}_t \right|_{p_t=p^{\rm st}}$. This implies that, in steady state, heat can flow in and out of the subsystem and that information flow drives heat flow. Furthermore, due to the antisymmetric relation [Eq.~\eqref{antisymmetric}], if the information flow is negative in one subsystem, the information flow in the other subsystem must be positive in the steady state. That is, only one subsystem can exhibit a negative apparent entropy change rate, and the system with a positive entropy change rate can be regarded as Maxwell's demon that pumps entropy from the other subsystem. By contrast, if we consider the transient dynamics, the apparent entropy change rate of both subsystems can become negative due to information flow. This creates a situation in which both subsystems can be regarded as Maxwell's demons during the transient dynamics.

As discussed in Ref.~\cite{matsumoto2025learning}, we can obtain a relationship between the conditional Fisher information and information flow. The expression of the information flow $\dot{I}_t^{\rm X}= \langle \boldsymbol{f}^{\rm X}_t,\nabla_{\boldsymbol{x}} i_t \rangle_{ p_t \mathsf{D}^{\rm X}}$ leads to the Cauchy-Schwarz inequality $(\langle \boldsymbol{f}^{\rm X}_t,\nabla_{\boldsymbol{x}} i_t \rangle_{ p_t \mathsf{D}^{\rm X}})^2 \leq \langle \boldsymbol{f}^{\rm X}_t,\boldsymbol{f}^{\rm X}_t \rangle_{ p_t \mathsf{D}^{\rm X}} \langle \nabla_{\boldsymbol{x}} i_t,\nabla_{\boldsymbol{x}} i_t \rangle_{ p_t \mathsf{D}^{\rm X}}$, or equivalently,
\begin{align}
(\dot{I}_t^{\rm X})^2 \leq \dot{\Sigma}_t^{\rm X} \mathcal{I}_t^{\rm Fisher;X}, 
\label{Fisherbound}
\end{align}
where $\mathcal{I}_t^{\rm Fisher;X}$ is defined as
\begin{align}
    \mathcal{I}_t^{\rm Fisher;X} = \langle \nabla_{\bs{x}} i_t, \nabla_{\bs{x}} i_t \rangle_{p_t \mathsf{D}^{\rm X}} = \text{tr} \big( \mathsf{D}^{\rm X} \mathsf{F}_t^{\rm Y \vert X} \big), 
\end{align}
with the conditional Fisher information matrix
\begin{align}    
    \big(\mathsf{F}_t^{\rm Y \vert X} \big)_{jk} := \int d\bs{x} \int d\bs{y} \ &\big(\partial_{x_j} \ln (p_t^{\rm Y \vert X}(\bs{y} \vert \bs{x}) \big) \nonumber\\
    &\times \big(\partial_{x_k} \ln (p_t^{\rm Y \vert X}(\bs{y} \vert \bs{x}) \big) p_t(\bs{x},\bs{y}).
\end{align}
Here, $p_t^{\rm Y \vert X}(\bs{y} \vert \bs{x})$ is the conditional probability defined as $ p_t^{\rm Y \vert X}(\bs{y} \vert \bs{x}):= p_t(\bs{x}, \bs{y}) /p^{\rm X}_t(\bs{x})$ and ${\rm tr}(\mathsf{A})$ is the trace of the matrix $\mathsf{A}$, and we used $\nabla_{\bs{x}} i_t(\bs{x},\bs{y}) = \nabla_{\bs{x}} \ln  p_t^{\rm Y \vert X}(\bs{y} \vert \bs{x})$.

The conditional Fisher information matrix $\mathsf{F}_t^{\rm Y \vert X}$~\cite{cover1999elements} measures the sensitivity of the state of ${\rm Y}$ with respect to changes in ${\rm X}$. The quantity $\mathcal{I}_t^{\rm Fisher;X}$ quantifies short-time fluctuations
of the stochastic mutual information induced by changes in $\rm X$.
We now consider the variance of $\Delta i_t (\boldsymbol{z} (t+dt), \boldsymbol{z} (t)):= i_t(\boldsymbol{x} (t+dt) ,\boldsymbol{y} (t+dt))- i_t(\boldsymbol{x} (t),\boldsymbol{y} (t)) \simeq d\boldsymbol{x} \circ \nabla_{\boldsymbol{x}} i_t(\boldsymbol{x} (t),\boldsymbol{y} (t))+d\boldsymbol{y} \circ \nabla_{\boldsymbol{y}} i_t(\boldsymbol{x} (t),\boldsymbol{y} (t))$, where $\circ$ stands for the Stratonovich discretization. Under the assumption that the diffusion matrix $\mathsf{D}$ is a block diagonal matrix [Eq.~\eqref{blockdiag}], the variance ${\rm Var}[\Delta i_t]$ to leading order in $dt$ is calculated as 
\begin{align}
    &{\rm Var}[\Delta i_t] \nonumber\\
    \simeq& {\rm Var}[\sqrt{2 T} \mathsf{B}^{\rm X}\boldsymbol{\xi}^{\rm X}_t dt \circ \nabla_{\boldsymbol{x}} i_t + \sqrt{2 T} \mathsf{B}^{\rm Y}\boldsymbol{\xi}^{\rm Y}_t dt \circ \nabla_{\boldsymbol{y}} i_t] \nonumber\\
\simeq & 2 dt\int d\boldsymbol{x} \int d\boldsymbol{y}  [\nabla_{\boldsymbol{x}} i_t(\boldsymbol{x},\boldsymbol{y})]^{\top} \mathsf{D}^{\rm X}[\nabla_{\boldsymbol{x}} i_t(\boldsymbol{x},\boldsymbol{y} ) ] p_t(\boldsymbol{x},\boldsymbol{y}) \nonumber\\
&+2 dt\int d\boldsymbol{x} \int d\boldsymbol{y}  [\nabla_{\boldsymbol{y}} i_t(\boldsymbol{x},\boldsymbol{y})]^{\top} \mathsf{D}^{\rm Y}[\nabla_{\boldsymbol{y}} i_t(\boldsymbol{x},\boldsymbol{y} ) ]  p_t(\boldsymbol{x},\boldsymbol{y}) \nonumber \\
=& 2 dt \Big(\langle \nabla_{\boldsymbol{x}} i_t, \nabla_{\boldsymbol{x}} i_t\rangle_{p_t \mathsf{D}^{\rm X}} + \langle \nabla_{\boldsymbol{y}} i_t, \nabla_{\boldsymbol{y}} i_t\rangle_{p_t \mathsf{D}^{\rm Y}} \Big) \nonumber \\
=&2dt( \mathcal{I}_t^{\rm Fisher;X} + \mathcal{I}_t^{\rm Fisher;Y} ),
\label{changestomuinfo}
\end{align}
where we used Ito calculus $\mathbb{E}[ (\boldsymbol{\xi}^{\rm Y}_t dt) (\boldsymbol{\xi}^{\rm Y}_t dt)^{\top}]\simeq \mathsf{I} dt$, $\mathbb{E}[ (\boldsymbol{\xi}^{\rm X}_t dt) (\boldsymbol{\xi}^{\rm X}_t dt)^{\top}]\simeq \mathsf{I} dt$, $\mathbb{E}[ (\boldsymbol{\xi}^{\rm X}_t dt) (\boldsymbol{\xi}^{\rm Y}_t dt)^{\top}]\simeq \mathsf{O}$,  $T\mathsf{B}^{\rm X} (\mathsf{B}^{\rm X})^{\top} =\mathsf{D}^{\rm X}$ and $T \mathsf{B}^{\rm Y}(\mathsf{B}^{\rm Y})^{\top} =\mathsf{D}^{\rm Y}$. Here, $\mathcal{I}_t^{\rm Fisher;Y}$ is also defined as $\mathcal{I}_t^{\rm Fisher;Y}:=\langle \nabla_{\boldsymbol{y}} i_t, \nabla_{\boldsymbol{y}} i_t\rangle_{p_t \mathsf{D}^{\rm Y}}$.
This expression means that the quantity $\mathcal{I}_t^{\rm Fisher;X}$ can be regarded as short-time fluctuations of the stochastic mutual information due to changes in $\rm X$. 

For the entropy change rate of the subsystem [Eq.~\eqref{apparentep}], we have
\begin{align}
    \sigma_t^{\rm X} = \dot{\Sigma}_t^{ X} + \dot{I}_t^{\rm X} \geq \frac{(\dot{I}_t^{\rm X})^2}{\mathcal{I}_t^{\rm Fisher;X}} + \dot{I}_t^{\rm X}, 
    \label{infothermo2nd_quadratic}
\end{align}
which provides a quadratic lower bound on the apparent violation of the second law in terms of information flow.
Interestingly, we see that an apparent negative entropy change rate can only be observed for $-\mathcal{I}_t^{\rm Fisher;X} < \dot{I}_t^{\rm X} < 0$.
While the second law of information thermodynamics [Eq.~\eqref{2ndlawinfothermo}] suggests that any negative information flow $\dot{I}_t^{\rm X}$ can lead to an apparently negative entropy change rate of subsystem $\rm X$, we instead find that an excessively large  information flow $-\dot{I}_t^{\rm X} \geq \mathcal{I}_t^{\rm Fisher;X}$ essentially prohibits any apparent second law violations.
In that sense, the influence $\mathcal{I}_t^{\rm Fisher;X}$ measures the maximal information flow that can be sustained by the system while beating the second law.
Moreover, minimizing the right-hand side of Eq.~\eqref{infothermo2nd_quadratic} with respect to the information flow $\dot{I}_t^{\rm X}$, we obtain the global lower bound
\begin{align}
    \sigma_t^{\rm X} \geq \frac{\Big( \dot{I}_t^{\rm X} + \frac{1}{2} \mathcal{I}_t^{\rm Fisher;X} \Big)^2}{\mathcal{I}_t^{\rm Fisher;X}}  - \frac{1}{4} \mathcal{I}_t^{\rm Fisher;X}\geq - \frac{1}{4} \mathcal{I}_t^{\rm Fisher;X} .
    \label{globallowerboundfisher}
\end{align}
Thus, the influence of $\rm X$ on the state of $\rm Y$, quantified in terms of $\mathcal{I}_t^{\rm Fisher;X}$, bounds the negativity of the apparent entropy change rate of $\rm X$.
We remark that, while the information flow explicitly depends on the dynamics via the thermodynamic forces, $\mathcal{I}_t^{\rm Fisher;X}$ only depends on the instantaneous probability density and the diffusion matrix.

\subsection{Geometric decomposition}
\label{subsec:geometricdecomposition}
Next, we consider a geometric decomposition of the entropy production rate into conservative and nonconservative contributions~\cite{nakazato2021geometrical,dechant2022geometric,ito2024geometric}. These contributions are called the excess and housekeeping entropy production rates, respectively. There are several ways to define these entropy production rates. Here, we consider a decomposition based on a geometric projection onto the conservative space.

We first consider the function $\phi_t(\boldsymbol{z})$ that satisfies the following equation
\begin{align}
- \nabla \cdot [p_t(\boldsymbol{z})\mathsf{D} \boldsymbol{f}_t(\boldsymbol{z}) ]= \nabla \cdot [p_t(\boldsymbol{z})\mathsf{D}(\nabla \phi_t (\boldsymbol{z})  )].
\label{geometricpotential}
\end{align}
If the thermodynamic force $\boldsymbol{f}_t(\boldsymbol{z})$ is conservative, $\boldsymbol{f}_t(\boldsymbol{z})=- \nabla \phi_t(\boldsymbol{z})$ holds. If the thermodynamic force is nonconservative, $\boldsymbol{f}_t(\boldsymbol{z}) \neq - \nabla \phi_t(\boldsymbol{z})$ and its difference $\boldsymbol{f}_t(\boldsymbol{z}) - [- \nabla \phi_t(\boldsymbol{z})]$ can be considered a nonconservative contribution to the thermodynamic force. We note that combining the Fokker-Planck equation [Eq.~\eqref{continuityeq}] with Eq.~\eqref{geometricpotential}, we obtain 
\begin{align}
\partial_t p_t(\boldsymbol{z})= -\nabla \cdot [p_t(\boldsymbol{z})\mathsf{D}(-\nabla \phi_t (\boldsymbol{z})  )].
\label{continuityoptimal}
\end{align}
This equation expresses the time evolution of the distribution $p_t(\boldsymbol{z})$ as driven by a conservative thermodynamic force $-\nabla \phi_t (\boldsymbol{z})$. Therefore, this quantity $-\nabla \phi_t (\boldsymbol{z})$ represents the time evolution of the distribution.

Although this function $\phi_t(\boldsymbol{z})$ allows for the addition of constants as a solution to Eq.~\eqref{geometricpotential}, the gradient  $\nabla \phi_t(\boldsymbol{z})$ is uniquely determined.  To confirm this fact, let $\phi^{(1)}_t(\boldsymbol{z})$ and $\phi^{(2)}_t(\boldsymbol{z})$ be two solutions of Eq.~\eqref{geometricpotential}. Using Eq.~\eqref{geometricpotential}, we can obtain
\begin{align}
&\langle \nabla \phi^{(1)}_t- \nabla \phi^{(2)}_t , \nabla \phi^{(1)}_t- \nabla \phi^{(2)}_t \rangle_{p_t\mathsf{D}} \nonumber \\
&= -\int d\boldsymbol{z} [\phi^{(1)}_t(\boldsymbol{z})-\phi^{(2)}_t(\boldsymbol{z})] \nabla \cdot [p_t \mathsf{D} \nabla (\phi_t^{(1)} (\boldsymbol{z})-  \phi^{(2)}_t (\boldsymbol{z}))] \nonumber\\
&=0,
\end{align}
where we performed integration by parts. Therefore, we confirm $\nabla \phi^{(1)}_t (\boldsymbol{z})=  \nabla \phi^{(2)}_t (\boldsymbol{z})$ on the support of $p_t$. In other words, the uniqueness holds in the sense of the inner product with the metric $p_t \mathsf{D}$.

 The conservative contribution $\boldsymbol{f}^{\rm ex}_t(\boldsymbol{z}) :=-\nabla \phi_t(\boldsymbol{z})$, namely the excess thermodynamic force, and the nonconservative contribution $\boldsymbol{f}^{\rm hk}_t(\boldsymbol{z}) := \boldsymbol{f}_t(\boldsymbol{z})-\boldsymbol{f}^{\rm ex}_t(\boldsymbol{z}) =\boldsymbol{f}_t(\boldsymbol{z})+ \nabla \phi_t(\boldsymbol{z})$, namely the housekeeping thermodynamic force, are geometrically orthogonal in terms of the inner product, i.e.,  $\langle \boldsymbol{f}^{\rm ex}_t, \boldsymbol{f}^{\rm hk}_t\rangle_{p_t \mathsf{D}} =0$. This orthogonality can be confirmed as follows,
 \begin{align}
&\langle \boldsymbol{f}^{\rm ex}_t, \boldsymbol{f}^{\rm hk}_t\rangle_{p_t \mathsf{D}} \nonumber\\
&= \int d\boldsymbol{z} \phi_t(\boldsymbol{z}) \nabla \cdot [p_t(\boldsymbol{z})\mathsf{D} [\boldsymbol{f}_t(\boldsymbol{z})+\nabla \phi_t (\boldsymbol{z})]] =0,
\end{align}
where we used Eq.~\eqref{geometricpotential} and integration by parts. Based on this orthogonality, we obtain the following generalized Pythagorean theorem,
 \begin{align}
\langle \boldsymbol{f}_t, \boldsymbol{f}_t\rangle_{p_t \mathsf{D}} &= \langle \boldsymbol{f}^{\rm ex}_t+\boldsymbol{f}^{\rm hk}_t, \boldsymbol{f}^{\rm ex}_t+\boldsymbol{f}^{\rm hk}_t \rangle_{p_t \mathsf{D}}\nonumber\\
&= \langle \boldsymbol{f}^{\rm ex}_t, \boldsymbol{f}^{\rm ex}_t \rangle_{p_t \mathsf{D}} + \langle \boldsymbol{f}^{\rm hk}_t, \boldsymbol{f}^{\rm hk}_t \rangle_{p_t \mathsf{D}}.
\label{pythagorean}
\end{align}
Because $\langle \boldsymbol{f}_t, \boldsymbol{f}_t\rangle_{p_t \mathsf{D}}$ is equivalent to the entropy production rate $\dot{\Sigma}_t$, this Pythagorean theorem can be interpreted as the decomposition of the entropy production rate into two nonnegative terms $\langle \boldsymbol{f}^{\rm ex}_t, \boldsymbol{f}^{\rm ex}_t \rangle_{p_t \mathsf{D}}$ and $\langle \boldsymbol{f}^{\rm hk}_t, \boldsymbol{f}^{\rm hk}_t \rangle_{p_t \mathsf{D}}$. 

We call the conservative contribution, 
 \begin{align}
 \dot{\Sigma}^{\rm ex}_t:=\langle \boldsymbol{f}^{\rm ex}_t, \boldsymbol{f}^{\rm ex}_t \rangle_{p_t \mathsf{D}} (\geq 0),
\end{align}
the excess entropy production rate, and the nonconservative contribution, 
\begin{align}
 \dot{\Sigma}^{\rm hk}_t:=\langle \boldsymbol{f}^{\rm hk}_t,\boldsymbol{f}^{\rm hk}_t \rangle_{p_t \mathsf{D}} (\geq 0),
\end{align}
the housekeeping entropy production rate. Thus, the generalized Pythagorean theorem [Eq.~\eqref{pythagorean}] can be rewritten as
\begin{align}
 \dot{\Sigma}_t=\dot{\Sigma}^{\rm ex}_t+ \dot{\Sigma}^{\rm hk}_t,
\end{align}
which is a geometric decomposition of the entropy production rate into the excess and housekeeping contributions.

The terms ``excess'' and ``housekeeping'' are based on Ref.~\cite{maes2014nonequilibrium}. Indeed, the authors in Ref.~\cite{maes2014nonequilibrium} use the terms ``excess'' and ``housekeeping'' which are widely used in steady-state thermodynamics~\cite{hatano2001steady, dechant2022geometric} because the framework is analogous to steady state thermodynamics. However, the definitions of the excess and housekeeping entropy production rates that we define differ from the conventional definitions in Refs.~\cite{hatano2001steady, Vanden2010} in steady-state thermodynamics~\cite{dechant2022geometric, dechant2022geometric2}. Our definitions are mathematically equivalent to those in Ref.~\cite{maes2014nonequilibrium} for overdamped Langevin equations~\cite{dechant2022geometric}. Unlike the geometric decomposition considered in this paper, other decompositions, such as the Hatano--Sasa decomposition~\cite{hatano2001steady, Vanden2010}, are based on different principles and do not generally contain a component that has the same time-evolution property as $-\nabla\phi_t(\boldsymbol{z})$. Therefore, the results derived in this paper, especially those relying on the identity $\partial_t p_t(\boldsymbol{z}) =-\nabla\cdot[ p_t(\boldsymbol{z}) \mathsf{D}(-\nabla\phi_t(\boldsymbol{z}))]$,  cannot be directly transferred to decompositions other than the geometric decomposition.
We note that our definitions of the excess and housekeeping entropy production rates are not based on the existence of the steady state, and thus these quantities can be generalized for nonlinear dynamics on chemical reaction networks~\cite{yoshimura2023housekeeping, kolchinsky2024generalized,nagayama2025geometric}, where a steady state may not necessarily be unique or stable. This geometric decomposition of the entropy production rate can be generalized to fluid dynamics~\cite{yoshimura2024two} and open quantum dynamics~\cite{yoshimura2025force}, making this decomposition universally well-defined.

If $\boldsymbol{F}_t(\boldsymbol{z}) =- \nabla U_t(\boldsymbol{z})$, we obtain $\phi_t(\boldsymbol{z}) = U_t(\boldsymbol{z})/T + \ln p_t(\boldsymbol{z})$ that satisfies Eq.~\eqref{geometricpotential} because $\boldsymbol{f}_t(\boldsymbol{z})= -\nabla \phi_t(\boldsymbol{z})$. Therefore, $\dot{\Sigma}_t= \dot{\Sigma}^{\rm ex}_t$ and $\dot{\Sigma}^{\rm hk}_t=0$ if the dynamics is given by a conservative thermodynamic force. If the system is in the steady state, $\left. \partial_t p_t(\boldsymbol{z}) \right|_{p_t=p^{\rm st}} = \left.- \nabla \cdot (p_t(\boldsymbol{z})\mathsf{D} \boldsymbol{f}_t(\boldsymbol{z}) ) \right|_{p_t=p^{\rm st}}=0$, and $\left. \nabla \phi_t(\boldsymbol{z}) \right|_{p_t=p^{\rm st}}$ satisfies the condition $\nabla \cdot [p^{\rm st} (\boldsymbol{z}) \mathsf{D} \left. \nabla \phi_t(\boldsymbol{z}) \right|_{p_t=p^{\rm st}}]=0$ in the steady state. Using this condition, we obtain
\begin{align}
&\left. \langle \nabla \phi_t, \nabla \phi_t\rangle_{p_t \mathsf{D}} \right|_{p_t=p^{\rm st}} \nonumber\\
&= -\int d\boldsymbol{z} [\left.\phi_t(\boldsymbol{z}) \right|_{p_t=p^{\rm st}}] \nabla \cdot [p^{\rm st} (\boldsymbol{z}) \mathsf{D} \left. \nabla \phi_t(\boldsymbol{z}) \right|_{p_t=p^{\rm st}}]\nonumber\\
&=0,
\end{align}
where we performed integration by parts. This means $-\left. \nabla \phi_t(\boldsymbol{z}) \right|_{p_t=p^{\rm st}} =0$ on the support of $p^{\rm st}$.
Therefore, $\dot{\Sigma}_t= \dot{\Sigma}^{\rm hk}_t$ and $\dot{\Sigma}^{\rm ex}_t=0$ if the system is in the steady state.

The excess entropy production rate can also generally be expressed in a manner analogous to the entropy production rate when the system is conservative. To discuss the correspondence, we introduce the pseudo energy function $U^*_t(\boldsymbol{z})$ that satisfies 
\begin{align}
 \phi_t(\boldsymbol{z}) =: \frac{U^*_t(\boldsymbol{z})}{T} +\ln p_t (\boldsymbol{z}).
\label{defpseudoenergy}
\end{align}
This pseudo energy $ U^*_t(\boldsymbol{z})$ becomes the potential energy $ U^*_t(\boldsymbol{z})= U_t(\boldsymbol{z})$ if the force is given by the potential force $\boldsymbol{F}_t (\boldsymbol{z})= -\nabla  U_t(\boldsymbol{z})$. Therefore, this pseudo energy can be regarded as the potential energy, which provides the same time evolution for the nonconservative force, i.e., $\partial_t p_t(\boldsymbol{z}) = -\nabla \cdot [\mathsf{\mu}(-\nabla U^*_t (\boldsymbol{z}) )p_t(\boldsymbol{z})] + \nabla \cdot [\mathsf{D} \nabla p_t(\boldsymbol{z})]$. We consider the correspondence of the first law of thermodynamics for the pseudo energy $U^*_t(\boldsymbol{z})$. If the excess heat is defined as $\dot{Q}^{\rm ex}_t:= \int d\boldsymbol{z} [\partial_t p_t (\boldsymbol{z})] U^*_t(\boldsymbol{z}) =-\langle \boldsymbol{f}^{\rm ex}_t,-\nabla U^*_t\rangle_{p_t \mathsf{D} }$ and the excess work is defined as $\dot{W}^{\rm ex}_t := \int d\boldsymbol{z} p_t (\boldsymbol{z}) [\partial_t  U^*_t(\boldsymbol{z})]$, the following relation corresponding to the first law of thermodynamics,
\begin{align}
d_t \mathbb{E}_{p_t}[U^*_t] = \dot{Q}^{\rm ex}_t +\dot{W}^{\rm ex}_t,
\end{align}
holds. Using this excess heat, the excess entropy production rate can be rewritten as
\begin{align}
\dot{\Sigma}^{\rm ex}_t= d_tS^{\rm sys}_t -\frac{\dot{Q}^{\rm ex}_t }{T} \: (\geq 0).
\label{excesseprheat}
\end{align}
This relation can be confirmed as follows, 
\begin{align}
 d_tS^{\rm sys}_t -\frac{\dot{Q}^{\rm ex}_t }{T} &=- \int d\boldsymbol{z} [\partial_t p_t(\boldsymbol{z})] \left[\ln p_t(\boldsymbol{z}) + \frac{U^*_t(\boldsymbol{z})}{T} \right] \nonumber\\
 &=\int d\boldsymbol{z} \nabla \cdot [p_t(\boldsymbol{z}) \mathsf{D} (-\nabla \phi_t(\boldsymbol{z}))] \phi_t(\boldsymbol{z}) \nonumber\\
 &= \langle -\nabla \phi_t, -\nabla \phi_t \rangle_{p_t \mathsf{D}} = \dot{\Sigma}^{\rm ex}_t,
\end{align}
where we used Eq.~\eqref{continuityoptimal}, Eq.~\eqref{defpseudoenergy} and integration by parts. Arguments corresponding to this expression [Eq.~\eqref{excesseprheat}] have been made in Ref.~\cite{maes2014nonequilibrium} as a study of a nonequilibrium extension of the Clausius heat
theorem [Eq.~\eqref{clausiusheat}].

These excess and housekeeping entropy production rates can also be formulated in the form of a dual optimization problem~\cite{dechant2022geometric} (see also Appendix~\ref{appendix:variationalformula}). The excess entropy production rate is given by
\begin{align}
 \dot{\Sigma}^{\rm ex}_t &= \inf_{\boldsymbol{f}'_t(\boldsymbol{z})|\partial_t p_t = -\nabla \cdot ( p_t\mathsf{D}\boldsymbol{f}'_t)} \langle \boldsymbol{f}'_t, \boldsymbol{f}'_t\rangle_{p_t \mathsf{D}} \label{BBformula} \\
 &= \sup_{\psi(\boldsymbol{z})} \frac{(\langle -\nabla \psi, \boldsymbol{f}_t\rangle_{p_t \mathsf{D}})^2}{\langle -\nabla \psi, -\nabla \psi\rangle_{p_t \mathsf{D}}}.  \label{TUR1} 
\end{align}
The notation $\sup_{\psi(\boldsymbol{z})}$ indicates that the optimization is over functions of $\boldsymbol{z}$, and does not imply optimization for a specific $\boldsymbol{z}$. In this paper, we adopt this notation to clearly distinguish between optimizations where the function depends on $\boldsymbol{z}$ versus those where it depends on $\boldsymbol{x}$, such as $\sup_{\psi(\boldsymbol{z})}$ and $\sup_{\psi^{\rm X}(\boldsymbol{x})}$.

Similarly, the housekeeping entropy production rate can also be expressed as
\begin{align}
 \dot{\Sigma}^{\rm hk}_t &= \inf_{\psi(\boldsymbol{z})} \langle \boldsymbol{f}_t +\nabla \psi, \boldsymbol{f}_t + \nabla \psi \rangle_{p_t \mathsf{D}} \label{variational formulas2}\\
 &= \sup_{\boldsymbol{f}'_t (\boldsymbol{z})|-\nabla \cdot ( p_t\mathsf{D}\boldsymbol{f}'_t) =0} \frac{(\langle \boldsymbol{f}_t', \boldsymbol{f}_t\rangle_{p_t \mathsf{D}})^2}{\langle \boldsymbol{f}_t', \boldsymbol{f}_t' \rangle_{p_t \mathsf{D}}}.  \label{TUR2} 
\end{align}
These are consequences of the fact that the image ${\rm im}[\nabla] := \{\nabla \psi | \psi(\boldsymbol{z}) \in \mathbb{R} \}$ and the kernel ${\rm ker}[\nabla \cdot p_t\mathsf{D}] :=\{ \boldsymbol{f}'_t| \nabla \cdot (p_t(\boldsymbol{z})\mathsf{D} \boldsymbol{f}'_t(\boldsymbol{z})) =0 \} $ are orthogonal when $p_t\mathsf{D}$ is considered as a metric. The point $-\nabla \phi_t$ is obtained via the projection of $\boldsymbol{f}_t$ onto ${\rm im}[\nabla]$~\cite{dechant2022geometric}. We also note that we can obtain another expression of $\dot{\Sigma}^{\rm ex}_t$ using $\dot{\Sigma}^{\rm ex}_t=\dot{\Sigma}_t- \dot{\Sigma}^{\rm hk}_t$ as follows, 
\begin{align}
 \dot{\Sigma}^{\rm ex}_t &= \langle \boldsymbol{f}_t, \boldsymbol{f}_t \rangle_{p_t \mathsf{D}} - \inf_{\psi(\boldsymbol{z})} \langle \boldsymbol{f}_t +\nabla \psi, \boldsymbol{f}_t + \nabla \psi \rangle_{p_t \mathsf{D}} \nonumber \\
 &= \sup_{\psi(\boldsymbol{z})} \left( 2\langle \boldsymbol{f}_t, -\nabla \psi \rangle_{p_t \mathsf{D}} -\langle \nabla \psi, \nabla \psi \rangle_{p_t \mathsf{D}}\right).
 \label{variationalanother}
\end{align}
The optimal value of $\nabla \psi$ in Eq.~\eqref{variationalanother} is uniquely determined whereas the optimal value of   $\nabla \psi$ in Eq.~\eqref{TUR1}  has degrees of freedom for the proportional coefficient. 

The expression in Eq.~\eqref{BBformula} is related to the Benamou-Brenier formula~\cite{benamou2000computational} in optimal transport theory and the thermodynamic speed limit~\cite{nakazato2021geometrical,ito2024geometric}. If $\mathsf{D}$ is the identity matrix $\mathsf{D}=\mathsf{I}$, Eq.~\eqref{BBformula} can be rewritten as
\begin{align}
 \dot{\Sigma}^{\rm ex}_t  &= \inf_{\boldsymbol{u}'_t(\boldsymbol{z})|\partial_t p_t = -\nabla \cdot ( p_t \boldsymbol{u}'_t)} \langle \boldsymbol{u}'_t, \boldsymbol{u}'_t\rangle_{p_t \mathsf{I}} \nonumber\\
 &=\lim_{\Delta t \to 0} \frac{[\mathcal{W}_2 (p_t, p_{t+\Delta t})]^2}{(\Delta t)^2},
\end{align}
where $\mathcal{W}_2 (p_t, p_{t+\Delta t})$ is the $2$-Wasserstein distance between $p_t$ and $p_{t+\Delta t}$ defined as
\begin{align}
 &\mathcal{W}_2 (p^{\rm ini}, p^{\rm fin}) \nonumber\\
 &= \sqrt{ \inf_{(\boldsymbol{u}_{s}(\boldsymbol{z}), \rho_{s}(\boldsymbol{z}))_{t \leq s\leq t+\Delta t}} (\Delta t) \int_t^{t+\Delta t} d{s} \langle \boldsymbol{u}_{s}, \boldsymbol{u}_{s}\rangle_{\rho_{s} \mathsf{I}} } \nonumber\\
 &{\rm s.t.} \: \: \partial_{s} \rho_{s} (\boldsymbol{z})= -\nabla \cdot ( \rho_{s} (\boldsymbol{z})\boldsymbol{u}_{s}(\boldsymbol{z})),  \nonumber\\
 &\:\:\:\:\:\:\:\:\:\rho_t(\boldsymbol{z})=p^{\rm ini}(\boldsymbol{z}),  \: \rho_{t+\Delta t}(\boldsymbol{z})=p^{\rm fin}(\boldsymbol{z}),
\end{align}
which is known as the Benamou-Brenier formula~\cite{benamou2000computational}. Unlike the 2-Wasserstein distance for discrete-state systems described by Markov jump processes~\cite{maas2011gradient, yoshimura2023housekeeping}, the $2$-Wasserstein distance $\mathcal{W}_2(p^{\rm ini}, p^{\rm fin})$ can also be expressed as an optimization problem for joint probability distributions in the Monge-Kantorovich problem~\cite{villani2008optimal},
\begin{align}
 &\mathcal{W}_2 (p^{\rm ini}, p^{\rm fin}) =\sqrt{ \inf_{\pi (\boldsymbol{z},\boldsymbol{z}')} \int d\boldsymbol{z} \int d\boldsymbol{z}' \| \boldsymbol{z}- \boldsymbol{z}' \|^2 \pi (\boldsymbol{z},\boldsymbol{z}') }\nonumber\\
 &{\rm s.t.} \: \: \pi (\boldsymbol{z},\boldsymbol{z}') \geq 0,  \: \int d\boldsymbol{z}'\pi (\boldsymbol{z},\boldsymbol{z}') =p^{\rm ini}(\boldsymbol{z}),\nonumber\\
 &\:\:\:\:\:\:\:\:\:  \: \int d\boldsymbol{z}\pi (\boldsymbol{z},\boldsymbol{z}') =p^{\rm fin}(\boldsymbol{z}').
\end{align}
We note that the equivalence between the Benamou–Brenier formula and the Monge–Kantorovich problem only applies to continuous-state distributions in Euclidean space. This equivalence does not apply to general Markov jump processes with discrete states.  To generalize the results to these processes, we must define the generalized $2$-Wasserstein distance in terms of the generalization of the Benamou–Brenier formula, as was done in Refs.~\cite{maas2011gradient, yoshimura2023housekeeping, maekawa2025geometric}.

Even when $\mathsf{D}$ is not the identity matrix $\mathsf{I}$, we can generalize the 2-Wasserstein distance $\tilde{\mathcal{W}}^{\mathsf{D}^{-1}}_2$ as
\begin{align}
 &\tilde{\mathcal{W}}^{\mathsf{D}^{-1}}_2 (p^{\rm ini}, p^{\rm fin}) \nonumber\\
 &= \sqrt{ \inf_{(\boldsymbol{f}'_{s}(\boldsymbol{z}), \rho_{s}(\boldsymbol{z}))_{t \leq {s} \leq t+\Delta t}}  
 (\Delta t)\int_t^{t+\Delta t} d{s} \langle \boldsymbol{f}'_{s}, \boldsymbol{f}'_{s}\rangle_{\rho_{s} \mathsf{D} } }\nonumber\\
 &{\rm s.t.} \: \: \partial_{s} \rho_{s} (\boldsymbol{z})= -\nabla \cdot ( \rho_{s} (\boldsymbol{z})\mathsf{D} \boldsymbol{f}'_{s}(\boldsymbol{z})),  \nonumber\\
 &\:\:\:\:\:\:\:\:\:\rho_t(\boldsymbol{z})=p^{\rm ini}(\boldsymbol{z}),  \: \rho_{t+\Delta t}(\boldsymbol{z})=p^{\rm fin}(\boldsymbol{z}),
 \label{generalBBformula}
\end{align}
and we obtain the following relation
\begin{align}
 \dot{\Sigma}^{\rm ex}_t  =\lim_{\Delta t \to 0} \frac{[\tilde{\mathcal{W}}^{\mathsf{D}^{-1}}_2 (p_t, p_{t+\Delta t})]^2}{(\Delta t)^2} .
 \label{instantaneous2-Wasser}
\end{align}
If we consider the coordinate transformation $\tilde{\boldsymbol{z}}:=\mathsf{D}^{-1/2} \boldsymbol{z}$, $\tilde{\mathcal{W}}^{\mathsf{D}^{-1}}_2 (p^{\rm ini}, p^{\rm fin})$ can be regarded as the $2$-Wasserstein distance for this new coordinate, and thus this generalized $2$-Wasserstein distance satisfies the axioms of the metric (see Appendix~\ref{appendix:generalized2-Wasserstein}). Because this generalized $2$-Wasserstein distance satisfies the triangle inequality, this expression [Eq.~\eqref{instantaneous2-Wasser}] leads to the lower bound on the excess entropy production $[\tilde{\mathcal{W}}^{\mathsf{D}^{-1}}_2 (p_0, p_{\tau})]^2/\tau (\leq \int_0^{\tau} dt \dot{\Sigma}^{\rm ex}_t )$, and this inequality is known as the thermodynamic speed limit for the excess entropy production~\cite{ito2024geometric}. 

We note that the optimal value of $\boldsymbol{f}'_s (\boldsymbol{z})$ in Eq.~\eqref{generalBBformula} is given by the conservative thermodynamic force $\boldsymbol{f}^*_s (\boldsymbol{z}) =-\nabla \psi^*_s (\boldsymbol{z})$. To confirm this fact, we consider two quantities $\boldsymbol{f}'_s (\boldsymbol{z})$ and $-\nabla \psi_s  (\boldsymbol{z})$ that satisfy the constraint in the optimization problem [Eq.~\eqref{generalBBformula}] as follows,
\begin{align}
-\nabla \cdot [ \rho_{s} (\boldsymbol{z})\mathsf{D} (-\nabla \psi_s (\boldsymbol{z}))] &=\partial_{s} \rho_{s} (\boldsymbol{z}) \nonumber\\
&= -\nabla \cdot ( \rho_{s} (\boldsymbol{z})\mathsf{D} \boldsymbol{f}'_{s}(\boldsymbol{z})),
\end{align}
or equivalently,
\begin{align}
&\langle -\nabla \psi_s,  \boldsymbol{f}_{s}' +\nabla \psi_s\rangle_{\rho_{s} \mathsf{D} } \nonumber \\
&= -\int d\boldsymbol{z} \nabla \psi_s (\boldsymbol{z})\rho_{s} (\boldsymbol{z})\mathsf{D} [ \boldsymbol{f}_{s}' (\boldsymbol{z})+\nabla \psi_s(\boldsymbol{z})]\nonumber \\
&= \int d\boldsymbol{z}  \psi_s(\boldsymbol{z}) \nabla \cdot [\rho_{s} (\boldsymbol{z})\mathsf{D} [ \boldsymbol{f}_{s}' (\boldsymbol{z})+\nabla \psi_s(\boldsymbol{z})]] =0,
\end{align}
where we performed integration by parts.
Therefore, we obtain the following inequality
\begin{align}
&\langle \boldsymbol{f}'_{s}, \boldsymbol{f}'_{s}\rangle_{\rho_{s} \mathsf{D} } \nonumber\\
&= \langle -\nabla \psi_s, -\nabla \psi_s \rangle_{\rho_{s} \mathsf{D} } +\langle \boldsymbol{f}_{s}' +\nabla \psi_s, \boldsymbol{f}_{s}' +\nabla \psi_s \rangle_{\rho_{s} \mathsf{D} } \nonumber\\
&\geq \langle -\nabla \psi_s , -\nabla \psi_s  \rangle_{\rho_{s} \mathsf{D} },
\label{inequalityopt}
\end{align}
and this inequality means that the optimal value $\boldsymbol{f}^*_s (\boldsymbol{z})$ in the minimization problem [Eq.~\eqref{generalBBformula}] should be given by the conservative thermodynamic force $-\nabla \psi^*_s(\boldsymbol{z})$. Using this fact, Eq.~\eqref{generalBBformula} can also be rewritten as
\begin{align}
 &\tilde{\mathcal{W}}^{\mathsf{D}^{-1}}_2 (p^{\rm ini}, p^{\rm fin}) \nonumber\\
 &= \sqrt{ \inf_{(\psi_{s}(\boldsymbol{z}), \rho_{s}(\boldsymbol{z}))_{t \leq {s} \leq t+\Delta t}}  \! \!
 (\Delta t)\int_t^{t+\Delta t} d{s} \langle -\nabla \psi_{s}, -\nabla \psi_{s} \rangle_{\rho_{s} \mathsf{D} } }\nonumber\\
 &{\rm s.t.} \: \: \partial_{s} \rho_{s} (\boldsymbol{z})= -\nabla \cdot [\rho_{s} (\boldsymbol{z})\mathsf{D}[  -\nabla \psi_{s}(\boldsymbol{z})]],  \nonumber\\
 &\:\:\:\:\:\:\:\:\:\rho_t(\boldsymbol{z})=p^{\rm ini}(\boldsymbol{z}),  \: \rho_{t+\Delta t}(\boldsymbol{z})=p^{\rm fin}(\boldsymbol{z}).
 \label{generalBBformula3}
\end{align}

We also note that this generalized $2$-Wasserstein distance has different physical units from the standard $2$-Wasserstein distance, as it involves the inner product of thermodynamic forces with respect to the metric $p_t(\boldsymbol{z}) \mathsf{D}$. 
Consequently, the diffusion coefficient does not appear in the thermodynamic speed limits as a separate prefactor, because it is incorporated into the generalized Wasserstein metric. If $\mathsf{D}=\mu T \mathsf{I}$, we obtain the relation $[\tilde{\mathcal{W}}^{\mathsf{D}^{-1}}_2 (p_0, p_{\tau})]^2 = [\mathcal{W}_2 (p_0, p_{\tau})]^2/(\mu T)$, and the contribution of the diffusion coefficient $\mu T$ appears in the thermodynamic speed limit as the conventional one~\cite{nakazato2021geometrical, ito2024geometric}.
This type of generalized $2$-Wasserstein distance and its corresponding thermodynamic speed limit were derived for Markov jump systems~\cite{yoshimura2023housekeeping} and subsequently extended to reaction-diffusion systems~\cite{nagayama2025geometric} containing diffusion components.

The expressions in Eqs.~\eqref{TUR1} and~\eqref{TUR2} are related to the thermodynamic uncertainty relations. Especially, for any function $\psi(\boldsymbol{z})$, the time derivative of $\mathbb{E}_{p_t} [\psi]=\int d\boldsymbol{z} \psi(\boldsymbol{z}) p_t (\boldsymbol{z})$ is given by
\begin{align}
\partial_t \mathbb{E}_{p_t} [\psi] &= -\int d\boldsymbol{z} \psi(\boldsymbol{z}) \nabla \cdot( p_t (\boldsymbol{z}) \mathsf{D} \boldsymbol{f}_t(\boldsymbol{z})) \nonumber\\
&=- \langle -\nabla \psi, \boldsymbol{f}_t\rangle_{p_t \mathsf{D}},
\end{align}
where we used Eq.~\eqref{continuityeq} and integration by parts.
Next, we consider the quantity $\Delta \psi( \boldsymbol{z}(t+dt),\boldsymbol{z}(t))=   \psi(\boldsymbol{z}(t+dt))-\psi(\boldsymbol{z}(t))$ in the Langevin description [Eq.~\eqref{Langevineq}]. We obtain $\Delta \psi(\boldsymbol{z}(t+dt),\boldsymbol{z}(t)) \simeq \dot{\boldsymbol{z}}(t) dt \circ \nabla \psi(\boldsymbol{z}(t))$. Similarly to Eq.~\eqref{changestomuinfo}, its variance to leading order in $dt$ is calculated as follows, 
\begin{align}
{\rm Var}[\Delta \psi] &\simeq {\rm Var}[\sqrt{2 T} \mathsf{B} \boldsymbol{\xi}_t dt \circ \nabla \psi] \nonumber\\
&\simeq 2 dt\int d\boldsymbol{z} [\nabla \psi(\boldsymbol{z})]^{\top} \mathsf{D} [\nabla \psi(\boldsymbol{z} ) ] p_t(\boldsymbol{z}) \nonumber\\
&= 2 dt\langle -\nabla \psi, -\nabla \psi\rangle_{p_t \mathsf{D}},
\end{align}
where we used Ito calculus $\mathbb{E}[ (\boldsymbol{\xi}_t dt) (\boldsymbol{\xi}_t dt)^{\top}]\simeq \mathsf{I} dt$ and $T\mathsf{B} \mathsf{B}^{\top} =\mathsf{D}$. Therefore, Eq.~\eqref{TUR1} can be rewritten as the inequality 
\begin{align}
 \dot{\Sigma}^{\rm ex}_t  \geq \frac{[\partial_t \mathbb{E}_{p_t} [\psi]]^2}{\lim_{dt \to 0} \frac{{\rm Var}[\Delta \psi]}{2 dt}},
 \label{turexcess}
\end{align}
which is the thermodynamic uncertainty relation for  the excess entropy production rate using a state-dependent observable $\psi(\boldsymbol{z})$. The expression of the thermodynamic uncertainty relation is related to the H\"{o}lder-type inequality between the $1$-Wasserstein distance and the $2$-Wasserstein distance~\cite{nagayama2025geometric}.
We note that Eq.~\eqref{TUR2} can also be regarded as the thermodynamic uncertainty relation for the housekeeping entropy production rate using a generalized current~\cite{dechant2022geometric2}.

Furthermore, the housekeeping entropy production rate can be decomposed in terms of intrinsic oscillation modes~\cite{sekizawa2025koopman}. If we define the housekeeping velocity field as $\boldsymbol{\nu}_t^{\rm hk}(\boldsymbol{z}):= \mathsf{D}\boldsymbol{f}^{\rm hk}_t(\boldsymbol{z})$, it satisfies $\nabla \cdot ( \boldsymbol{\nu}_t^{\rm hk}(\boldsymbol{z})p_t(\boldsymbol{z}) )=0$, which means that the flux $\boldsymbol{\nu}_t^{\rm hk}(\boldsymbol{z})p_t(\boldsymbol{z})$ can be written in oscillatory modes. To introduce independent
oscillatory modes, we consider
the Koopman generator $\mathcal{K}$ for the housekeeping velocity field $\boldsymbol{\nu}_t^{\rm hk}(\boldsymbol{z})$, which is defined as 
\begin{align}
\mathcal{K} g(\boldsymbol{z}) := \nabla g(\boldsymbol{z}) \cdot \boldsymbol{\nu}_t^{\rm hk}(\boldsymbol{z}),
\end{align}
for any function $g(\boldsymbol{z})$. This Koopman generator gives the time evolution of any function $d_{s}g(\boldsymbol{z}(s))= \mathcal{K} g(\boldsymbol{z}(s)) $ when the state $\boldsymbol{z}(s)$ is thought to be moving using this housekeeping velocity field, i.e., $d_{s} \boldsymbol{z}(s) = \boldsymbol{\nu}_t^{\rm hk}(\boldsymbol{z}(s))$. Therefore, the Koopman generator $\mathcal{K}$ is defined for a fixed time $t$.
For any functions $g_1(\boldsymbol{z})$ and $g_2(\boldsymbol{z})$, an antisymmetric relation $\langle g_1, \mathcal{K}g_2 \rangle_{p_t} :=\int d\boldsymbol{z} g_1 (\boldsymbol{z})   p_t(\boldsymbol{z}) \mathcal{K} g_2 (\boldsymbol{z}) =-\int d\boldsymbol{z} [\mathcal{K}g_1 (\boldsymbol{z}) ]  p_t(\boldsymbol{z})  g_2 (\boldsymbol{z}) = -\langle \mathcal{K}g_1, g_2 \rangle_{p_t}$ holds because $\nabla \cdot (p_t (\boldsymbol{z}) \boldsymbol{\nu}_t^{\rm hk}(\boldsymbol{z})) =0$. If we consider the eigenvalue and eigenfunction of $\mathcal{K}$,  $\lambda^{\mathcal{K}}_i$ and $g^{\mathcal{K}}_i$, that satisfy 
\begin{align}
\mathcal{K} g^{\mathcal{K}}_i(\boldsymbol{z}) =\lambda^{\mathcal{K}}_i g^{\mathcal{K}}_i (\boldsymbol{z}),
\end{align}
we obtain the identity $((\lambda^{\mathcal{K}}_i)^\dagger + \lambda^{\mathcal{K}}_j)\langle (g^{\mathcal{K}}_i)^\dagger  , g^{\mathcal{K}}_j \rangle_{p_t} = 0$ from the antisymmetric relation, where the symbol $^\dagger$ stands for the complex conjugate. This identity implies that the eigenvalue is purely imaginary $-(\lambda^{\mathcal{K}}_i)^\dagger = \lambda^{\mathcal{K}}_i =\boldsymbol{{\rm i}}\Im[\lambda^{\mathcal{K}}_i] $ if $\langle (g^{\mathcal{K}}_i)^\dagger  , g^{\mathcal{K}}_i \rangle_{p_t } \neq 0$, and $\langle (g^{\mathcal{K}}_i)^\dagger  , g^{\mathcal{K}}_j \rangle_{p_t} = 0$ if $-(\lambda^{\mathcal{K}})_i^\dagger \neq \lambda^{\mathcal{K}}_j$, where $\boldsymbol{{\rm i}}$ is the imaginary unit and $\Im[\lambda_i^{\mathcal{K}}]$ stands for the imaginary part of $\lambda_i^{\mathcal{K}}$. Here, we assume that the identity function ${\rm Id}(\boldsymbol{z}) = \boldsymbol{z}$ is spanned by 
\begin{align}
{\rm Id}(\boldsymbol{z}) = \sum_i \boldsymbol{d}_i g^{\mathcal{K}}_i(\boldsymbol{z}).
\label{expansion}
\end{align}
For simplicity, we also assume that the Koopman eigenvalues $\{ \lambda^{\mathcal{K}}_i \}$ corresponding to the eigenfunctions $\{ g^{\mathcal{K}}_i(\boldsymbol{z}) \}$ in this expansion are non-degenerate.

Generally, it is not obvious that the identity function can be approximately expanded using discrete spectra. For example, if deterministic dynamics $d_{s} \boldsymbol{z}(s) = \boldsymbol{\nu}_t^{\rm hk}(\boldsymbol{z}(s))$ show chaotic behavior, the identity function can be expressed using the continuous spectra, and the approximation using discrete spectra may not be justified. Therefore, our treatment using the Koopman generator may only be effective in regimes where chaotic behavior is not apparent.

Under these assumptions, the housekeeping velocity field $\boldsymbol{\nu}_t^{\rm hk}(\boldsymbol{z})$ can be calculated as $\boldsymbol{\nu}_t^{\rm hk}(\boldsymbol{z}) = \mathcal{K}{\rm Id}(\boldsymbol{z}) = \sum_i \lambda^{\mathcal{K}}_i \boldsymbol{d}_i g^{\mathcal{K}}_i(\boldsymbol{z})$. Therefore, the housekeeping entropy production rate can be expressed as 
\begin{align}
 \dot{\Sigma}^{\rm hk}_t  &=\langle \mathsf{D}^{-1}\boldsymbol{\nu}^{\rm hk}_t   , \mathsf{D}^{-1}\boldsymbol{\nu}^{\rm hk}_t\rangle_{p_t \mathsf{D}} \nonumber\\
 &=\langle \mathsf{D}^{-1} \sum_i (\lambda^{\mathcal{K}}_i)^\dagger \boldsymbol{d}_i^\dagger (g^{\mathcal{K}}_i)^\dagger  , \mathsf{D}^{-1} \sum_j \lambda^{\mathcal{K}}_j \boldsymbol{d}_j g^{\mathcal{K}}_j \rangle_{p_t \mathsf{D}}\nonumber\\
 &= \sum_i (2\pi)^2 \left(\frac{\Im[\lambda^{\mathcal{K}}_i]}{2 \pi} \right)^2 (\boldsymbol{d}_i^{\mathrm H}  \mathsf{D}^{-1}  \boldsymbol{d}_i) \langle (g^{\mathcal{K}}_i)^\dagger, g^{\mathcal{K}}_i \rangle_{p_t},
 \label{koopmandecomp}
\end{align}
where we used the fact that $-(\lambda^{\mathcal{K}}_i)^\dagger = \lambda^{\mathcal{K}}_i =\boldsymbol{{\rm i}}\Im[\lambda^{\mathcal{K}}_i] $ if $\langle (g^{\mathcal{K}}_i)^\dagger  , g^{\mathcal{K}}_i \rangle_{p_t } \neq 0$, and $\langle (g^{\mathcal{K}}_i)^\dagger  , g^{\mathcal{K}}_j \rangle_{p_t} = 0$ if $-(\lambda^{\mathcal{K}}_i)^\dagger \neq \lambda^{\mathcal{K}}_j$. Here, the symbol $^{\mathrm H}$ stands for the Hermitian transpose.
Because $(\boldsymbol{d}_i^{\mathrm H}  \mathsf{D}^{-1}  \boldsymbol{d}_i) \langle (g^{\mathcal{K}}_i)^\dagger, g^{\mathcal{K}}_i \rangle_{p_t}$ corresponds to the intensity of the $i$-th oscillatory mode in $\boldsymbol{\nu}_t^{\rm hk} (\boldsymbol{z})$, and $\Im[\lambda^{\mathcal{K}}_i]/(2\pi)$ corresponds to the frequency of the $i$-th oscillatory mode, this expression is regarded as the oscillatory mode decomposition of the housekeeping entropy production rate.

\section{Main results}
\label{sec3}
In this section, we will generalize the results from Sec.~\ref{sec2} using the decomposition of information flow. 

\subsection{Decomposition of information flow}
\label{sec3a}
As the Langevin-system counterpart to the results in Ref.~\cite{maekawa2025geometric}, we first decompose information flow into excess contributions and housekeeping contributions. 

To introduce the decomposition of information flow, we consider the decomposition of the excess and housekeeping thermodynamic forces into $\rm X$ and $\rm Y$ components as follows,
\begin{align}
\boldsymbol{f}^{\rm ex}_t  (\boldsymbol{x}, \boldsymbol{y})& = \begin{pmatrix}
   \boldsymbol{f}^{\rm ex; X}_t   (\boldsymbol{x}, \boldsymbol{y}) \\
   \boldsymbol{f}^{\rm ex; Y}_t   (\boldsymbol{x}, \boldsymbol{y})
\end{pmatrix}=\begin{pmatrix}
   - \nabla_{\boldsymbol{x}} \phi_t   (\boldsymbol{x}, \boldsymbol{y}) \\
   - \nabla_{\boldsymbol{y}} \phi_t  (\boldsymbol{x}, \boldsymbol{y})
\end{pmatrix}, \nonumber\\
\boldsymbol{f}^{\rm hk}_t  (\boldsymbol{x}, \boldsymbol{y}) &= \begin{pmatrix}
   \boldsymbol{f}^{\rm hk; X}_t   (\boldsymbol{x}, \boldsymbol{y}) \\
   \boldsymbol{f}^{\rm hk; Y}_t   (\boldsymbol{x}, \boldsymbol{y})
\end{pmatrix}
=\begin{pmatrix}
   \boldsymbol{f}^{\rm X}_t(\boldsymbol{x}, \boldsymbol{y}) + \nabla_{\boldsymbol{x}} \phi_t   (\boldsymbol{x}, \boldsymbol{y}) \\
   \boldsymbol{f}^{\rm Y}_t(\boldsymbol{x}, \boldsymbol{y}) + \nabla_{\boldsymbol{y}} \phi_t   (\boldsymbol{x}, \boldsymbol{y})
\end{pmatrix}
.
\end{align}
Similarly to the relation $\boldsymbol{f}_t(\boldsymbol{z})= \boldsymbol{f}^{\rm ex}_t(\boldsymbol{z})+ \boldsymbol{f}^{\rm hk}_t(\boldsymbol{z})$ for the total system, the relations $\boldsymbol{f}^{\rm X}_t(\boldsymbol{x}, \boldsymbol{y})= \boldsymbol{f}^{\rm ex;X}_t(\boldsymbol{x}, \boldsymbol{y})+ \boldsymbol{f}^{\rm hk;X}_t(\boldsymbol{x}, \boldsymbol{y})$ and $\boldsymbol{f}^{\rm Y}_t(\boldsymbol{x}, \boldsymbol{y})= \boldsymbol{f}^{\rm ex;Y}_t(\boldsymbol{x}, \boldsymbol{y})+ \boldsymbol{f}^{\rm hk;Y}_t(\boldsymbol{x}, \boldsymbol{y})$ hold for each subsystem. 

Using this decomposition of the thermodynamic force, we consider the following decomposition of information flow into excess information flow and housekeeping information flow,
\begin{align}
\dot{I}^{\rm X}_t &= \dot{I}^{\rm ex; X}_t +\dot{I}^{\rm hk; X}_t, \nonumber\\
\dot{I}^{\rm Y}_t &= \dot{I}^{\rm ex; Y}_t +\dot{I}^{\rm hk; Y}_t, 
\label{infodecomposition}
\end{align}
where excess information flow is defined as
\begin{align}
\dot{I}^{\rm ex; X}_t := \langle \boldsymbol{f}^{\rm ex; X}_t,\nabla_{\boldsymbol{x}} i_t \rangle_{ p_t \mathsf{D}^{\rm X}},  \: \dot{I}^{\rm ex; Y}_t:= \langle \boldsymbol{f}^{\rm ex; Y}_t,\nabla_{\boldsymbol{y}} i_t \rangle_{ p_t \mathsf{D}^{\rm Y}},
\label{infodecompositiondef}
\end{align}
and housekeeping information flow is defined as
\begin{align}
\dot{I}^{\rm hk; X}_t := \langle \boldsymbol{f}^{\rm hk; X}_t,\nabla_{\boldsymbol{x}} i_t \rangle_{ p_t \mathsf{D}^{\rm X}},  \: \dot{I}^{\rm hk; Y}_t:= \langle \boldsymbol{f}^{\rm hk; Y}_t,\nabla_{\boldsymbol{y}} i_t \rangle_{ p_t \mathsf{D}^{\rm Y}}.
\label{infodecompositiondef2}
\end{align}
The decomposition [Eq.~\eqref{infodecomposition}] can be confirmed as $\dot{I}^{\rm X}_t=\langle \boldsymbol{f}^{\rm X}_t,\nabla_{\boldsymbol{x}} i_t \rangle_{ p_t \mathsf{D}^{\rm X}}= \langle \boldsymbol{f}^{\rm ex;X}_t+ \boldsymbol{f}^{\rm hk;X}_t,\nabla_{\boldsymbol{x}} i_t \rangle_{ p_t \mathsf{D}^{\rm X}}$ and $\dot{I}^{\rm Y}_t=\langle \boldsymbol{f}^{\rm Y}_t,\nabla_{\boldsymbol{y}} i_t \rangle_{ p_t \mathsf{D}^{\rm Y}}= \langle \boldsymbol{f}^{\rm ex;Y}_t+ \boldsymbol{f}^{\rm hk;Y}_t,\nabla_{\boldsymbol{y}} i_t \rangle_{ p_t \mathsf{D}^{\rm Y}}$. 

This decomposition of information flow is simplified in steady-state and conservative cases. In the steady state, 
\begin{align}
\left. \dot{I}^{\rm ex;X}_t \right|_{p_t=p^{\rm st}}=0, \: \:
\left.\dot{I}^{\rm hk;X}_t \right|_{p_t=p^{\rm st}}=\left. \dot{I}^{\rm X}_t \right|_{p_t=p^{\rm st}} ,\nonumber\\
\left. \dot{I}^{\rm ex;Y}_t \right|_{p_t=p^{\rm st}}=0, \: \:
\left.\dot{I}^{\rm hk;Y}_t \right|_{p_t=p^{\rm st}}=\left. \dot{I}^{\rm Y}_t \right|_{p_t=p^{\rm st}},
\end{align}
holds since $\left. \boldsymbol{f}^{\rm ex}_t (\boldsymbol{z}) \right|_{p_t=p^{\rm st}} = \boldsymbol{0}$ in the steady state. Conversely, if the thermodynamic force is conservative, the condition $\boldsymbol{f}^{\rm hk}_t=\boldsymbol{0}$ provides $\dot{I}^{\rm hk;X}_t=\dot{I}^{\rm hk;Y}_t =0$, $\dot{I}^{\rm ex;X}_t=\dot{I}^{\rm X}_t$ and $\dot{I}^{\rm ex;Y}_t=\dot{I}^{\rm Y}_t$.

Excess information flow has the same property as information flow in a non-steady state [Eq.~\eqref{decompositioninfo}]. We obtain the following relation corresponding to Eq.~\eqref{decompositioninfo},
\begin{align}
d_t I(\hat{\rm X}_t; \hat{\rm Y}_t)= \dot{I}^{\rm ex;X}_t + \dot{I}^{\rm ex;Y}_t.
\label{decompositionexinfo}
\end{align}
This result can be proved as follows,
\begin{align}
 &\dot{I}^{\rm ex;X}_t + \dot{I}^{\rm ex;Y}_t = \langle \boldsymbol{f}^{\rm ex;X}_t,\nabla_{\boldsymbol{x}} i_t \rangle_{ p_t \mathsf{D}^{\rm X}} + \langle \boldsymbol{f}^{\rm ex;Y}_t,\nabla_{\boldsymbol{y}} i_t \rangle_{ p_t \mathsf{D}^{\rm Y}} \nonumber\\
 &= \int d\boldsymbol{x} \int d\boldsymbol{y} [-\nabla \cdot (p_t(\boldsymbol{x}, \boldsymbol{y})\mathsf{D} \boldsymbol{f}^{\rm ex}_t (\boldsymbol{x}, \boldsymbol{y}) ) ]i_t(\boldsymbol{x}, \boldsymbol{y}) \nonumber\\
 &= \int d\boldsymbol{x} \int d\boldsymbol{y} [\partial_t p_t(\boldsymbol{x}, \boldsymbol{y})] i_t(\boldsymbol{x}, \boldsymbol{y}) = d_t I(\hat{\rm X}_t; \hat{\rm Y}_t),
\end{align}
where we used the normalization of probability and integration by parts.
Therefore, excess information flow includes the contribution of the time variation of mutual information. This equation implies that the housekeeping information flow does not contribute to the net time variation of mutual information.

On the other hand, housekeeping information flow has the same property as information flow in the steady state [Eq.~\eqref{antisymmetric}]. We obtain the following antisymmetric relation corresponding to Eq.~\eqref{antisymmetric},
\begin{align}
 \dot{I}^{\rm hk;X}_t = -\dot{I}^{\rm hk;Y}_t.
 \label{antisymhk}
\end{align}
Remarkably, unlike Eq.~\eqref{antisymmetric}, this antisymmetric relation holds even far from the steady state. The antisymmetric relation [Eq.~\eqref{antisymhk}] can be confirmed as follows,
\begin{align}
 \dot{I}^{\rm hk;X}_t &=\langle \boldsymbol{f}^{\rm hk; X}_t,\nabla_{\boldsymbol{x}} i_t \rangle_{ p_t \mathsf{D}^{\rm X}} \nonumber\\
 &=\int d\boldsymbol{x} \int d\boldsymbol{y} [-\nabla_{\boldsymbol{x}} \cdot (p_t(\boldsymbol{x}, \boldsymbol{y})\mathsf{D}^{\rm X} \boldsymbol{f}^{\rm hk;X}_t (\boldsymbol{x}, \boldsymbol{y}) )]i_t(\boldsymbol{x}, \boldsymbol{y}) \nonumber\\
 &=\int d\boldsymbol{x} \int d\boldsymbol{y} [\nabla_{\boldsymbol{y}} \cdot (p_t(\boldsymbol{x}, \boldsymbol{y})\mathsf{D}^{\rm Y} \boldsymbol{f}^{\rm hk;Y}_t (\boldsymbol{x}, \boldsymbol{y}) )]i_t(\boldsymbol{x}, \boldsymbol{y}) \nonumber\\
 &=- \langle \boldsymbol{f}^{\rm hk; Y}_t,\nabla_{\boldsymbol{y}} i_t \rangle_{ p_t \mathsf{D}^{\rm Y}} = - \dot{I}^{\rm hk;Y}_t,
\end{align}
where we used $-\nabla_{\boldsymbol{x}} \cdot (p_t(\boldsymbol{x}, \boldsymbol{y})\mathsf{D}^{\rm X} \boldsymbol{f}^{\rm hk;X}_t (\boldsymbol{x}, \boldsymbol{y}) )-\nabla_{\boldsymbol{y}} \cdot (p_t(\boldsymbol{x}, \boldsymbol{y})\mathsf{D}^{\rm Y} \boldsymbol{f}^{\rm hk;Y}_t (\boldsymbol{x}, \boldsymbol{y}))=0$ and integration by parts.

\subsection{Generalized second laws of information thermodynamics}
\label{sec3b}
Here we discuss how the nonnegativity of the excess entropy production rate
and that of the housekeeping entropy production rate
can be formulated separately for each subsystem. Just as the nonnegativity of the partial entropy production rate can be regarded as the second law of information thermodynamics, the nonnegativity of the excess entropy production rate and the nonnegativity of the housekeeping entropy production rate for each subsystem become generalizations of the second law of information thermodynamics, respectively. We also show that excess information flow and housekeeping information flow naturally emerge in the generalized second laws of information thermodynamics.

We introduce the partial excess entropy production rates defined as
\begin{align}
 \dot{\Sigma}^{\rm ex;X}_t := \langle \boldsymbol{f}^{\rm ex; X}_t,\boldsymbol{f}^{\rm ex; X}_t \rangle_{ p_t \mathsf{D}^{\rm X}},  \: \dot{\Sigma}^{\rm ex;Y}_t := \langle \boldsymbol{f}^{\rm ex; Y}_t,\boldsymbol{f}^{\rm ex; Y}_t \rangle_{ p_t \mathsf{D}^{\rm Y}},
\end{align}
and the partial housekeeping entropy production rates defined as
\begin{align}
 \dot{\Sigma}^{\rm hk;X}_t := \langle \boldsymbol{f}^{\rm hk; X}_t,\boldsymbol{f}^{\rm hk; X}_t \rangle_{ p_t \mathsf{D}^{\rm X}},  \: \dot{\Sigma}^{\rm hk;Y}_t := \langle \boldsymbol{f}^{\rm hk; Y}_t,\boldsymbol{f}^{\rm hk; Y}_t \rangle_{ p_t \mathsf{D}^{\rm Y}}.
\end{align}
We can confirm that these quantities satisfy 
\begin{align}
\dot{\Sigma}^{\rm ex}_t=\dot{\Sigma}^{\rm ex;X}_t+\dot{\Sigma}^{\rm ex;Y}_t,
\end{align}
and 
\begin{align}
\dot{\Sigma}^{\rm hk}_t=\dot{\Sigma}^{\rm hk;X}_t+\dot{\Sigma}^{\rm hk;Y}_t,
\end{align}
because $\mathsf{D} = T\mathsf{\mu}$ is a block diagonal matrix [Eq.~\eqref{blockdiag}]. These quantities are nonnegative, i.e., $\dot{\Sigma}^{\rm ex;X}_t \geq 0$, $\dot{\Sigma}^{\rm ex;Y}_t \geq 0$,  $\dot{\Sigma}^{\rm hk;X}_t \geq 0$ and $\dot{\Sigma}^{\rm hk;Y}_t \geq 0$. We note that these partial housekeeping and excess entropy production rates do not generally provide the decomposition of the partial entropy production rate, i.e.,
$\dot{\Sigma}^{\rm X}_t \neq \dot{\Sigma}^{\rm ex;X}_t+\dot{\Sigma}^{\rm hk;X}_t$.

These partial excess entropy production rates and partial housekeeping entropy production rates can be considered generalizations of the partial entropy production rate.
In the steady state,  we obtain
\begin{align}
 \left. \dot{\Sigma}^{\rm ex;X}_t \right|_{p_t=p^{\rm st}} &= 0, \: \left. \dot{\Sigma}^{\rm ex;Y}_t \right|_{p_t=p^{\rm st}}= 0 ,\nonumber\\
 \left. \dot{\Sigma}^{\rm hk;X}_t \right|_{p_t=p^{\rm st}}&= \left. \dot{\Sigma}^{\rm X}_t \right|_{p_t=p^{\rm st}}, \: \left. \dot{\Sigma}^{\rm hk;Y}_t \right|_{p_t=p^{\rm st}}= \left. \dot{\Sigma}^{\rm Y}_t \right|_{p_t=p^{\rm st}},
\end{align}
because $\left. \boldsymbol{f}^{\rm ex}_t (\boldsymbol{z}) \right|_{p_t=p^{\rm st}}= \boldsymbol{0}$ is satisfied.
If the thermodynamic force $\boldsymbol{f}_t (\boldsymbol{z})$ is conservative, $\boldsymbol{f}^{\rm hk}_t (\boldsymbol{z})=\boldsymbol{0}$ implies $\dot{\Sigma}^{\rm ex;X}_t=\dot{\Sigma}^{\rm X}_t$, $\dot{\Sigma}^{\rm ex;Y}_t=\dot{\Sigma}^{\rm Y}_t$ and $\dot{\Sigma}^{\rm hk;X}_t=\dot{\Sigma}^{\rm hk;Y}_t=0$.

First, we consider a physical interpretation of the nonnegativity $\dot{\Sigma}^{\rm ex;X}_t \geq 0$ and  $\dot{\Sigma}^{\rm ex;Y}_t \geq 0$. We present the results for subsystem $\rm X$; those for $\rm Y$ follow analogously. Using the pseudo energy [Eq.~\eqref{defpseudoenergy}], we can consider the decomposition of the excess heat as follows,
\begin{align}
\dot{Q}^{\rm ex}_t = \dot{Q}^{\rm ex;X}_t+\dot{Q}^{\rm ex;Y}_t,
\label{decompexheat}
\end{align}
where $\dot{Q}^{\rm ex;X}_t:=- \langle \boldsymbol{f}^{\rm ex;X}_t ,-\nabla_{\boldsymbol{x}} U_t^* \rangle_{p_t \mathsf{D}^X}$ and $\dot{Q}^{\rm ex;Y}_t:=- \langle \boldsymbol{f}^{\rm ex;Y}_t ,-\nabla_{\boldsymbol{y}} U_t^* \rangle_{p_t \mathsf{D}^Y}$. The decomposition [Eq.~\eqref{decompexheat}] holds because $\dot{Q}^{\rm ex}_t = - \langle \boldsymbol{f}^{\rm ex}_t ,-\nabla U_t^* \rangle_{p_t \mathsf{D}}$ and $\mathsf{D} = T\mathsf{\mu}$ is a block diagonal matrix [Eq.~\eqref{blockdiag}]. Using this excess heat, the partial excess entropy production rate can be rewritten as
\begin{align}
\dot{\Sigma}^{\rm ex;X}_t =d_t S^{\rm sys;X}_t -\frac{\dot{Q}^{\rm ex; X}_t}{T}  - \dot{I}^{\rm ex;X}_t.
\end{align}
This equation can be confirmed as follows,
\begin{align}
&\dot{\Sigma}^{\rm ex;X}_t \nonumber \\
&= \langle \boldsymbol{f}^{\rm ex;X}_t,  -\nabla_{\boldsymbol{x}} U^*_t/T - \nabla_{\boldsymbol{x}}  \ln p_t \rangle_{p_t \mathsf{D}^{\rm X}} \nonumber\\
&= -\frac{\dot{Q}^{\rm ex; X}_t}{T}  + \langle \boldsymbol{f}^{\rm ex;X}_t,  - \nabla_{\boldsymbol{x}}  (i_t +\ln p^{\rm X}_t) \rangle_{p_t \mathsf{D}^{\rm X}} \nonumber\\
&=d_t S^{\rm sys;X}_t -\frac{\dot{Q}^{\rm ex; X}_t}{T}  - \dot{I}^{\rm ex;X}_t,
\end{align}
where we used $\nabla_{\boldsymbol{x}}\ln p^{\rm Y}_t(\boldsymbol{y})=\boldsymbol{0}$ and $d_t S^{\rm sys;X}_t$ can be calculated as follows,
\begin{align}
&\langle \boldsymbol{f}^{\rm ex;X}_t,  - \nabla_{\boldsymbol{x}}  \ln p^{\rm X}_t \rangle_{p_t \mathsf{D}^{\rm X}} \nonumber\\
&=\int d\boldsymbol{x} \int d\boldsymbol{y} \nabla_{\boldsymbol{x}}  \cdot ( p_t(\boldsymbol{x},\boldsymbol{y}) \mathsf{D}^{\rm X} \boldsymbol{f}^{\rm ex;X}_t (\boldsymbol{x},\boldsymbol{y}) ) \ln p^{\rm X}_t (\boldsymbol{x}) \nonumber \\
&= \int d\boldsymbol{x} \int d\boldsymbol{y} \nabla  \cdot ( p_t(\boldsymbol{x},\boldsymbol{y}) \mathsf{D} \boldsymbol{f}^{\rm ex}_t (\boldsymbol{x},\boldsymbol{y}) ) \ln p^{\rm X}_t (\boldsymbol{x}) \nonumber\\
&=- \int d\boldsymbol{x} \int d\boldsymbol{y} [\partial_t  p_t (\boldsymbol{x},\boldsymbol{y})] \ln p^{\rm X}_t (\boldsymbol{x}) \nonumber\\
&=- \int d\boldsymbol{x}  [\partial_t  p_t^{\rm X} (\boldsymbol{x})] \ln p^{\rm X}_t (\boldsymbol{x}) = d_t S^{\rm sys;X}_t.
\end{align}
Here, we used $\int d\boldsymbol{x}\partial_t  p_t^{\rm X} (\boldsymbol{x})=0$, $\int d\boldsymbol{y} p_t (\boldsymbol{x},\boldsymbol{y})=  p_t^{\rm X} (\boldsymbol{x})$ and integration by parts. Here, we define the apparent excess entropy rate as  
\begin{align}
\sigma_t^{\rm ex;X}:=d_t S^{\rm sys;X}_t -\frac{\dot{Q}^{\rm ex; X}_t}{T}. \label{excess_apparent}
\end{align}
We can also define $\sigma_t^{\rm ex;Y}$ as
$\sigma_t^{\rm ex;Y}:=d_t S^{\rm sys;Y}_t -\dot{Q}^{\rm ex; Y}_t/T$. Using these quantities, the nonnegativity of the partial excess entropy production rates, $\dot{\Sigma}^{\rm ex;X}_t \geq 0$ and $\dot{\Sigma}^{\rm ex;Y}_t \geq 0$, can be interpreted as inequalities
\begin{align}
\sigma_t^{\rm ex;X} \geq \dot{I}^{\rm ex;X}_t,\nonumber\\
\sigma_t^{\rm ex;Y} \geq \dot{I}^{\rm ex;Y}_t,
\label{excessinfothermo2nd}
\end{align}
which can be regarded as generalized second laws of information thermodynamics in terms of the excess dissipation. These inequalities imply that $\sigma_t^{\rm ex;X}$ and $\sigma_t^{\rm ex;Y}$ can be negative. This negativity must be compensated for by a negative value of the excess information flow. If $\sigma_t^{\rm ex;X}$ ($\sigma_t^{\rm ex;Y}$) is negative, the system $\rm Y$ ($\rm X$) can be regarded as Maxwell's demon in terms of excess dissipation. Before proceeding, we introduce terminology that clarifies the physical interpretation
of negative apparent entropy change rates in subsystems. We propose using the term ``excess demon in $\rm Y$ ($\rm X$)'' to describe the negativity of $\sigma^{\rm ex;X}_t$ ($\sigma^{\rm ex;Y}_t$) in the subsystem. For example,  we say that the subsystem $\rm Y$ acts as the excess demon for the subsystem $\rm X$ if $\sigma^{\rm ex;X}_t <0$.

Furthermore, we note that $\sigma_t^{\rm ex;X}  =\langle \boldsymbol{f}^{\rm ex;X}_t,  -\nabla_{\boldsymbol{x}} U^*_t/T - \nabla_{\boldsymbol{x}}  \ln p^{\rm X}_t \rangle_{p_t \mathsf{D}^{\rm X}}$ vanishes in the steady state because $\left. \boldsymbol{f}^{\rm ex;X}_t (\boldsymbol{z})\right|_{p_t=p^{\rm st}} = \boldsymbol{0}$ in the steady state. In that sense, the excess demon only emerges in transient dynamics. If the thermodynamic force $\boldsymbol{f}_t(\boldsymbol{z})$ is conservative, $ \dot{I}_t^{\rm ex;X} = \dot{I}^{\rm X}_t $ and $\dot{\Sigma}^{\rm ex;X}_t =  \dot{\Sigma}^{\rm X}_t$ hold, and thus $\sigma^{\rm ex;X}_t =  \sigma^{\rm X}_t$. Hence, $\sigma^{\rm ex;X}_t$ is negative if and only if $\sigma^{\rm X}_t$ is negative. Thus, the excess demon can be regarded as a conventional Maxwell's demon if the thermodynamic force is conservative.

Next, we consider a physical interpretation of the nonnegativity $\dot{\Sigma}^{\rm hk;X}_t \geq 0$ and  $\dot{\Sigma}^{\rm hk;Y}_t \geq 0$. We present the results for subsystem $\rm X$; those for $\rm Y$ follow analogously. We now introduce an apparent thermodynamic force in the subsystem as follows,
\begin{align}
\boldsymbol{f}^{\rm ap;X}_t (\boldsymbol{x},\boldsymbol{y}) := \frac{\boldsymbol{F}^{\rm X}_t(\boldsymbol{x},\boldsymbol{y})}{T}  - \nabla_{\boldsymbol{x}} \ln p^{\rm X}_t   (\boldsymbol{x}).
\end{align}
This apparent thermodynamic force provides the entropy change rate of the subsystem as follows, $\langle \boldsymbol{f}^{\rm X}_t, \boldsymbol{f}^{\rm ap;X}_t \rangle_{p_t \mathsf{D}^{\rm X}} =\sigma^{X}_t$. Here, we introduce the difference  $\boldsymbol{f}^{\rm ap;X}_t (\boldsymbol{x},\boldsymbol{y}) -\boldsymbol{f}^{\rm ex;X}_t (\boldsymbol{x},\boldsymbol{y})$ as an apparent nonconservative contribution of the thermodynamic force. The current $p_{t} (\boldsymbol{x},\boldsymbol{y}) \mathsf{D}^{\rm X}\boldsymbol{f}^{\rm hk;X}_t (\boldsymbol{x},\boldsymbol{y})$ can be regarded as the nonconservative current, and the quantity
\begin{align}
\sigma^{\rm hk;X}_t:= \langle \boldsymbol{f}^{\rm hk;X}_t,\boldsymbol{f}^{\rm ap;X}_t  -\boldsymbol{f}^{\rm ex;X}_t   \rangle_{p_{t} \mathsf{D}^{\rm X}},
\label{housekeepingepchange}
\end{align}
can be regarded as an apparent nonconservative dissipation. Using $\sigma^{\rm hk;X}_t$, the partial housekeeping entropy production rate can be rewritten as
\begin{align}
\dot{\Sigma}^{\rm hk;X}_t &=  \langle \boldsymbol{f}^{\rm hk;X}_t,\boldsymbol{f}^{\rm X}_t -\boldsymbol{f}^{\rm ex;X}_t + \boldsymbol{f}^{\rm ap;X}_t  -\boldsymbol{f}^{\rm ap;X}_t \rangle_{p_{t} \mathsf{D}^{\rm X}} \nonumber\\
&= \sigma^{\rm hk;X}_t
- \langle \boldsymbol{f}^{\rm hk;X}_t,
\nabla_{\boldsymbol{x}} i_t \rangle_{p_{t} \mathsf{D}^{\rm X}}\nonumber\\
&= \sigma^{\rm hk;X}_t - \dot{I}_t^{\rm hk;X},
\end{align}
where we used $\nabla_{\boldsymbol{x}}\ln p^{\rm Y}_t(\boldsymbol{y})=\boldsymbol{0}$. Similarly, defining $\boldsymbol{f}^{\rm ap;Y}_t (\boldsymbol{x},\boldsymbol{y}) := \boldsymbol{F}^{\rm Y}_t(\boldsymbol{x},\boldsymbol{y})/T  - \nabla_{\boldsymbol{y}} \ln p^{\rm Y}_t   (\boldsymbol{y})$ and $\sigma^{\rm hk;Y}_t:= \langle \boldsymbol{f}^{\rm hk;Y}_t,\boldsymbol{f}^{\rm ap;Y}_t -\boldsymbol{f}^{\rm ex;Y}_t   \rangle_{p_{t} \mathsf{D}^{\rm Y}}$ for system {\rm Y} yields the same result $\dot{\Sigma}^{\rm hk;Y}_t= \sigma^{\rm hk;Y}_t - \dot{I}_t^{\rm hk;Y}$. The nonnegativity of the partial housekeeping entropy production rates, $\dot{\Sigma}^{\rm hk;X}_t \geq 0$ and $\dot{\Sigma}^{\rm hk;Y}_t \geq 0$, can also be interpreted as inequalities
\begin{align}
\sigma_t^{\rm hk;X} \geq \dot{I}^{\rm hk;X}_t,\nonumber\\
\sigma_t^{\rm hk;Y} \geq \dot{I}^{\rm hk;Y}_t,
\label{housekeeping2ndlawofinfothermo}
\end{align}
which can be regarded as generalized second laws of information thermodynamics in terms of the housekeeping dissipation. These inequalities imply that $\sigma_t^{\rm hk;X}$ and $\sigma_t^{\rm hk;Y}$ can be negative, and their negativity must be compensated for by a negative value of the housekeeping information flow. If $\sigma_t^{\rm hk;X}$ ($\sigma_t^{\rm hk;Y}$) is negative, the system $\rm Y$ ($\rm X$) can be regarded as Maxwell's demon in terms of housekeeping dissipation. We propose using the term ``housekeeping demon in $\rm Y$ ($\rm X$)'' to describe the negativity of $\sigma^{\rm hk;X}_t$ ($\sigma^{\rm hk;Y}_t$) in the subsystem. Interestingly, due to the antisymmetric relation [Eq.~\eqref{antisymhk}], the housekeeping information flow cannot be negative in both systems simultaneously, and only one of $\sigma^{\rm hk;X}_t$ or $\sigma^{\rm hk;Y}_t$ can be negative. Regarding the housekeeping demon, it can only exist in one system at a time.

Furthermore, we note that $\sigma_t^{\rm hk;X}$ vanishes if the thermodynamic force is conservative $\boldsymbol{f}^{\rm ex;X}_t (\boldsymbol{z})=\boldsymbol{f}^{\rm X}_t (\boldsymbol{z})$. In that sense, the housekeeping demon only emerges if the thermodynamic force is nonconservative. In the steady state,  $\left. \dot{I}_t^{\rm hk;X} \right|_{p_t=p^{\rm st}}= \left. \dot{I}^{\rm X}_t \right|_{p_t=p^{\rm st}}$ and $\left.\dot{\Sigma}^{\rm hk;X}_t \right|_{p_t=p^{\rm st}}= \left. \dot{\Sigma}^{\rm X}_t \right|_{p_t=p^{\rm st}}$ hold, and thus $\left.\sigma^{\rm hk;X}_t \right|_{p_t=p^{\rm st}}= \left. \sigma^{\rm X}_t \right|_{p_t=p^{\rm st}}$. This implies that the negativity of $\sigma^{\rm hk;X}_t$ in the steady state is equivalent to the negativity of $\sigma^{\rm X}_t$ in the steady state. Thus, the housekeeping demon can be regarded as a conventional autonomous demon in the steady state.

In summary, the excess demon is transient by nature, and may be relevant to the conventional Maxwell's demon in a feedback-measurement protocol driven by a conservative force. By contrast, the housekeeping demon can only emerge in the presence of a nonconservative force, and persist in the steady state, which corresponds to an autonomous demon.

\subsection{Koopman mode decomposition for the partial housekeeping entropy production rate}
\label{sec3c}
Here, we decompose the partial housekeeping entropy production rate using Koopman modes discussed in Sec.~\ref{subsec:geometricdecomposition}. The Koopman mode decomposition of the housekeeping entropy production rate is analogous to the cycle decomposition of the housekeeping entropy production rate in discrete-state Markov jump systems~\cite{maekawa2025geometric}.
We present the results for subsystem $\rm X$; those for $\rm Y$ follow analogously. Because the housekeeping velocity field is given by $\boldsymbol{\nu}_t^{\rm hk}(\boldsymbol{x}, \boldsymbol{y}) = \mathcal{K}{\rm Id}(\boldsymbol{z}) = \sum_i \lambda^{\mathcal{K}}_i \boldsymbol{d}_i g^{\mathcal{K}}_i(\boldsymbol{x}, \boldsymbol{y})$ if the identity function is expanded as ${\rm Id}(\boldsymbol{z}) = \sum_i \boldsymbol{d}_i g^{\mathcal{K}}_i(\boldsymbol{z})$, the housekeeping thermodynamic force in subsystem $\rm X$, $\boldsymbol{f}^{\rm hk;X}_t (\boldsymbol{x},\boldsymbol{y} )$, can be written as $\boldsymbol{f}^{\rm hk;X}_t (\boldsymbol{x},\boldsymbol{y} ) =\sum_i \lambda^{\mathcal{K}}_i (\mathsf{D}^{\rm X})^{-1} \boldsymbol{d}^{\rm X}_i g^{\mathcal{K}}_i(\boldsymbol{x}, \boldsymbol{y})$, where $\boldsymbol{d}^{\rm X}_i$ is the $\rm X$ component of $\boldsymbol{d}_i$, i.e., $\boldsymbol{d}_i^{\top}= ((\boldsymbol{d}^{\rm X}_i)^{\top} , (\boldsymbol{d}^{\rm Y}_i)^{\top})$. For simplicity, we also assume that the Koopman eigenvalues $\{ \lambda^{\mathcal{K}}_i \}$ corresponding to the eigenfunctions $\{ g^{\mathcal{K}}_i(\boldsymbol{z}) \}$ in this expansion are non-degenerate.

As an expression parallel to Eq.~\eqref{koopmandecomp}, the partial housekeeping entropy production rate can be written using the Koopman modes as follows, 
\begin{align}
 &\dot{\Sigma}^{\rm hk;X}_t =\sigma^{\rm hk;X}_t - \dot{I}^{\rm hk;X}_t=\langle ( \boldsymbol{f}^{\rm hk;X}_t )^{\dagger}  , \boldsymbol{f}^{\rm hk;X}_t  \rangle_{p_t \mathsf{D}^{\rm X}} \nonumber\\
 &= \sum_{i|\boldsymbol{d}_i^{\rm X} \neq \boldsymbol{0}} (2\pi)^2 \left(\frac{\Im[\lambda^{\mathcal{K}}_i]}{2 \pi} \right)^2 [(\boldsymbol{d}_i^{\rm X})^{\rm H}  (\mathsf{D}^{\rm X})^{-1}  \boldsymbol{d}^{\rm X}_i] \langle (g^{\mathcal{K}}_i)^\dagger, g^{\mathcal{K}}_i \rangle_{p_t},
 \label{koopmanpartial}
\end{align}
 where we used the fact that $-(\lambda^{\mathcal{K}}_i)^\dagger = \lambda^{\mathcal{K}}_i =\boldsymbol{{\rm i}}\Im[\lambda^{\mathcal{K}}_i] $ if $\langle (g^{\mathcal{K}}_i)^\dagger  , g^{\mathcal{K}}_i \rangle_{p_t } \neq 0$, and $\langle (g^{\mathcal{K}}_i)^\dagger  , g^{\mathcal{K}}_j \rangle_{p_t} = 0$ if $-(\lambda^{\mathcal{K}}_i)^\dagger \neq \lambda^{\mathcal{K}}_j$.
We note that the partial housekeeping entropy production rate is not the sum of all modes in the identity function ${\rm Id}(\boldsymbol{z}) = \sum_i \boldsymbol{d}_i g^{\mathcal{K}}_i(\boldsymbol{z})$, and the contribution of components satisfying $\boldsymbol{d}_i^{\rm X}=\boldsymbol{0}$ does not appear. Housekeeping information flow can also be described using the same Koopman modes as $\dot{I}^{\rm hk; X}_t = \sum_{i| \boldsymbol{d}_i^{\rm X} \neq \boldsymbol{0} }  (\lambda^{\mathcal{K}}_i/2)  \langle   \nabla_{\boldsymbol{x}} i_t,\boldsymbol{d}^{\rm X}_i g^{\mathcal{K}}_i \rangle_{ p_t \mathsf{I}} + {\rm c.c.}$, where we used the fact that the same quantity can be described using complex conjugation ${\rm c.c.}$ because $\boldsymbol{f}_t^{\rm hk;X}(\boldsymbol{x},\boldsymbol{y}) = [\boldsymbol{f}_t^{\rm hk;X}(\boldsymbol{x},\boldsymbol{y})]^{\dagger}$.

\subsection{Thermodynamic uncertainty relations for the partial excess entropy production rate}
\label{sec3d}
The fact that the partial excess entropy production rate $\dot{\Sigma}^{\rm ex;X}_t$ contains information about the time evolution of the subsystem $\rm X$ can be demonstrated by the fact that it yields the lower bound in the form of the thermodynamic uncertainty relation analogous to Eq.~\eqref{turexcess}. We present the results for subsystem $\rm X$; those for $\rm Y$ follow analogously.

The lower bound can be obtained using the Cauchy-Schwarz inequality as follows,
\begin{align}
 \dot{\Sigma}^{\rm ex;X}_t &= \langle \boldsymbol{f}^{\rm ex;X}_t, \boldsymbol{f}^{\rm ex;X}_t\rangle_{p_t \mathsf{D}^{\rm X}} \nonumber\\
 &\geq \frac{(\langle -\nabla_{\boldsymbol{x}} \psi^{\rm X}, \boldsymbol{f}^{\rm ex;X}_t\rangle_{p_t \mathsf{D}^{\rm X}})^2}{\langle -\nabla_{\boldsymbol{x}} \psi^{\rm X}, -\nabla_{\boldsymbol{x}} \psi^{\rm X} \rangle_{p_t \mathsf{D}^{\rm X}}},
 \label{csX}
\end{align}
where $\psi^{\rm X}(\boldsymbol{x})$ is a function that depends only on $\boldsymbol{x}$.

Here, the quantity $\langle -\nabla_{\boldsymbol{x}} \psi^{\rm X}, \boldsymbol{f}^{\rm ex;X}_t\rangle_{p_t \mathsf{D}^{\rm X}}$ is calculated as 
\begin{align}
& \langle -\nabla_{\boldsymbol{x}} \psi^{\rm X}, \boldsymbol{f}^{\rm ex;X}_t\rangle_{p_t \mathsf{D}^{\rm X}} \nonumber\\
&= \int d \boldsymbol{x} \int d\boldsymbol{y} \nabla_{\boldsymbol{x}} \cdot [p_t(\boldsymbol{x},\boldsymbol{y}) \mathsf{D}^{\rm X} \boldsymbol{f}^{\rm ex;X}_t(\boldsymbol{x},\boldsymbol{y})] \psi^{\rm X}(\boldsymbol{x}) \nonumber\\
&= \int d \boldsymbol{x} \int d\boldsymbol{y} \nabla \cdot [p_t(\boldsymbol{x},\boldsymbol{y}) \mathsf{D} \boldsymbol{f}^{\rm ex}_t(\boldsymbol{x},\boldsymbol{y})] \psi^{\rm X}(\boldsymbol{x}) \nonumber\\
&= -\int d \boldsymbol{x} \int d\boldsymbol{y} [ \partial_t p_t(\boldsymbol{x},\boldsymbol{y}) ] \psi^{\rm X}(\boldsymbol{x}) \nonumber\\
&= -\int d \boldsymbol{x}  \partial_t p^{\rm X}_t(\boldsymbol{x})  \psi^{\rm X}(\boldsymbol{x}) = - \partial_t \mathbb{E}_{p^{\rm X}_t}[\psi^{\rm X}],
\label{expectedX}
\end{align}
where $\mathbb{E}_{p^{\rm X}_t}[\psi^{\rm X}]:=\int d \boldsymbol{x} p^{\rm X}_t(\boldsymbol{x})  \psi^{\rm X}(\boldsymbol{x})$. Here we used $\partial_t p_t(\boldsymbol{x},\boldsymbol{y}) = -\nabla \cdot [p_t(\boldsymbol{x},\boldsymbol{y}) \mathsf{D} \boldsymbol{f}^{\rm ex}_t(\boldsymbol{x},\boldsymbol{y})]$, $\int d\boldsymbol{y} p_t(\boldsymbol{x},\boldsymbol{y})=p^{\rm X}_t(\boldsymbol{x})$ and integration by parts.

Next, we consider the quantity $\Delta \psi^{\rm X}(\boldsymbol{x}(t+dt), \boldsymbol{x}(t)):=   \psi^{\rm X}(\boldsymbol{x}(t+dt))-\psi^{\rm X}(\boldsymbol{x}(t))\simeq \boldsymbol{\dot{x}}(t) dt \circ \nabla_{\boldsymbol{x}} \psi^{\rm X}(\boldsymbol{x}(t))$ in the Langevin description [Eq.~\eqref{LangevinX}]. Its variance to leading order in $dt$ is calculated as follows, 
\begin{align}
{\rm Var}[\Delta \psi^{\rm X}] &\simeq {\rm Var}[\sqrt{2 T} \mathsf{B}^{\rm X}\boldsymbol{\xi}^{\rm X}_t dt \circ \nabla_{\boldsymbol{x}} \psi^{\rm X}] \nonumber\\
&= 2 dt\int d\boldsymbol{x} [\nabla_{\boldsymbol{x}} \psi^{\rm X}(\boldsymbol{x})]^{\top} \mathsf{D}^{\rm X}[\nabla_{\boldsymbol{x}} \psi^{\rm X}(\boldsymbol{x} ) ] p^{\rm X}_t(\boldsymbol{x}) \nonumber\\
&= 2 dt\langle -\nabla_{\boldsymbol{x}} \psi^{\rm X}, -\nabla_{\boldsymbol{x}} \psi^{\rm X}\rangle_{p_t \mathsf{D}^{\rm X}},
\label{VarianceX}
\end{align}
where we used Ito calculus $\mathbb{E}[ (\boldsymbol{\xi}^{\rm X}_t dt) (\boldsymbol{\xi}^{\rm X}_t dt)^{\top}]\simeq \mathsf{I} dt$ and $T\mathsf{B}^{\rm X} (\mathsf{B}^{\rm X})^{\top} =\mathsf{D}^{\rm X}$. 

Using the expressions in Eqs.~\eqref{expectedX} and~\eqref{VarianceX}, the Cauchy-Schwarz inequality [Eq.~\eqref{csX}] can be rewritten as 
\begin{align}
 \dot{\Sigma}^{\rm ex;X}_t &\geq \frac{(\partial_t \mathbb{E}_{p^{\rm X}_t} [\psi^{\rm X}])^2}{\lim_{dt \to 0}\frac{{\rm Var}[\Delta \psi^{\rm X}]}{2dt}},
 \label{TURpartialexcess}
\end{align}
which can be regarded as the thermodynamic uncertainty relation for the partial excess entropy production rate. This relation states that the partial excess entropy production rate of $\rm X$ is an upper bound on the ratio of the squared rate of change of the mean of any function depending only on $\rm X$ to its short-time fluctuations.  This result can also be regarded as an upper bound on excess information flow,
\begin{align}
 \sigma^{\rm ex;X}_t -\frac{(\partial_t \mathbb{E}_{p^{\rm X}_t} [\psi^{\rm X}])^2}{\lim_{dt \to 0}\frac{{\rm Var}[\Delta \psi^{\rm X}]}{2dt}} &\geq \dot{I}^{\rm ex;X}_t.
\end{align}
\subsection{Local excess and housekeeping entropy production rates}
\label{sec3e}
Here, we consider the geometric decomposition of the partial entropy production rate $\dot{\Sigma}^{\rm X}_t$ into excess and housekeeping contributions. We emphasize that ``partial” refers to subsystem-wise decomposition,
whereas ``local” refers to decomposition with respect to the marginal dynamics.
We present the results for subsystem $\rm X$; those for $\rm Y$ follow analogously.

We first define the function $\phi^{\rm X}_t(\boldsymbol{x})$ as a solution of the following equation
\begin{align}
-\nabla_{\boldsymbol{x}} \cdot [p_t^{\rm X} (\boldsymbol{x}) \mathsf{D}^{\rm X} \bar{\boldsymbol{f}}^{\rm X}_t (\boldsymbol{x}) ] =\nabla_{\boldsymbol{x}} \cdot [p^{\rm X}_t(\boldsymbol{x})\mathsf{D}^{\rm X}(\nabla_{\boldsymbol{x}} \phi^{\rm X}_t (\boldsymbol{x})  )],
\label{solutionphiX}
\end{align}
where $\bar{\boldsymbol{f}}^{\rm X}_t (\boldsymbol{x})$ is defined as 
\begin{align}
\bar{\boldsymbol{f}}^{\rm X}_t (\boldsymbol{x}) := \int d\boldsymbol{y} \boldsymbol{f}^{\rm X}_t (\boldsymbol{x},\boldsymbol{y}) p_t^{\rm Y \vert X}(\bs{y} \vert \bs{x}).
\label{apparentXforce}
\end{align}
From the Fokker-Planck equation [Eq.~\eqref{continuityeq}], we can obtain the time evolution of the marginal distribution $p^{\rm X}_t(\boldsymbol{x})$ as follows,
\begin{align}
\partial_t p^{\rm X}_t(\boldsymbol{x}) &= \partial_t \int d\boldsymbol{y} p_t(\boldsymbol{x},\boldsymbol{y}) \nonumber\\
&= -\nabla_{\boldsymbol{x}} \cdot [p_t^{\rm X} (\boldsymbol{x}) \mathsf{D}^{\rm X} \bar{\boldsymbol{f}}^{\rm X}_t (\boldsymbol{x}) ],
\end{align}
where we used Eq.~\eqref{apparentXforce}, and assumed boundary terms vanish $-\int d\boldsymbol{y} \nabla_{\boldsymbol{y}} \cdot (p_t(\boldsymbol{x},\boldsymbol{y}) \mathsf{D}^{\rm Y} \boldsymbol{f}^{\rm Y}_t(\boldsymbol{x},\boldsymbol{y}))=0$. Therefore, the function $\phi^{\rm X}_t(\boldsymbol{x})$ provides the time evolution of the marginal distribution as follows,
\begin{align}
\partial_t p^{\rm X}_t(\boldsymbol{x}) &= -\nabla_{\boldsymbol{x}} \cdot [p^{\rm X}_t(\boldsymbol{x})\mathsf{D}^{\rm X}(-\nabla_{\boldsymbol{x}} \phi^{\rm X}_t (\boldsymbol{x})  )].
\label{continuityX}
\end{align}

We next show the orthogonality $\langle -\nabla_{\boldsymbol{x}}\phi^{\rm X}_t, \boldsymbol{f}^{\rm X}_t +\nabla_{\boldsymbol{x}} \phi^{\rm X}_t\rangle_{p_t\mathsf{D}^{\rm X}}=0$ as follows,
\begin{align}
&\langle -\nabla_{\boldsymbol{x}}\phi^{\rm X}_t, \boldsymbol{f}^{\rm X}_t +\nabla_{\boldsymbol{x}} \phi^{\rm X}_t \rangle_{p_t\mathsf{D}^{\rm X}} \nonumber\\
&= \int d\boldsymbol{x} \phi^{\rm X}_t (\boldsymbol{x}) \nabla_{\boldsymbol{x}} \cdot [p_t^{\rm X}(\boldsymbol{x}) \mathsf{D}^{\rm X}(\bar{\boldsymbol{f}}^{\rm X}_t (\boldsymbol{x})+ \nabla_{\boldsymbol{x}} \phi^{\rm X}_t(\boldsymbol{x}))]\nonumber\\
&=0,
\label{orthogonalloc}
\end{align}
where we used Eqs.~\eqref{solutionphiX} and ~\eqref{apparentXforce} and integration by parts.

Based on orthogonality [Eq.~\eqref{orthogonalloc}], we obtain the geometric decomposition of the partial entropy production rate as follows.
\begin{align}
\dot{\Sigma}_t^{\rm X} = \dot{\Sigma}_t^{\rm localex; X}+\dot{\Sigma}_t^{\rm localhk; X},
\label{geometricdecomploc}
\end{align}
where $\dot{\Sigma}_t^{\rm localex; X}$ and $\dot{\Sigma}_t^{\rm localhk; X}$ are the local excess entropy production rate and  the local housekeeping entropy production rate, defined as
\begin{align}
\dot{\Sigma}_t^{\rm localex; X} &:= \langle  -\nabla_{\boldsymbol{x}}\phi^{\rm X}_t,  -\nabla_{\boldsymbol{x}}\phi^{\rm X}_t\rangle_{p_t \mathsf{D}^{\rm X}} \: (\geq 0),
\end{align}
and
\begin{align}
\dot{\Sigma}_t^{\rm localhk; X} &:=\langle  \boldsymbol{f}^{\rm X}_t +\nabla_{\boldsymbol{x}} \phi^{\rm X}_t , \boldsymbol{f}^{\rm X}_t +\nabla_{\boldsymbol{x}} \phi^{\rm X}_t \rangle_{p_t \mathsf{D}^{\rm X}} \:(\geq 0),
\end{align}
respectively. This geometric decomposition [Eq.~\eqref{geometricdecomploc}] holds because the identity $\langle  \boldsymbol{f}^{\rm X}_t , \boldsymbol{f}^{\rm X}_t \rangle_{p_t \mathsf{D}^{\rm X}} = \langle  -\nabla_{\boldsymbol{x}}\phi^{\rm X}_t,  -\nabla_{\boldsymbol{x}}\phi^{\rm X}_t\rangle_{p_t \mathsf{D}^{\rm X}} + \langle  \boldsymbol{f}^{\rm X}_t +\nabla_{\boldsymbol{x}} \phi^{\rm X}_t , \boldsymbol{f}^{\rm X}_t +\nabla_{\boldsymbol{x}} \phi^{\rm X}_t \rangle_{p_t \mathsf{D}^{\rm X}} + 2 \langle -\nabla_{\boldsymbol{x}}\phi^{\rm X}_t, \boldsymbol{f}^{\rm X}_t +\nabla_{\boldsymbol{x}} \phi^{\rm X}_t \rangle_{p_t\mathsf{D}^{\rm X}}$ and Eq.~\eqref{orthogonalloc} hold.

This geometric decomposition is related to local conservativeness and local stationarity.
The local housekeeping entropy production rate becomes zero if the subsystem is locally conservative and the thermodynamic force acting on the subsystem does not depend on the state of the other subsystem.
 The subsystem $\rm X$ is called locally conservative when there exists a potential $\phi^{\rm X}_t(\boldsymbol{x})$ such that $\bar{\boldsymbol{f}}^{\rm X}_t (\boldsymbol{x})=-\nabla_{\boldsymbol{x}} \phi^{\rm X}_t(\boldsymbol{x})$. The condition that the thermodynamic force acting on $\rm X$ does not depend on $\boldsymbol{y}$ is expressed as $\boldsymbol{f}^{\rm X}_t (\boldsymbol{x},\boldsymbol{y})=\bar{\boldsymbol{f}}^{\rm X}_t (\boldsymbol{x})$. Therefore, the local housekeeping entropy production rate vanishes when $\boldsymbol{f}^{\rm X}_t (\boldsymbol{x},\boldsymbol{y})=-\nabla_{\boldsymbol{x}} \phi^{\rm X}_t(\boldsymbol{x})$.
On the other hand, the local excess entropy production rate becomes zero if the subsystem is in a local steady state $\left. \partial_t p^{\rm X}_t(\boldsymbol{x}) \right|_{p^{\rm X}_t=p^{\rm st;X}}= \left. -\nabla_{\boldsymbol{x}} \cdot [p^{\rm X}_t(\boldsymbol{x})\mathsf{D}^{\rm X}(-\nabla_{\boldsymbol{x}} \phi^{\rm X}_t (\boldsymbol{x})  )] \right|_{p^{\rm X}_t=p^{\rm st;X}}=0$. This fact can be confirmed because we can rewrite $\dot{\Sigma}_t^{\rm localex; X}$ as follows,
\begin{align}
\dot{\Sigma}_t^{\rm localex; X} &= \langle  -\nabla_{\boldsymbol{x}}\phi^{\rm X}_t,  -\nabla_{\boldsymbol{x}}\phi^{\rm X}_t\rangle_{p_t \mathsf{D}^{\rm X}} \nonumber\\
&= \int d\boldsymbol{x} \phi_t^{\rm X} (\boldsymbol{x}) \nabla_{\boldsymbol{x}} \cdot [p^{\rm X}_t(\boldsymbol{x})\mathsf{D}^{\rm X}(-\nabla_{\boldsymbol{x}} \phi^{\rm X}_t (\boldsymbol{x})  )]\nonumber \\
&=-\int d\boldsymbol{x} \phi_t^{\rm X} (\boldsymbol{x}) [\partial_t p_t^{\rm X} (\boldsymbol{x})],
\end{align}
where we used Eq.~\eqref{continuityX}, $\int d\boldsymbol{y} p_t(\boldsymbol{x},\boldsymbol{y})=p^{\rm X}_t(\boldsymbol{x})$ and integration by parts.

\subsection{Variational formulas}
\label{sec3f}
We discuss variational formulas for the local excess and local housekeeping entropy production rates in parallel with Eqs.~\eqref{BBformula}, ~\eqref{TUR1}, ~\eqref{variational formulas2} and ~\eqref{TUR2}.
The local excess and local housekeeping entropy production rates can be formulated in the form of a dual optimization problem
(see also Appendix~\ref{appendix:variationalformula}). We present the results for subsystem $\rm X$; those for $\rm Y$ follow analogously.

The local excess entropy production rate is given by
\begin{align}
 \dot{\Sigma}^{\rm localex;X}_{t} &= \inf_{{\boldsymbol{f}^{\rm X}_t}' (\boldsymbol{x},\boldsymbol{y})|\partial_t p^{\rm X}_t = -\int d\boldsymbol{y}\nabla_{\boldsymbol{x}} \cdot ( p_t\mathsf{D}^{\rm X}{\boldsymbol{f}^{\rm X}_t}')} \langle {\boldsymbol{f}^{\rm X}_t}', {\boldsymbol{f}^{\rm X}_t}' \rangle_{p_t \mathsf{D}^{\rm X}} \label{BBformula2}\\
 &= \sup_{\psi^{\rm X}(\boldsymbol{x})} \frac{(\langle -\nabla_{\boldsymbol{x}} \psi^{\rm X}, \boldsymbol{f}^{\rm X}_t\rangle_{p_t \mathsf{D}^{\rm X}})^2}{\langle -\nabla_{\boldsymbol{x}} \psi^{\rm X}, -\nabla_{\boldsymbol{x}} \psi^{\rm X}\rangle_{p_t \mathsf{D}^{\rm X}}}.  \label{TUR3}
\end{align}
Similarly, the local housekeeping entropy production rate can also be expressed as
\begin{align}
 &\dot{\Sigma}^{\rm localhk;X}_t \nonumber\\
 &= \inf_{\psi^{\rm X}(\boldsymbol{x})} \langle \boldsymbol{f}^{\rm X}_t +\nabla_{\boldsymbol{x}} \psi^{\rm X}, \boldsymbol{f}^{\rm X}_t + \nabla_{\boldsymbol{x}} \psi^{\rm X} \rangle_{p_t \mathsf{D}^{\rm X}}\label{variationallocal} \\
 &= \sup_{{\boldsymbol{f}^{\rm X}_t}' (\boldsymbol{x},\boldsymbol{y})|-\int d\boldsymbol{y} \nabla_{\boldsymbol{x}} \cdot ( p_t\mathsf{D}^{\rm X}{\boldsymbol{f}^{\rm X}_t}') =0} \frac{(\langle {\boldsymbol{f}^{\rm X}_t}', \boldsymbol{f}^{\rm X}_t\rangle_{p_t \mathsf{D}^{\rm X}})^2}{\langle {\boldsymbol{f}^{\rm X}_t}', {\boldsymbol{f}^{\rm X}_t}' \rangle_{p_t \mathsf{D}^{\rm X}}}. 
 \label{TUR4}
\end{align}
These are consequences of the fact that the image ${\rm im}[\nabla_{\boldsymbol{x}}] := \{\nabla_{\boldsymbol{x}} \psi^{\rm X}| \psi^{\rm X}(\boldsymbol{x}) \in \mathbb{R} \}$ and the kernel ${\rm ker}[\int d\boldsymbol{y}  \nabla_{\boldsymbol{x}} \cdot p_t\mathsf{D}^{\rm X}] :=\{ {\boldsymbol{f}^{\rm X}_t}'| \int d\boldsymbol{y}  \nabla_{\boldsymbol{x}} \cdot [p
_t(\boldsymbol{x},\boldsymbol{y})\mathsf{D}^{\rm X} {\boldsymbol{f}^{\rm X}_t}'(\boldsymbol{x},\boldsymbol{y})] =0 \} $ are orthogonal when $p_t\mathsf{D}^{\rm X}$ is considered as a metric and that a point $-\nabla_{\boldsymbol{x}} \phi^{\rm X}_t$ is obtained via the projection of $\boldsymbol{f}^{\rm X}_t$ onto ${\rm im}[\nabla_{\boldsymbol{x}}]$. We also note that we can obtain another expression of $\dot{\Sigma}^{\rm localex;X}_t$ using $\dot{\Sigma}^{\rm localex;X}_t=\dot{\Sigma}^{\rm X}_t- \dot{\Sigma}^{\rm localhk;X}_t$ as follows, 
\begin{align}
 &\dot{\Sigma}^{\rm localex;X}_t\nonumber\\
 &= \sup_{\psi^{\rm X}(\boldsymbol{x})} \left( 2\langle \boldsymbol{f}^{\rm X}_t, -\nabla_{\boldsymbol{x}} \psi^{\rm X} \rangle_{p_t \mathsf{D}^{\rm X}} -\langle\nabla_{\boldsymbol{x}} \psi^{\rm X},\nabla_{\boldsymbol{x}} \psi^{\rm X} \rangle_{p_t \mathsf{D}^{\rm X}}\right).
\end{align}

Based on the variational formula [Eq.~\eqref{BBformula2}], we obtain the following inequality
\begin{align}
 &\dot{\Sigma}^{\rm localex;X}_t \leq  \langle \boldsymbol{f}^{\rm ex;X}_t, \boldsymbol{f}^{\rm ex;X}_t\rangle_{p_t \mathsf{D}^{\rm X}} =\dot{\Sigma}^{\rm ex;X}_t, 
 \label{localvspartial}
\end{align}
because $\boldsymbol{f}^{\rm ex;X}_t (\boldsymbol{x},\boldsymbol{y})$
 satisfies the constraint of the optimization problem [Eq.~\eqref{BBformula2}] as follows,
\begin{align}
 \partial_t p^{\rm X}_t (\boldsymbol{x}) =& \partial_t \int d\boldsymbol{y} p_t (\boldsymbol{x},\boldsymbol{y}) \nonumber\\
 =& -\int d\boldsymbol{y} \nabla_{\boldsymbol{x}} \cdot  ( p_t(\boldsymbol{x}, \boldsymbol{y})\mathsf{D}^{\rm X}\boldsymbol{f}^{\rm ex;X}_t(\boldsymbol{x}, \boldsymbol{y})) \nonumber\\
 &-\int d\boldsymbol{y} \nabla_{\boldsymbol{y}} \cdot  ( p_t(\boldsymbol{x}, \boldsymbol{y})\mathsf{D}^{\rm Y}\boldsymbol{f}^{\rm ex;Y}_t(\boldsymbol{x}, \boldsymbol{y})) \nonumber\\
 =& -\int d\boldsymbol{y} \nabla_{\boldsymbol{x}} \cdot  ( p_t(\boldsymbol{x}, \boldsymbol{y})\mathsf{D}^{\rm X}\boldsymbol{f}^{\rm ex;X}_t(\boldsymbol{x}, \boldsymbol{y})),
 \label{partialpxex}
\end{align}
where we used $p^{\rm X}_t (\boldsymbol{x})=\int d\boldsymbol{y} p_t (\boldsymbol{x},\boldsymbol{y}) $, $\partial_t p_t (\boldsymbol{z}) =-\nabla \cdot [p_t (\boldsymbol{z}) \mathsf{D} \boldsymbol{f}_t^{\rm ex} (\boldsymbol{z})]$ and integration by parts. If $\boldsymbol{f}^{\rm ex;X}_t (\boldsymbol{x}, \boldsymbol{y}) = -\nabla_{\boldsymbol{x}} \phi_t(\boldsymbol{x}, \boldsymbol{y})$ does not depend on $\boldsymbol{y}$, $\boldsymbol{f}^{\rm ex;X}_t (\boldsymbol{x}, \boldsymbol{y})$ can be interpreted as $-\nabla_{\boldsymbol{x}}\phi^{\rm X}_t (\boldsymbol{x})$ that satisfies Eq.~\eqref{solutionphiX}, and the equality 
$\dot{\Sigma}^{\rm ex;X}_t=\dot{\Sigma}^{\rm localex;X}_t$ can hold. In the steady state, $\dot{\Sigma}^{\rm ex;X}_t$ vanishes and thus $\dot{\Sigma}^{\rm localex;X}_t$ vanishes due to Eq.~\eqref{localvspartial}. Therefore, $\left. \dot{\Sigma}^{\rm X}_t \right|_{p_t=p^{\rm st}}=\left. \dot{\Sigma}^{\rm localhk;X}_t \right|_{p_t=p^{\rm st}}$ in the steady state.

The variational formula [Eq.~\eqref{TUR3}] provides the thermodynamic uncertainty relations for the local excess entropy production rate, which gives the same lower bound in Eq.~\eqref{TURpartialexcess}. By performing calculations similar to those in Eq.~\eqref{expectedX}, we obtain 
\begin{align}
&\langle -\nabla_{\boldsymbol{x}} \psi^{\rm X}, \boldsymbol{f}^{\rm X}_t\rangle_{p_t \mathsf{D}^{\rm X}} \nonumber\\
&= \int d\boldsymbol{x} \psi^{\rm X}(\boldsymbol{x}) \int d\boldsymbol{y} \nabla_{\boldsymbol{x}} \cdot [p_t(\boldsymbol{x},\boldsymbol{y}) \mathsf{D}^{\rm X} \boldsymbol{f}^{\rm X}_t (\boldsymbol{x},\boldsymbol{y})] \nonumber\\
&= -\int d\boldsymbol{x} \psi^{\rm X}(\boldsymbol{x})  \partial_t p_t^{\rm X}(\boldsymbol{x}) = -\partial_t \mathbb{E}_{p_t^{\rm X}}[\psi^{\rm X} ].
\label{expectedX2}
\end{align}
Using the expressions in Eqs.~\eqref{expectedX2} and~\eqref{VarianceX}, the variational formula [Eq.~\eqref{TUR3}] and Eq.~\eqref{localvspartial} can be rewritten as
\begin{align}
\dot{\Sigma}^{\rm ex;X}_t \geq \dot{\Sigma}^{\rm localex;X}_t &\geq \frac{(\partial_t \mathbb{E}_{p^{\rm X}_t} [\psi^{\rm X}])^2}{\lim_{dt \to 0}\frac{{\rm Var}[\Delta \psi^{\rm X}]}{2dt}}.
\end{align}
This implies that the inequality using $\dot{\Sigma}^{\rm localex;X}_t$ is tighter than the inequality using $\dot{\Sigma}^{\rm ex;X}_t$ [Eq.~\eqref{TURpartialexcess}].

\subsection{Optimal transport for the subsystem}
\label{sec3g}
The local excess entropy production rate is related to the generalized $2$-Wasserstein distance for the marginal distributions. This relationship provides the information-thermodynamic speed limit, which is the lower bound on the partial excess entropy production. This result implies that the optimal transport cost in the subsystem provides a lower bound on the partial excess dissipation.
We present the results for subsystem $\rm X$; those for $\rm Y$ follow analogously.

We discuss the relationship between $\dot{\Sigma}^{\rm localex;X}_t$ and the generalized $2$-Wasserstein distance. We consider another variational formula for the local excess entropy production rate (see also Appendix~\ref{appendix:variationalformula}) as
\begin{align}
 &\dot{\Sigma}^{\rm localex;X}_{t} = \inf_{\psi^{\rm X} (\boldsymbol{x})} \langle - \nabla_{\boldsymbol{x}} \psi^{\rm X}, - \nabla_{\boldsymbol{x}} \psi^{\rm X} \rangle_{p^{\rm X}_t \mathsf{D}^{\rm X}} \nonumber\\
 &{\rm s.t.} \: \:  \partial_t p^{\rm X}_t(\boldsymbol{x}) = -\nabla_{\boldsymbol{x}} \cdot [p_t^{\rm X} (\boldsymbol{x}) \mathsf{D}^{\rm X} (-\nabla_{\boldsymbol{x}} \psi^{\rm X} (\boldsymbol{x}) )].
 \label{BBformula3}
\end{align}
By comparing Eq.~\eqref{BBformula3} with Eq.~\eqref{generalBBformula3}, we obtain an expression of the local excess entropy production rate as follows,
\begin{align}
\dot{\Sigma}_t^{\rm localex;X}= \lim_{\Delta t \to 0} \frac{[\tilde{\mathcal{W}}_2^{(\mathsf{D}^{\rm X})^{-1}} (p_t^{\rm X}, p_{t+\Delta t}^{\rm X})]^2}{(\Delta t)^2}.
\label{localwasserstein}
\end{align}

We note that the relationship between the generalized $2$-Wasserstein distance and the local excess entropy production rate is generalized using the inequality for the general Markov jump systems~\cite{maekawa2025geometric}. Because the Onsager matrix is not proportional to the probability distribution for general Markov jump systems, we need to consider a new definition of the generalized $2$-Wasserstein distance for the subsystem~\cite{maekawa2025geometric}, which is analogous to the Maas' generalization of the $2$-Wasserstein distance~\cite{maas2011gradient,yoshimura2023housekeeping}. As shown here, the relationship between the well-defined generalized 2-Wasserstein distance and the local excess entropy production rate holds for the overdamped Langevin system.

\subsection{Information-thermodynamic speed limit for the partial excess entropy production}
\label{sec3h}
We discuss the information-thermodynamic speed limit as a generalization of the result in Ref.~\cite{nakazato2021geometrical}. 
We present the results for subsystem $\rm X$; those for $\rm Y$ follow analogously.

We discuss the hierarchy of the entropy production rates. The hierarchy of the entropy production rates can provide the inequalities
\begin{align}
\dot{\Sigma}_t \geq \dot{\Sigma}^{\rm ex}_t \geq  \dot{\Sigma}^{\rm ex; X}_t \geq \dot{\Sigma}^{\rm localex; X}_t, \nonumber\\
\dot{\Sigma}_t \geq \dot{\Sigma}^{\rm X}_t \geq \dot{\Sigma}^{\rm localex; X}_t.
\end{align}
In Ref.~\cite{nakazato2021geometrical}, the inequality $\dot{\Sigma}^{\rm X}_t \geq \dot{\Sigma}^{\rm localex; X}_t$ has been discussed as a generalization of the second law of information thermodynamics using the following expression, 
\begin{align}
\sigma^{X}_t -\dot{I}^{\rm X}_t \geq \lim_{\Delta t \to 0} \frac{[\tilde{\mathcal{W}}_2^{(\mathsf{D}^{\rm X})^{-1}} (p_t^{\rm X}, p_{t+\Delta t}^{\rm X})]^2}{(\Delta t)^2}. 
\label{nakazatoito}
\end{align}
This expression has been used to derive the information-thermodynamic speed limit in Ref.~\cite{nakazato2021geometrical}.

To generalize the information-thermodynamic speed limit, we consider the expression based on the inequality $\dot{\Sigma}^{\rm ex; X}_t \geq \dot{\Sigma}^{\rm localex; X}_t$. 
The inequality $\dot{\Sigma}^{\rm ex; X}_t \geq \dot{\Sigma}^{\rm localex; X}_t$ can be rewritten as 
\begin{align}
\sigma^{\rm ex;X}_t -\dot{I}^{\rm ex; X}_t \geq \lim_{\Delta t \to 0} \frac{[\tilde{\mathcal{W}}_2^{(\mathsf{D}^{\rm X})^{-1}} (p_t^{\rm X}, p_{t+\Delta t}^{\rm X})]^2}{(\Delta t)^2}.
\end{align}
Using this expression, we obtain a generalized information-thermodynamic speed limit as follows, 
\begin{align}
&\int_0^{\tau} dt \sigma^{\rm ex;X}_t - \int_0^{\tau} dt \dot{I}^{\rm ex; X}_t \nonumber\\
&\geq \int_0^{\tau} dt \left[\lim_{\Delta t \to 0} \frac{[\tilde{\mathcal{W}}_2^{(\mathsf{D}^{\rm X})^{-1}} (p_t^{\rm X}, p_{t+\Delta t}^{\rm X})]^2}{(\Delta t)^2} \right] \nonumber \\
&\geq \frac{(\mathcal{L}_{\tau}^{(\mathsf{D}^{\rm X})^{-1}} )^2}{\tau} \geq \frac{[\tilde{\mathcal{W}}_2^{(\mathsf{D}^{\rm X})^{-1}} (p_0^{\rm X}, p_{\tau}^{\rm X})]^2}{\tau},
\label{inequalityinfospeedlimit}
\end{align}
where the generalized $2$-Wasserstein path length $\mathcal{L}_{\tau}^{(\mathsf{D}^{\rm X})^{-1}}$ is defined as
\begin{align}
\mathcal{L}_{\tau}^{(\mathsf{D}^{\rm X})^{-1}} := \int_0^{\tau} dt \left[\lim_{\Delta t \to 0} \frac{\tilde{\mathcal{W}}_2^{(\mathsf{D}^{\rm X})^{-1}} (p_t^{\rm X}, p_{t+\Delta t}^{\rm X})}{\Delta t} \right], 
\end{align}
and we used the Cauchy-Schwarz inequality and the triangle inequality $\mathcal{L}_{\tau}^{(\mathsf{D}^{\rm X})^{-1}} \geq \tilde{\mathcal{W}}_2^{(\mathsf{D}^{\rm X})^{-1}} (p_0^{\rm X}, p_{\tau}^{\rm X})$.
This result suggests that the time integral of excess information flow has the following upper bound:
\begin{align}
&\int_0^{\tau} dt \sigma^{\rm ex;X}_t - \frac{[\tilde{\mathcal{W}}_2^{(\mathsf{D}^{\rm X})^{-1}} (p_0^{\rm X}, p_{\tau}^{\rm X})]^2}{\tau} \geq \int_0^{\tau} dt \dot{I}^{\rm ex; X}_t.
\label{excessinformationflowwassersteinbound}
\end{align}
This result imposes a tighter constraint on excess information flow than the generalized second law of information thermodynamics [Eq.~\eqref{excessinfothermo2nd}].

\subsection{Information flows and Fisher information}
\label{sec3i}
We can obtain relationships between the conditional Fisher information and excess and housekeeping information flows as a generalization of the result in Ref.~\cite{matsumoto2025learning}. We present the results for subsystem $\rm X$; those for $\rm Y$ follow analogously.

Applying the Cauchy-Schwarz inequality to the excess and housekeeping information flows defined in Eqs.~\eqref{infodecompositiondef} and~\eqref{infodecompositiondef2}, we then obtain 
\begin{subequations}
\begin{align}
    \big(\dot{I}_t^{\rm ex;X} \big)^2 &\leq \dot{\Sigma}_t^{\rm ex;X} \mathcal{I}_t^{\rm Fisher;X}, \label{cauchy-scharz-loose}\\
    \big(\dot{I}_t^{\rm hk;X} \big)^2 &\leq \dot{\Sigma}_t^{\rm hk;X} \mathcal{I}_t^{\rm Fisher;X},
\end{align}
\end{subequations}
which are generalizations of Eq.~\eqref{Fisherbound}.
For the apparent excess entropy change rate [Eq.~\eqref{excess_apparent}] and the apparent housekeeping entropy change rate [Eq.~\eqref{housekeepingepchange}], these bounds can be rewritten as
\begin{subequations}
\begin{align}
    \sigma_t^{\rm ex;X} = \dot{\Sigma}_t^{\rm ex; X} + \dot{I}_t^{\rm ex;X} \geq \frac{(\dot{I}_t^{\rm ex;X})^2}{\mathcal{I}_t^{\rm Fisher;X}} + \dot{I}_t^{\rm ex;X}, \label{excessinfothermo2nd_quadratic}\\
    \sigma_t^{\rm hk;X} = \dot{\Sigma}_t^{\rm hk; X} + \dot{I}_t^{\rm hk;X} \geq \frac{(\dot{I}_t^{\rm hk;X})^2}{\mathcal{I}_t^{\rm Fisher;X}} + \dot{I}_t^{\rm hk;X}.
\label{housekeepinginfothermo2nd_quadratic}
\end{align} \label{exhkinfothermo2nd_quadratic}%
\end{subequations}
Interestingly, we see that the same quantity $\mathcal{I}_t^{\rm Fisher;X}  = \text{tr} ( \mathsf{D}^{\rm X} \mathsf{F}_t^{\rm Y \vert X} )$ appears in both generalizations. The apparent negative excess and housekeeping entropy change rates can only be observed for  $-\mathcal{I}_t^{\rm Fisher;X} < \dot{I}_t^{\rm ex;X} < 0$ and  $-\mathcal{I}_t^{\rm Fisher;X} < \dot{I}_t^{\rm hk;X} < 0$, respectively. Too large excess and housekeeping information flows, i.e., $-\dot{I}_t^{\rm ex;X} \geq  \mathcal{I}_t^{\rm Fisher;X}$ and $-\dot{I}_t^{\rm hk;X} \geq \mathcal{I}_t^{\rm Fisher;X}$,  will essentially prohibit the negativity of $\sigma_t^{\rm ex;X}$ and $\sigma_t^{\rm hk;X}$, respectively.
Moreover, we also obtain the global lower bound
\begin{subequations}
\begin{align}
    \sigma_t^{\rm ex;X} \geq - \frac{1}{4} \mathcal{I}_t^{\rm Fisher;X}, \label{ex_lower_global}\\
    \sigma_t^{\rm hk;X} \geq - \frac{1}{4} \mathcal{I}_t^{\rm Fisher;X}, \label{hk_lower_global}
\end{align}
\label{exhk_lower_global}%
\end{subequations}
by considering the minimization of the right-hand sides in Eqs.~\eqref{excessinfothermo2nd_quadratic} and~\eqref{housekeepinginfothermo2nd_quadratic}. 
We note that $\sigma_t^{\rm X} \geq - \mathcal{I}_t^{\rm Fisher;X}/4$ [Eq.~\eqref{globallowerboundfisher}] also holds. These inequalities provide universal,
although generally non-tight, lower bounds on the apparent
entropy change rates.

Furthermore, the excess bound in Eq.~\eqref{ex_lower_global} can be sharpened by incorporating the local excess entropy production rate $\dot{\Sigma}_t^{\mathrm{localex};X} =\langle  -\nabla_{\boldsymbol{x}}\phi^{\rm X}_t,  -\nabla_{\boldsymbol{x}}\phi^{\rm X}_t\rangle_{p_t \mathsf{D}^{\rm X}}$. The quantity $\nabla_{\boldsymbol{x}} \phi^{\rm X}_t (\boldsymbol{x}) $ depending only on $\boldsymbol{x}$ satisfies 
\begin{align}
&\langle \nabla_{\boldsymbol{x}} i_t, \nabla_{\boldsymbol{x}} \phi^{\rm X}_t \rangle_{p_t \mathsf{D}^{\rm X}} \nonumber\\
&= \int d\boldsymbol{x}  \left[ \nabla_{\boldsymbol{x}}\int d\boldsymbol{y} p_t^{\rm Y \vert X}(\bs{y} \vert \bs{x}) \right] \cdot \mathsf{D}^{\rm X} \nabla_{\boldsymbol{x}} \phi^{\rm X}_t (\boldsymbol{x}) p^{\rm X}_t(\boldsymbol{x})=0,
\end{align}
where we used $\nabla_{\boldsymbol{x}} i_t(\boldsymbol{x},\boldsymbol{y})=\nabla_{\boldsymbol{x}} \ln p_t^{\rm Y \vert X}(\bs{y} \vert \bs{x}) = [p_t^{\rm Y \vert X}(\bs{y} \vert \bs{x})]^{-1} \nabla_{\boldsymbol{x}}  p_t^{\rm Y \vert X}(\bs{y} \vert \bs{x})$ and $\int d\boldsymbol{y} p_t^{\rm Y \vert X}(\bs{y} \vert \bs{x})=1$. Therefore, the excess information flow is also given by $\dot{I}^{\rm ex; X}_t = \langle \boldsymbol{f}^{\rm ex; X}_t +\nabla_{\boldsymbol{x}} \phi^{\rm X}_t ,\nabla_{\boldsymbol{x}} i_t \rangle_{ p_t \mathsf{D}^{\rm X}}$. Here, we consider the quantity $\langle \boldsymbol{f}^{\rm ex; X}_t +\nabla_{\boldsymbol{x}} \phi^{\rm X}_t , \boldsymbol{f}^{\rm ex; X}_t +\nabla_{\boldsymbol{x}} \phi^{\rm X}_t\rangle_{p_t \mathsf{D}^{\rm X}}$ that appears when discussing the Cauchy-Schwarz inequality for $\dot{I}^{\rm ex; X}_t=\langle \boldsymbol{f}^{\rm ex; X}_t +\nabla_{\boldsymbol{x}} \phi^{\rm X}_t ,\nabla_{\boldsymbol{x}} i_t \rangle_{ p_t \mathsf{D}^{\rm X}}$. Using a derivation similar to that for Eq.~\eqref{orthogonalloc}, the orthogonality $\langle \nabla_{\boldsymbol{x}} \phi^{\rm X}_t  , \boldsymbol{f}^{\rm ex; X}_t + \nabla_{\boldsymbol{x}} \phi^{\rm X}_t \rangle_{p_t \mathsf{D}^{\rm X}} =0$ can be shown as follows,
\begin{align}
&\langle \nabla_{\boldsymbol{x}}\phi^{\rm X}_t, \boldsymbol{f}^{\rm ex; X}_t + \nabla_{\boldsymbol{x}} \phi^{\rm X}_t  \rangle_{p_t\mathsf{D}^{\rm X}} \nonumber\\
&= -\int d\boldsymbol{x} \phi^{\rm X}_t (\boldsymbol{x}) \int d\boldsymbol{y} \nabla_{\boldsymbol{x}} \cdot [p_t (\boldsymbol{x}, \boldsymbol{y}) \mathsf{D}^{\rm X}\boldsymbol{f}^{\rm ex;X}_t (\boldsymbol{x}, \boldsymbol{y})] \nonumber \\
&-\int d\boldsymbol{x} \phi^{\rm X}_t (\boldsymbol{x}) \nabla_{\boldsymbol{x}} \cdot [p_t^{\rm X}(\boldsymbol{x}) \mathsf{D}^{\rm X} \nabla_{\boldsymbol{x}} \phi^{\rm X}_t(\boldsymbol{x})]\nonumber\\
&=0,
\label{orthogonalloc2}
\end{align}
where we used Eqs.~\eqref{continuityX} and~\eqref{partialpxex} and integration by parts.
This orthogonality leads to the generalized Pythagorean theorem,
\begin{align}
\langle \boldsymbol{f}^{\rm ex; X}_t  , \boldsymbol{f}^{\rm ex; X}_t \rangle_{p_t \mathsf{D}^{\rm X}}
=&\langle \boldsymbol{f}^{\rm ex; X}_t +\nabla_{\boldsymbol{x}} \phi^{\rm X}_t , \boldsymbol{f}^{\rm ex; X}_t +\nabla_{\boldsymbol{x}} \phi^{\rm X}_t\rangle_{p_t \mathsf{D}^{\rm X}} \nonumber\\
&+ \langle -\nabla_{\boldsymbol{x}} \phi^{\rm X}_t , -\nabla_{\boldsymbol{x}} \phi^{\rm X}_t\rangle_{p_t \mathsf{D}^{\rm X}},
\end{align}
that is $\langle \boldsymbol{f}^{\rm ex; X}_t +\nabla_{\boldsymbol{x}} \phi^{\rm X}_t , \boldsymbol{f}^{\rm ex; X}_t +\nabla_{\boldsymbol{x}} \phi^{\rm X}_t\rangle_{p_t \mathsf{D}^{\rm X}} = \dot{\Sigma}^{\rm ex;X}_t-\dot{\Sigma}^{\rm localex;X}_t$. Thus, the Cauchy-Schwarz inequality for $\dot{I}^{\rm ex; X}_t=\langle \boldsymbol{f}^{\rm ex; X}_t +\nabla_{\boldsymbol{x}} \phi^{\rm X}_t ,\nabla_{\boldsymbol{x}} i_t \rangle_{ p_t \mathsf{D}^{\rm X}}$ yields the following inequality,
\begin{align}
(\dot I_t^{\rm ex;X })^2 \leq ( \dot{\Sigma}^{\rm ex;X}_t -\dot{\Sigma}^{\rm localex;X}_t ) \mathcal{I}_t^{\rm Fisher;X},
\label{eq:local-fisher-excess}
\end{align}
which is tighter than Eq.~\eqref{cauchy-scharz-loose}.
Consequently, the apparent excess entropy change rate also satisfies the tighter lower bound
\begin{align}
\sigma_t^{\rm ex;X} &\geq \dot{\Sigma}_t^{\rm localex;X} +\dot{I}_t^{\rm ex;X} +\frac{(\dot {I}_t^{\rm ex;X})^2} {\mathcal{I}_t^{\rm Fisher;X}} \nonumber\\ &\geq \dot\Sigma_t^{\rm localex;X} -\frac{1}{4}\mathcal{I}_t^{\rm Fisher;X}.
\label{eq:local-excess-demon-bound}
\end{align}
Thus, a necessary condition for the emergence of an excess demon (i.e., $\sigma_t^{\rm ex;X}<0$) is $(1/4)\mathcal{I}_t^{\rm Fisher;X}>\dot\Sigma_t^{\rm localex;X} = \lim_{\Delta t \to 0} [\tilde{\mathcal{W}}_2^{(\mathsf{D}^{\rm X})^{-1}} (p_t^{\rm X}, p_{t+\Delta t}^{\rm X})]^2/(\Delta t)^2$. This condition compares the conditional Fisher information with the squared generalized $2$-Wasserstein speed of the marginal distribution $p_t^X$. Furthermore, integrating the inequalities in Eq.~\eqref{eq:local-excess-demon-bound} and using the Wasserstein bound in Eq.~\eqref{inequalityinfospeedlimit}, we obtain
\begin{align}
&\int_0^{\tau} dt \sigma^{\rm ex;X}_t - \frac{[\tilde{\mathcal{W}}_2^{(\mathsf{D}^{\rm X})^{-1}} (p_0^{\rm X}, p_{\tau}^{\rm X})]^2}{\tau}\nonumber\\
&\geq \int_0^{\tau} dt \dot{I}_t^{\rm ex;X} +\int_0^{\tau} dt  \frac{(\dot {I}_t^{\rm ex;X})^2} {\mathcal{I}_t^{\rm Fisher;X}} \geq - \frac{1}{4}\int_0^{\tau} dt \mathcal{I}^{\rm Fisher; X}_t.
\end{align}
The first inequality strengthens Eq.~\eqref{excessinformationflowwassersteinbound}, whereas the second provides a finite-time bound in terms of the marginal transport cost and the conditional Fisher information.

\section{examples}
\label{sec4}
\subsection{Gaussian case}
\label{sec4a}
We illustrate the results using the Gaussian case. We consider the linear force $\boldsymbol{F}_t(\boldsymbol{z})/T = \mathsf{A}_t \boldsymbol{z} +\boldsymbol{b}_t$, and the initial distribution is given by the Gaussian distribution $p_0 (\boldsymbol{z})\sim \mathcal{N}(\boldsymbol{m}_0, \mathsf{V}_0)$, where $\boldsymbol{m}_0$ is the vector of means, and $\mathsf{V}_0$ is the variance-covariance matrix. Under the conditions of linear forces and an initial state following a Gaussian distribution, the distribution at time $t>0$ is given by a Gaussian distribution $p_t (\boldsymbol{z})\sim \mathcal{N}(\boldsymbol{m}_t, \mathsf{V}_t)$, where $\boldsymbol{m}_t$ is the vector of means, and $\mathsf{V}_t$ is the variance-covariance matrix. These matrices and vectors can be written using the components corresponding to system ${\rm X}$ and system ${\rm Y}$ as follows,
\begin{align}
\boldsymbol{m}_t = \begin{pmatrix}
   \boldsymbol{m}_t^{\rm X} \\
   \boldsymbol{m}_t^{\rm Y} \end{pmatrix},\: \: 
  \mathsf{V}_t = \begin{pmatrix}
   \mathsf{V}_t^{\rm XX}& \mathsf{V}^{\rm XY}_t \\
   (\mathsf{V}^{\rm XY}_t)^{\top} & \mathsf{V}^{\rm YY}_t
\end{pmatrix} = \begin{pmatrix}
   \mathsf{V}_t^{\rm X} \\
   \mathsf{V}^{\rm Y}_t
\end{pmatrix},
\end{align}
\begin{align}
\boldsymbol{b}_t = \begin{pmatrix}
   \boldsymbol{b}_t^{\rm X} \\
   \boldsymbol{b}_t^{\rm Y} \end{pmatrix},\: \: 
  \mathsf{A}_t = \begin{pmatrix}
   \mathsf{A}_t^{\rm XX}& \mathsf{A}_t^{\rm XY} \\
   \mathsf{A}_t^{\rm YX} & \mathsf{A}_t^{\rm YY}
\end{pmatrix} = \begin{pmatrix}
   \mathsf{A}_t^{\rm X}\\
   \mathsf{A}_t^{\rm Y} 
\end{pmatrix}.
\label{decompmatrix}
\end{align}
Here, $\boldsymbol{m}_t^{\rm X}$ and $\boldsymbol{b}_t^{\rm X}$ ($\boldsymbol{m}_t^{\rm Y}$ and $\boldsymbol{b}_t^{\rm Y}$) are $d^{\rm X}$-dimensional ($d^{\rm Y}$-dimensional) column vectors,  $\mathsf{V}_t^{\rm XX}$ and $\mathsf{A}_t^{\rm XX}$ ($\mathsf{V}_t^{\rm YY}$ and $\mathsf{A}_t^{\rm YY}$) are $d^{\rm X} \times d^{\rm X}$ ($d^{\rm Y} \times d^{\rm Y}$) matrices,  $\mathsf{V}^{\rm XY}_t$ and $\mathsf{A}_t^{\rm XY}$ are $d^{\rm X} \times d^{\rm Y}$ matrices, $\mathsf{A}_t^{\rm YX}$ is a $d^{\rm Y} \times d^{\rm X}$ matrix, and $\mathsf{A}_t^{\rm X}$ and $\mathsf{V}_t^{\rm X}$ ($\mathsf{A}_t^{\rm Y}$ and $\mathsf{V}_t^{\rm Y}$) are $d^{\rm X} \times d$ ($d^{\rm Y} \times d$) matrices. The time evolution of the mean and the variance-covariance matrix are given by
\begin{align}
    \partial_t \boldsymbol{m}_t &= \mathsf{D} \mathsf{A}_t \boldsymbol{m}_t + \mathsf{D}\boldsymbol{b}_t, \nonumber\\
    \partial_t \mathsf{V}_t &= \mathsf{D} \mathsf{A}_t \mathsf{V}_t + \mathsf{V}_t ( \mathsf{D} \mathsf{A}_t)^{\top} +2 \mathsf{D}.
\end{align}

The geometric decomposition for the Gaussian case has been discussed in Ref.~\cite{sekizawa2024decomposing}. The entropy production rate $\dot{\Sigma}_t$, the excess entropy production rate $\dot{\Sigma}^{\rm ex}_t$ and the housekeeping entropy production rate $\dot{\Sigma}^{\rm hk}_t$ are calculated as
\begin{align}
 \dot{\Sigma}_t =& (\mathsf{A}_t \boldsymbol{m}_t + \boldsymbol{b}_t)^{\top}\mathsf{D} (\mathsf{A}_t \boldsymbol{m}_t + \boldsymbol{b}_t)\nonumber\\
 &+ {\rm tr} [(\mathsf{A}_t + \mathsf{V}_t^{-1})^{\top} \mathsf{D} (\mathsf{A}_t + \mathsf{V}_t^{-1})\mathsf{V}_t],\nonumber\\
 \dot{\Sigma}^{\rm ex}_t =& (\mathsf{A}_t^{\rm ex} \boldsymbol{m}_t + \boldsymbol{b}^{\rm ex}_t)^{\top}\mathsf{D} (\mathsf{A}_t^{\rm ex} \boldsymbol{m}_t + \boldsymbol{b}^{\rm ex}_t)\nonumber\\
 &+ {\rm tr} [(\mathsf{A}_t^{\rm ex} + \mathsf{V}_t^{-1})^{\top} \mathsf{D} (\mathsf{A}_t^{\rm ex} + \mathsf{V}_t^{-1})\mathsf{V}_t],\nonumber\\
 \dot{\Sigma}^{\rm hk}_t =&  {\rm tr} [(\mathsf{A}_t^{\rm hk} )^{\top} \mathsf{D} \mathsf{A}_t^{\rm hk}\mathsf{V}_t ],
\end{align}
where we consider the decompositions $\boldsymbol{b}_t=\boldsymbol{b}_t^{\rm ex}+\boldsymbol{b}_t^{\rm hk}$ and $\mathsf{A}_t=\mathsf{A}_t^{\rm ex}+\mathsf{A}_t^{\rm hk}$ that satisfy
\begin{align}
    \partial_t \boldsymbol{m}_t &= \mathsf{D} \mathsf{A}_t^{\rm ex} \boldsymbol{m}_t + \mathsf{D}\boldsymbol{b}_t^{\rm ex}, \nonumber\\
    \boldsymbol{0} &= \mathsf{D} \mathsf{A}_t^{\rm hk} \boldsymbol{m}_t + \mathsf{D}\boldsymbol{b}_t^{\rm hk}, \nonumber\\
    \partial_t \mathsf{V}_t &= \mathsf{D} \mathsf{A}_t^{\rm ex} \mathsf{V}_t + \mathsf{V}_t ( \mathsf{D} \mathsf{A}_t^{\rm ex})^{\top} +2 \mathsf{D},\nonumber\\
    \mathsf{O} &= \mathsf{D} \mathsf{A}_t^{\rm hk} \mathsf{V}_t + \mathsf{V}_t ( \mathsf{D} \mathsf{A}_t^{\rm hk})^{\top}, \nonumber\\
    (\mathsf{A}_t^{\rm ex})^{\top} &=\mathsf{A}_t^{\rm ex}.
    \label{defAexAhk}
\end{align}
Furthermore, not only the excess entropy production rate but also the generalized $2$-Wasserstein distance can be calculated analytically for Gaussian distributions (see Appendix~\ref{appendix:generalized2-Wasserstein}).

For the decompositions $\boldsymbol{b}_t=\boldsymbol{b}_t^{\rm ex}+\boldsymbol{b}_t^{\rm hk}$ and $\mathsf{A}_t=\mathsf{A}_t^{\rm ex}+\mathsf{A}_t^{\rm hk}$, we also introduce the following expressions using the components corresponding to system $\rm X$ and system $\rm Y$ as follows,
\begin{align}
\boldsymbol{b}_t^{\rm ex} = \begin{pmatrix}
   \boldsymbol{b}_t^{\rm ex;X} \\
   \boldsymbol{b}_t^{\rm ex; Y} \end{pmatrix},\: \: \boldsymbol{b}_t^{\rm hk} = \begin{pmatrix}
   \boldsymbol{b}_t^{\rm hk;X} \\
   \boldsymbol{b}_t^{\rm hk; Y} \end{pmatrix},
\end{align}
\begin{align}
 \mathsf{A}_t^{\rm ex} &= \begin{pmatrix}
   \mathsf{A}_t^{\rm ex;XX}& \mathsf{A}_t^{\rm ex;XY} \\
   \mathsf{A}_t^{\rm ex;YX} & \mathsf{A}_t^{\rm ex;YY} \end{pmatrix} =\begin{pmatrix}
   \mathsf{A}_t^{\rm ex;X} \\
   \mathsf{A}_t^{\rm ex;Y} \end{pmatrix}, \nonumber\\
   \mathsf{A}_t^{\rm hk} &= \begin{pmatrix}
   \mathsf{A}_t^{\rm hk;XX}& \mathsf{A}_t^{\rm hk;XY} \\
   \mathsf{A}_t^{\rm hk;YX} & \mathsf{A}_t^{\rm hk;YY} \end{pmatrix}= \begin{pmatrix}
   \mathsf{A}_t^{\rm hk;X} \\
   \mathsf{A}_t^{\rm hk;Y} \end{pmatrix}.
\end{align}
Here, the sizes of each matrix and vector are taken as in Eq.~\eqref{decompmatrix}.

In the Gaussian case, the spectral decomposition representation of the housekeeping entropy production rate~\cite{sekizawa2024decomposing} is the same as the Koopman mode decomposition~\cite{sekizawa2025koopman}. To obtain the Koopman mode decomposition, we consider the spectral decomposition $\mathsf{D} \mathsf{A}^{\rm hk}_t = \sum_i \lambda_i \mathsf{P}_i$ where $\mathsf{P}_i$ is the projection matrix. The Koopman mode decomposition of the housekeeping entropy production rate~\cite{sekizawa2024decomposing} is obtained as
\begin{align}
    \dot{\Sigma}^{\rm hk}_t =& \sum_i (2\pi)^2 \left(\frac{\Im[\lambda_i]}{2 \pi} \right)^2 {\rm tr}[(\mathsf{P}_i)^{\mathrm{H}} \mathsf{D}^{-1} \mathsf{P}_i \mathsf{V}_t ].
    \label{koopmangausshk}
\end{align}

Using the fact that $\mathsf{D}$ is a block diagonal matrix, the partial entropy production rates, the partial excess entropy production rates and the partial housekeeping entropy production rates are calculated as
\begin{align}
 \dot{\Sigma}^{\rm X}_t =& (\mathsf{A}_t^{\rm X} \boldsymbol{m}_t + \boldsymbol{b}_t^{\rm X})^{\top}\mathsf{D}^{\rm X} (\mathsf{A}_t^{\rm X} \boldsymbol{m}_t + \boldsymbol{b}_t^{\rm X})\nonumber\\
 &+ {\rm tr} [(\mathsf{A}_t^{\rm X} + \mathsf{\Theta}_t^{\rm X} )^{\top} \mathsf{D}^{\rm X} (\mathsf{A}_t^{\rm X} + \mathsf{\Theta}_t^{\rm X})\mathsf{V}_t],\nonumber\\
 \dot{\Sigma}^{\rm ex;X}_t =& (\mathsf{A}_t^{\rm ex;X} \boldsymbol{m}_t + \boldsymbol{b}_t^{\rm ex;X})^{\top}\mathsf{D}^{\rm X} (\mathsf{A}_t^{\rm ex;X} \boldsymbol{m}_t + \boldsymbol{b}_t^{\rm ex;X})\nonumber\\
 &+ {\rm tr} [(\mathsf{A}_t^{\rm ex;X} + \mathsf{\Theta}_t^{\rm X})^{\top} \mathsf{D}^{\rm X} (\mathsf{A}_t^{\rm ex;X} + \mathsf{\Theta}_t^{\rm X} )\mathsf{V}_t],\nonumber\\
 \dot{\Sigma}^{\rm hk;X}_t =&  {\rm tr} [(\mathsf{A}_t^{\rm hk;X} )^{\top} \mathsf{D}^{\rm X} \mathsf{A}_t^{\rm hk;X}\mathsf{V}_t ],
\end{align}
where we used the following notation
\begin{align}
\mathsf{V}_t^{\rm -1}=\mathsf{\Theta}_t = \begin{pmatrix}
 \mathsf{\Theta}_t^{\rm XX}& \mathsf{\Theta}_t^{\rm XY}\\
    (\mathsf{\Theta}_t^{\rm XY})^{\top}& \mathsf{\Theta}_t^{\rm YY}
\end{pmatrix}  = \begin{pmatrix}
   \mathsf{\Theta}_t^{\rm X} \\
   \mathsf{\Theta}_t^{\rm Y}
\end{pmatrix}.
\end{align}
Here, the sizes of each matrix and vector are taken as in Eq.~\eqref{decompmatrix}. Due to symmetry, the following results also hold for system ${\rm Y}$:
\begin{align}
 \dot{\Sigma}^{\rm Y}_t =& (\mathsf{A}_t^{\rm Y} \boldsymbol{m}_t + \boldsymbol{b}_t^{\rm Y})^{\top}\mathsf{D}^{\rm Y} (\mathsf{A}_t^{\rm Y} \boldsymbol{m}_t + \boldsymbol{b}_t^{\rm Y})\nonumber\\
 &+ {\rm tr} [(\mathsf{A}_t^{\rm Y} + \mathsf{\Theta}_t^{\rm Y} )^{\top} \mathsf{D}^{\rm Y} (\mathsf{A}_t^{\rm Y} + \mathsf{\Theta}_t^{\rm Y})\mathsf{V}_t],\nonumber\\
 \dot{\Sigma}^{\rm ex;Y}_t =& (\mathsf{A}_t^{\rm ex;Y} \boldsymbol{m}_t + \boldsymbol{b}_t^{\rm ex;Y})^{\top}\mathsf{D}^{\rm Y} (\mathsf{A}_t^{\rm ex;Y} \boldsymbol{m}_t + \boldsymbol{b}_t^{\rm ex;Y})\nonumber\\
 &+ {\rm tr} [(\mathsf{A}_t^{\rm ex;Y} + \mathsf{\Theta}_t^{\rm Y})^{\top} \mathsf{D}^{\rm Y} (\mathsf{A}_t^{\rm ex;Y} + \mathsf{\Theta}_t^{\rm Y} )\mathsf{V}_t],\nonumber\\
 \dot{\Sigma}^{\rm hk;Y}_t =&  {\rm tr} [(\mathsf{A}_t^{\rm hk;Y} )^{\top} \mathsf{D}^{\rm Y} \mathsf{A}_t^{\rm hk;Y}\mathsf{V}_t ].
\end{align}

The Koopman mode decomposition of the partial housekeeping entropy production rate can also be given by the spectral decomposition $\mathsf{D} \mathsf{A}^{\rm hk}_t = \sum_i \lambda_i \mathsf{P}_i$. We here introduce the selection matrix  $\mathsf{\Pi}^{\rm X}$ defined as
\begin{align}
\mathsf{\Pi}^{\rm X} := \begin{pmatrix}
   \mathsf{I} & \mathsf{O}
\end{pmatrix},
\end{align}
where $\mathsf{\Pi}^{\rm X}$ is a $d^{\rm X}\times d$ matrix. Using $\mathsf{\Pi}^{\rm X}$, we obtain $\mathsf{D}^{\rm X} \mathsf{A}_t^{\rm hk;X}= \mathsf{\Pi}^{\rm X} \mathsf{D} \mathsf{A}_t^{\rm hk}$. Therefore, similar to the derivation of Eq.~\eqref{koopmangausshk}, the Koopman mode decomposition of the partial housekeeping entropy production rate can be expressed as follows,
\begin{align}
    &\dot{\Sigma}^{\rm hk;X}_t \nonumber\\
    &= \sum_{i|\mathsf{\Pi}^{\rm X} \mathsf{P}_i \neq \mathsf{O}} (2\pi)^2 \left(\frac{\Im[\lambda_i]}{2 \pi} \right)^2 {\rm tr}[(\mathsf{\Pi}^{\rm X} \mathsf{P}_i)^{\mathrm{H}} (\mathsf{D}^{\rm X})^{-1} \mathsf{\Pi}^{\rm X} \mathsf{P}_i \mathsf{V}_t ].
\end{align}

Finally, we discuss analytical expressions of information flow. Because $\nabla_{\boldsymbol{x}}i_t(\boldsymbol{x}, \boldsymbol{y})$ is calculated as
\begin{align}
\nabla_{\boldsymbol{x}}i_t(\boldsymbol{x}, \boldsymbol{y})=& -\mathsf{\Theta}^{\rm X}_t (\boldsymbol{z}- \boldsymbol{m}_t) + (\mathsf{V}_t^{\rm XX})^{-1}(\boldsymbol{x}- \boldsymbol{m}_t^{\rm X}),
 \label{scoregauss}
\end{align}
information flow, excess information flow and housekeeping information flow are calculated as
\begin{align}
 \dot{I}^{\rm X}_t =& \langle \mathsf{A}_t^{\rm X} \boldsymbol{m}_t + \boldsymbol{b}_t^{\rm X} + (\mathsf{A}_t^{\rm X} +\mathsf{\Theta}^{\rm X}_t) (\boldsymbol{z}- \boldsymbol{m}_t) ,\nabla_{\boldsymbol{x}}i_t  \rangle_{p_t \mathsf{D}^{\rm X}} \nonumber\\
 =& {\rm tr}[(\mathsf{A}_t^{\rm XY} +\mathsf{\Theta}^{\rm XY}_t)^{\top} \mathsf{D}^{\rm X} (\mathsf{V}_t^{\rm XX})^{-1}\mathsf{V}_t^{\rm XY}], \nonumber \\
  \dot{I}^{\rm ex;X}_t =& \langle  \mathsf{A}^{\rm ex;X}_t\boldsymbol{m}_t  \!+  \!\boldsymbol{b}^{\rm ex;X}_t  \!+  \!(\mathsf{A}_t^{\rm ex;X}  \!+ \!\mathsf{\Theta}^{\rm X}_t) (\boldsymbol{z}- \boldsymbol{m}_t) ,\nabla_{\boldsymbol{x}}i_t  \rangle_{p_t \mathsf{D}^{\rm X}} \nonumber\\
 =& {\rm tr}[(\mathsf{A}_t^{\rm ex;XY} +\mathsf{\Theta}^{\rm XY}_t)^{\top} \mathsf{D}^{\rm X} (\mathsf{V}_t^{\rm XX})^{-1}\mathsf{V}_t^{\rm XY}], \nonumber \\
  \dot{I}^{\rm hk; X}_t =& \langle \mathsf{A}^{\rm hk;X}_t (\boldsymbol{z} - \boldsymbol{m}_t) ,\nabla_{\boldsymbol{x}}i_t  \rangle_{p_t \mathsf{D}^{\rm X}} \nonumber\\
 =& {\rm tr}[(\mathsf{A}_t^{\rm hk;XY} )^{\top} \mathsf{D}^{\rm X} (\mathsf{V}_t^{\rm XX})^{-1}\mathsf{V}_t^{\rm XY}],
 \label{informationflowsgaussX}
\end{align}
where we used $\mathsf{\Theta}_t \mathsf{V}_t = \mathsf{V}_t \mathsf{\Theta}_t=\mathsf{I}$, or equivalently,
\begin{align}
 &  \begin{pmatrix}
   \!\mathsf{V}_t^{\rm XX} & \mathsf{V}_t^{\rm XY} \\
   \!(\mathsf{V}_t^{\rm XY})^{\top} & \mathsf{V}_t^{\rm YY}
\end{pmatrix} \!  \begin{pmatrix}\!
    (\mathsf{V}_t^{\rm XX})^{-1}\! -\!\mathsf{\Theta}_t^{\rm XX} \!\\ -(\mathsf{\Theta}_t^{\rm XY})^{\top}
\end{pmatrix}\! =\!\begin{pmatrix}
   \mathsf{O} \\ 
   \!(\mathsf{V}_t^{\rm XY})^{\top} (\mathsf{V}_t^{\rm XX})^{-1}\!
\end{pmatrix}.
\end{align}
Due to symmetry, the following results also hold for system ${\rm Y}$:
\begin{align}
 \dot{I}^{\rm Y}_t =& {\rm tr}[(\mathsf{A}_t^{\rm YX} +(\mathsf{\Theta}^{\rm XY}_t)^{\top})^{\top} \mathsf{D}^{\rm Y} (\mathsf{V}_t^{\rm YY})^{-1}(\mathsf{V}_t^{\rm XY})^{\top}], \nonumber \\
  \dot{I}^{\rm ex;Y}_t =& {\rm tr}[(\mathsf{A}_t^{\rm ex;YX} +(\mathsf{\Theta}^{\rm XY}_t)^{\top})^{\top} \mathsf{D}^{\rm Y} (\mathsf{V}_t^{\rm YY})^{-1}(\mathsf{V}_t^{\rm XY})^{\top}], \nonumber \\
  \dot{I}^{\rm hk; Y}_t =& {\rm tr}[(\mathsf{A}_t^{\rm hk;YX} )^{\top} \mathsf{D}^{\rm Y} (\mathsf{V}_t^{\rm YY})^{-1}(\mathsf{V}_t^{\rm XY})^{\top}].
  \label{informationflowsgaussY}
\end{align}
Based on the expressions [Eqs.~\eqref{informationflowsgaussX} and~\eqref{informationflowsgaussY}], we can confirm that the antisymmetric relation $\dot{I}^{\rm hk;X}_t  =-\dot{I}^{\rm hk;Y}_t$ [Eq.~\eqref{antisymhk}] holds by using the condition $\mathsf{D} \mathsf{A}_t^{\rm hk} \mathsf{V}_t =- \mathsf{V}_t ( \mathsf{D} \mathsf{A}_t^{\rm hk})^{\top}$ in Eq.~\eqref{defAexAhk}. Because we can easily confirm $\dot{I}^{\rm X}_t= \dot{I}^{\rm ex;X}_t+\dot{I}^{\rm hk;X}_t$ and $\dot{I}^{\rm Y}_t= \dot{I}^{\rm ex;Y}_t+\dot{I}^{\rm hk;Y}_t$ based on the expressions [Eqs.~\eqref{informationflowsgaussX} and~\eqref{informationflowsgaussY}], $(d_t I(\hat{\rm X}_t;\hat{\rm Y}_t) =)\dot{I}^{\rm X}_t +\dot{I}^{\rm Y}_t =\dot{I}^{\rm ex;X}_t +\dot{I}^{\rm ex;Y}_t$ [Eq.~\eqref{decompositionexinfo}] can also be confirmed using the antisymmetric relation $\dot{I}^{\rm hk;X}_t  =-\dot{I}^{\rm hk;Y}_t$ .

Finally, the quantities $\mathcal{I}_t^{\rm Fisher;X}$ and $\mathcal{I}_t^{\rm Fisher;Y}$ providing the quadratic and global lower bounds on the emergence of the conventional, excess and housekeeping demons (see Eqs.~\eqref{infothermo2nd_quadratic}, \eqref{globallowerboundfisher}, \eqref{exhkinfothermo2nd_quadratic} and \eqref{exhk_lower_global}) are explicitly given by
\begin{align}
    \mathcal{I}_t^{\rm Fisher;X} &= {\rm tr}[\mathsf{D}^{\rm X} (\mathsf{\Theta}^{XX}_t - (\mathsf{V}^{XX}_t)^{-1}) ], \nonumber \\
    \mathcal{I}_t^{\rm Fisher;Y} &= {\rm tr}[\mathsf{D}^{\rm Y} (\mathsf{\Theta}^{YY}_t - (\mathsf{V}^{YY}_t)^{-1}) ] .
\end{align}
To obtain this expression, we used Eq.~\eqref{scoregauss} and $\mathsf{\Theta}_t \mathsf{V}_t =\mathsf{I}$.
As mentioned in Sec.~\ref{sec3i}, $\mathcal{I}_t^{\rm Fisher;X}$ and $\mathcal{I}_t^{\rm Fisher;Y}$ only depend on the probability density, which in the Gaussian case is characterized by its covariance matrix. In order to prevent the quantity of $\mathcal{I}_t^{\rm Fisher;X}$ from becoming too small, nonzero correlations between $\rm X$ and $\rm Y$ ($\mathsf{\Theta}^{XX}_t \neq  (\mathsf{V}^{XX}_t)^{-1}$) are necessary, while stronger correlations or smaller fluctuations of $\rm X$ can enhance it.
From Eqs.~\eqref{globallowerboundfisher} and \eqref{exhk_lower_global}, this expression implies that, in the presence of nonzero correlations, $\rm Y$ can act more effectively as Maxwell's demon for $\rm X$ when the correlations are stronger or the fluctuations in $\rm X$ are smaller.

\subsection{Numerical examples}
\label{sec4b}
We numerically discuss the conditions under which the excess and housekeeping demons arise in the $2$-dimensional Gaussian case ($d^{\rm X}=d^{\rm Y}=1$). We use the notation $\boldsymbol{x}=x$ and $\boldsymbol{y}=y$ because $\boldsymbol{x}$ and $\boldsymbol{y}$ are $1$-dimensional states. We numerically track the time evolution of $\boldsymbol{m}_t$ and $\mathsf{V}_t$ with $\boldsymbol{m}_0 =\boldsymbol{0}$ and $\mathsf{V}_0$.
We set parameters to $\mathsf{D}^X=\mathsf{D}^Y=1$ ($\mathsf{D} =\mathsf{I}$) and $\boldsymbol{b}_t=\boldsymbol{0}$. We describe how to determine $\mathsf{V}_0$ and $\mathsf{A}_t$ in the next paragraph. 
We further obtain $\mathsf{A}_t^\mathrm{ex}$, $\mathsf{A}_t^\mathrm{hk}$, $\boldsymbol{b}_t^{\mathrm{ex}}$ and $\boldsymbol{b}_t^{\mathrm{hk}}$ by solving Eq.~\eqref{defAexAhk}. 
Finally, we obtain the time series of $\sigma_t^{\mathrm{ex;X}}$, $\sigma_t^{\mathrm{hk;X}}$,  $\dot{I}_t^{\mathrm{ex;X}}$ and $\dot{I}_t^{\mathrm{hk;X}}$. 

\begin{figure}
    \centering
    \includegraphics[width=\linewidth]{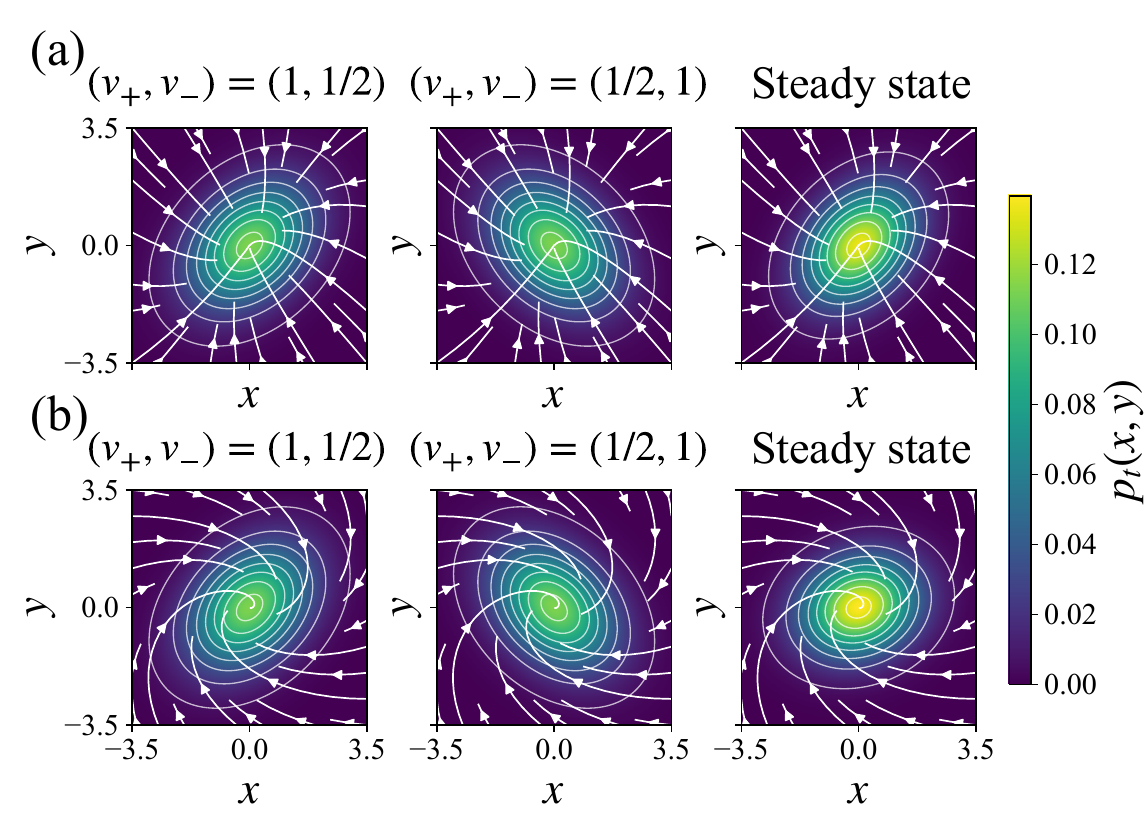}
    \caption{The probability distributions and streamlines for (a) $r=0.1$ and (b) $r=-1$. (Left) The initial distribution for $(v_+, v_-) =(1, 1/2)$ and its streamlines. (Center) The initial distribution for $(v_+, v_-) =(1/2, 1)$ and its streamlines. (Right) The steady-state distribution and its streamlines.}
    \label{fig:heatmap}
\end{figure}
To analyze the behavior of $(\sigma_t^{\mathrm{ex;X}}, \sigma_t^{\mathrm{hk;X}},  \dot{I}_t^{\mathrm{ex;X}}, \dot{I}_t^{\mathrm{hk;X}})$, we introduce a few parameters that can characterize the system. 
We now define the following fundamental matrices as 
\begin{align}
    \mathsf{M}_+ := \begin{pmatrix}
        1&1\\1&1
    \end{pmatrix},\;\;
    \mathsf{M}_- := \begin{pmatrix}
        1&-1\\-1&1
    \end{pmatrix},\;\;
    \mathsf{R} := \begin{pmatrix}
        0&-1\\1&0
    \end{pmatrix},
\end{align}
where $\mathsf{M}_+$ and $\mathsf{M}_-$ are symmetric matrices, and $\mathsf{R}$ is an antisymmetric matrix. Then, we parametrize the matrix $\mathsf{A}_t $ as 
\begin{align}
    \mathsf{A}_t = - k_{+}\mathsf{M}_+ - k_{-}\mathsf{M}_- + r\mathsf{R}. 
\end{align}
Here, $k_{+}$ and $k_{-}$ represent the spring coefficients in directions parallel to the lines $y=+x$ and $y=-x$, respectively. Because $\mathsf{M}_{+}\boldsymbol{z}$ and $\mathsf{M}_{-}\boldsymbol{z}$ can be expressed as the gradient of a potential, $k_{+}$ and $k_{-}$ are the parameters for the conservative contribution of $\boldsymbol{F}_t(\boldsymbol{z})/T$. On the other hand, $r$ characterizes the nonconservative contribution of $\boldsymbol{F}_t(\boldsymbol{z})/T$ because $\mathsf{R}$ is an antisymmetric matrix and $\mathsf{R}\boldsymbol{z}$ cannot be expressed as the gradient of a potential.  For the following numerical calculations, we set $k_{+}=0.3$ and $k_{-}=0.6$, while allowing $r$ to vary.

We also provide the initial covariance $\mathsf{V}_0$ in the form 
\begin{align}
    \mathsf{V}_0 = v_+\mathsf{M}_+ + v_-\mathsf{M}_-,
\end{align}
where $v_+$ and $v_-$ indicate the spatial dispersion in the respective directions. The initial fluctuation is chosen from one of the following two options: (i) $v_+=1$ and $v_-=1/2$, or (ii) $v_+=1/2$ and $v_-=1$. 
Figure~\ref{fig:heatmap} shows the initial probability distributions and streamlines of $\dot{\boldsymbol{z}}=\mathsf{A}\boldsymbol{z}$ for (a) $r=0.1$ or (b) $r=-1$, as well as the corresponding steady-state distributions and their streamlines. As shown in the figure, $r$ determines the rotation of the streamlines, while $(v_+, v_-)$ indicates whether the distribution spreads along the lines $x=+y$ or $x=-y$. Furthermore, the steady-state distribution spreads along the line $x=+y$, similar to the initial distribution when $(v_+, v_-) = (1,1/2)$.

\begin{figure}
    \centering
    \includegraphics[width=\linewidth]{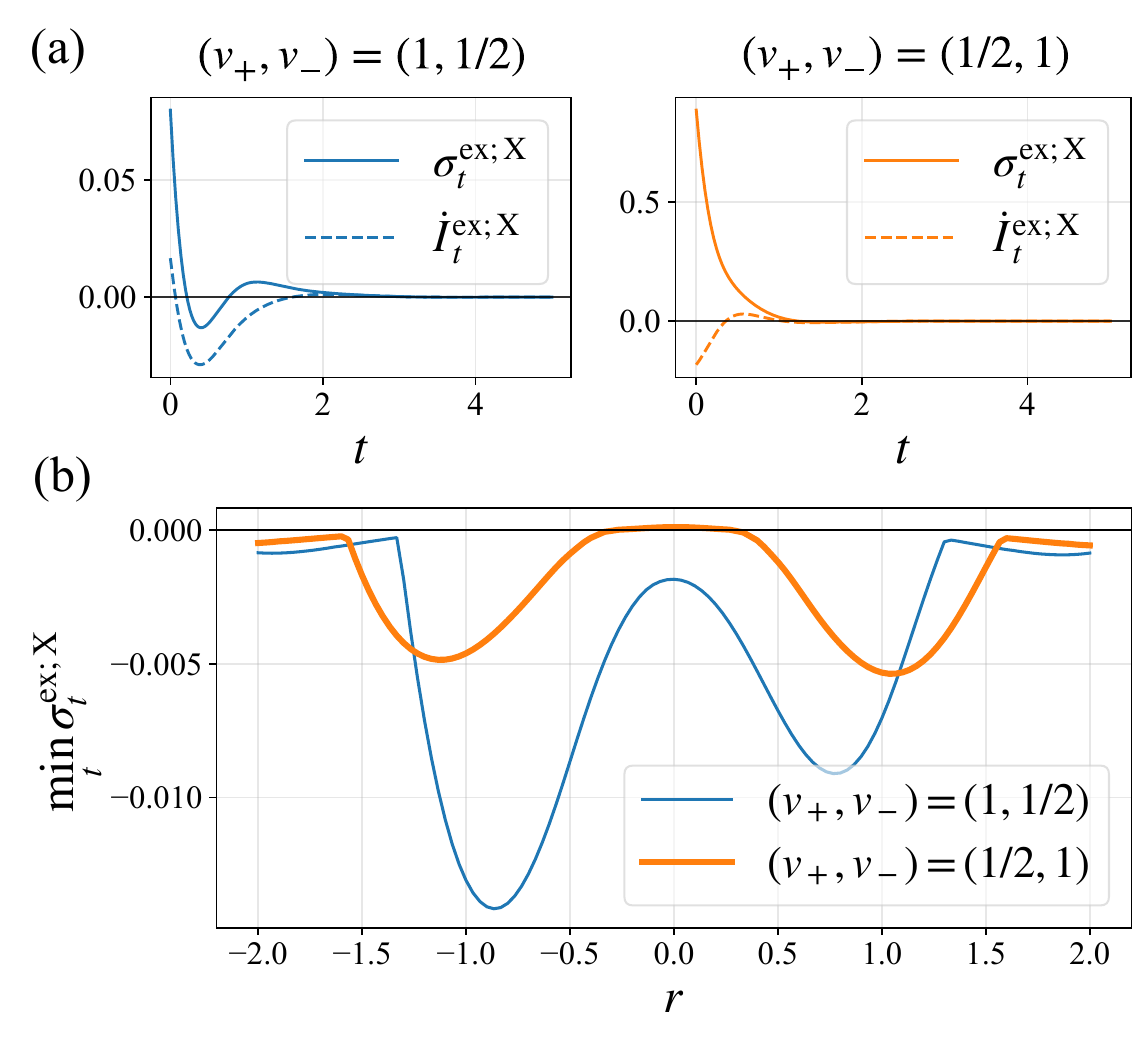}
    \caption{(a) Time evolution of $\sigma_t^{\mathrm{ex};\mathrm{X}}$ and $\dot{I}_t^{\mathrm{ex};\mathrm{X}}$ for $r=-1.0$. (b) Minimum value of $\sigma_t^{\mathrm{ex};\mathrm{X}}$ over time $t$. }
    \label{fig:excess}
\end{figure}
We first discuss the emergence of the excess demon. The typical behavior of $\sigma_t^{\mathrm{ex};\mathrm{X}}$ and $\dot{I}_t^{\mathrm{ex};\mathrm{X}}$ for $r=- 1.0$ is shown in Fig.~\ref{fig:excess}(a). 
We confirm that the generalized second law of information thermodynamics in terms of the excess dissipation, i.e., Eq.~\eqref{excessinfothermo2nd}, holds. 
Due to the nonconservative force, $\sigma_t^{\mathrm{ex};\mathrm{X}}$ and $\dot{I}_t^{\mathrm{ex};\mathrm{X}}$ exhibit damped oscillations, and finally converge to zero. 
During the oscillation, $\sigma_t^{\mathrm{ex};\mathrm{X}}$ repeatedly crosses the $x$-axis, which means the emergence of the excess demon. 
Negative values of $\sigma_t^{\mathrm{ex};\mathrm{X}}$ can be observed during the relaxation process for the current initial distributions over a very wide range of $r$ (Fig.~\ref{fig:excess}(b)). The region where the value $\sigma_t^{\mathrm{ex};\mathrm{X}}$ is always positive is very limited, occurring only when $r$ is small and the initial distribution differs significantly from the steady-state distribution, i.e., $(v_+, v_-)=(1/2,1)$.
Under such conditions, a small nonconservative force can result in a small oscillation and significantly shift the distribution during the relaxation process. Therefore, a large value of $d_t S_t^{\rm sys;X}$ may be responsible for the positivity of $\sigma_t^{\mathrm{ex};\mathrm{X}}$ if an oscillation is small enough. Conversely, if the steady-state and initial distributions are similar, i.e., $(v_+, v_-)=(1,1/2)$,  a negative value of $\sigma^{\rm ex;X}_t$ can occur because $d_t S^{\rm sys;X}_t$ cannot be responsible for the positivity of $\sigma^{\rm ex;X}_t$.
\begin{figure}
    \centering
    \includegraphics[width=\linewidth]{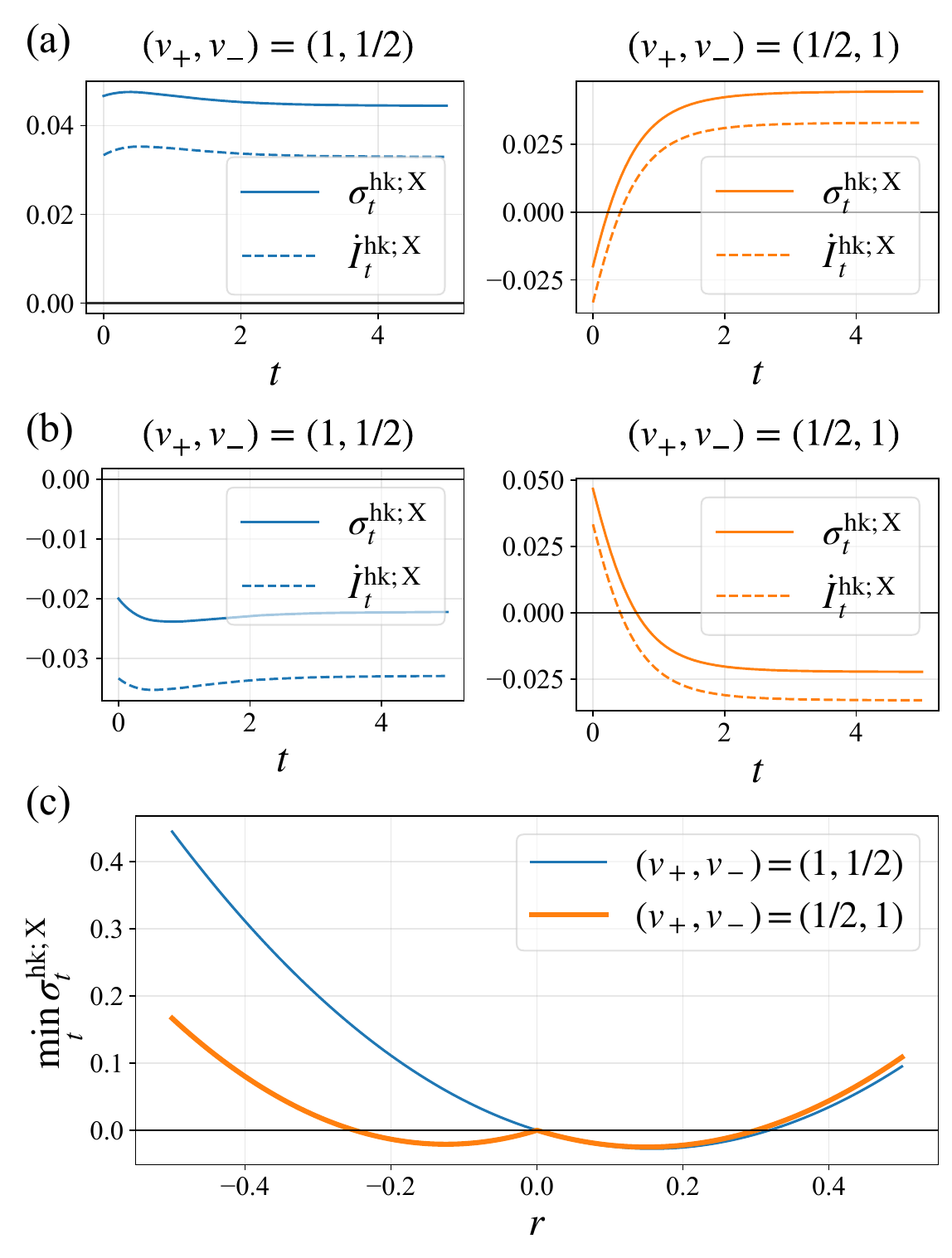}
    \caption{(a,b) Time evolution of $\sigma_t^{\mathrm{hk};\mathrm{X}}$ and $\dot{I}_t^{\mathrm{hk};\mathrm{X}}$ for (a) $r=-0.1$ and (b) $r=0.1$. (c) Minimum value of $\sigma_t^{\mathrm{hk};\mathrm{X}}$ over time $t$.}
    \label{fig:housekeeping}
\end{figure}

We next discuss the emergence of the housekeeping demon. 
The typical behavior of $\sigma_t^{\mathrm{hk};\mathrm{X}}$ and $\dot{I}_t^{\mathrm{hk};\mathrm{X}}$ is shown in Fig.~\ref{fig:housekeeping}(a) and (b), where $r$ is set to either (a) $r=-0.1$ or (b) $r=0.1$. We confirm that the generalized second law of information thermodynamics in terms of the housekeeping dissipation, i.e., Eq.~\eqref{housekeeping2ndlawofinfothermo}, holds. 
When $r=-0.1$, the initial condition $(v_+,v_-)=(1/2,1)$ yields transient negative values at early times, but the apparent entropy change rate does not become negative at late times. 
Conversely, the housekeeping demon is observed for $r=0.1$ at later times. In the steady state, the housekeeping demon can be considered a conventional autonomous demon for $r=0.1$, where
the counterclockwise nonconservative force reduces the fluctuations of $\rm X$'s degree of freedom (see also Fig.~\ref{fig:heatmap}(a)). If the steady-state and initial distributions are similar, i.e., $(v_+, v_-)=(1,1/2)$, $\sigma^{\rm hk;X}_t$ can be negative and the housekeeping demon can emerge only when $r>0$ and the counterclockwise nonconservative force exists (see Fig.~\ref{fig:housekeeping}(c)). Conversely, the housekeeping demon can also emerge during transient dynamics even if $r<0$ (see Fig.~\ref{fig:housekeeping}(c)). The emergence of the housekeeping demon during transient dynamics (see also Fig.~\ref{fig:housekeeping}(a)) differs from the conventional autonomous demon, which is only observed in the steady state. 

\section{discussions}
\label{sec5}
In this paper, we investigated the decomposition of information flow proposed in Ref.~\cite{maekawa2025geometric} in an overdamped Langevin system with continuous state variables. This enabled us to derive continuous-state counterparts of the generalized second law of information thermodynamics, thermodynamic uncertainty relations, and information-thermodynamic speed limits, which were originally established for Markov jump systems with discrete degrees of freedom. Furthermore, relative to the discrete case, the key differences are as follows. The problem can be formulated elegantly as an optimal transport problem for the continuous degrees of freedom of the subsystem without requiring a new generalization of the 2-Wasserstein distance for the subsystems as discussed in Ref.~\cite{maekawa2025geometric}. We also derive trade-off relations expressed in terms of the Fisher information matrix and a Koopman mode decomposition of the partial housekeeping entropy production rate, leveraging methods specific to continuous degrees of freedom.

This decomposition of information flow introduces the concepts of the excess demon and the housekeeping demon into information thermodynamics, because excess information flow and housekeeping information flow lead to an apparent violation of the generalized second law of thermodynamics in terms of excess dissipation and housekeeping dissipation, respectively. A key advantage of this paper over discrete-state Markov jump systems~\cite{maekawa2025geometric} is that, when the dynamics can be described by Gaussian distributions, both housekeeping and excess information flows admit an analytical treatment, as demonstrated in the examples. Based on this analytical formulation, we were also able to numerically investigate the conditions under which the housekeeping demon and the excess demon emerge.

It is important to note that the emergence of an excess demon or a housekeeping demon does not, by itself, imply any particular physical implementation of the underlying dynamics. The excess demon is
 associated with nonstationary dynamics, as exemplified by measurement-and-feedback procedures, whereas the housekeeping demon is associated with the behavior of an autonomous demon that can operate in a nonequilibrium steady state without changing the probability
 distribution of the entire system. However, these correspondences concern the dynamical characteristics captured by the present classification and do not determine whether the physical implementation itself is autonomous or explicitly involves measurement and feedback. Furthermore, some classes of autonomous demons, including those inspired by the work of Mandal and Jarzynski~\cite{mandal2012work} and implemented using information reservoirs~\cite{barato2014stochastic}, are defined in terms of specific physical implementations. These classes of autonomous demons are generally not directly related to the classification of demons introduced in the present work.

Furthermore, these results can be regarded as a fundamental framework for optimal transport in subsystems, the thermodynamics of subsystems, and information exchange between subsystems. Optimal transport in subsystems has been studied in terms of finite-time information erasure~\cite{aurell2012refined,proesmans2020,nakazato2021geometrical,Zhen2021,Lee2022,oikawa2025experimentally}, its optimal protocols~\cite{aurell2011optimal,nakazato2021geometrical, kamijima2025optimal}, and a generalization of the second law of information thermodynamics. Our results are also related to thermodynamic trade-off relations in information thermodynamics~\cite{otsubo2020estimating,wolpert2020uncertainty,nakazato2021geometrical,  tasnim2021thermodynamic,tanogami2023universal, matsumoto2025learning, dechant2025precision}. Our results are closely related to these studies, and we believe that the concepts of excess information flow and housekeeping information flow introduced here are useful both for obtaining a more refined understanding of existing results and for deriving tighter thermodynamic bounds. In addition, building on the present work, it would be interesting to explore extensions to settings involving game-theoretic conflicts between subsystems, such as those considered in Refs.~\cite{fujimoto2024game, nagase2024thermodynamically, nagase2025thermodynamic, kamijima2025finite, kamijima2025optimal}.

The results presented in this paper can be naturally extended to situations in which $\mathsf{D}$ varies over time. In that case, however, the generalized $2$-Wasserstein distance can no longer be interpreted solely in terms of variable transformations involving $\mathsf{D}$, as discussed in Appendix~\ref{appendix:generalized2-Wasserstein}. In such cases, only the instantaneous speed of the generalized $2$-Wasserstein distance and its time integral can be defined in general. If $\mathsf{D}= \mu T_t\mathsf{I}$ with a time-dependent temperature $T_t$, we may instead consider the corresponding expression in terms of the conventional $2$-Wasserstein distance. In this setting, the bound based on the $2$-Wasserstein distance is no longer a lower bound on the entropy production itself, but rather a lower bound on the time integral of the product of the entropy production rate and the temperature~\cite{ikeda2025}. 
A representative Maxwell’s demon-like phenomenon in which $\mathsf{D}$ changes over time is sensory adaptation~\cite{lan2012energy,ito2015maxwell,hartich2016sensory}, which is described by the chemical Langevin equations. Indeed, Ref.~\cite{ito2015maxwell} considers the magnitude of time-varying noise, that is, time-varying $\mathsf{D}$, to discuss stimulus-induced transient conditions and stimulus-driven periodic steady states, and examines Maxwell’s demon-like behavior in sensory adaptation via the second law of information thermodynamics. The concept of the excess demon may be particularly well suited to characterizing such Maxwell’s demon-like properties of sensory adaptation, which is a transient signal-response phenomenon. It is therefore an interesting problem to investigate how excess and housekeeping information flows behave in the presence of external stimuli, taking into account the temporal evolution of $\mathsf{D}$ or periodic steady states, and how these behaviors can be interpreted in terms of an excess demon or a housekeeping demon.

Furthermore, the results presented here can be extended to cases involving contact with multiple heat baths at different temperatures. For example, the discussion of the general Markov jump process~\cite{maekawa2025geometric} can be applied to such cases, as can the corresponding result for overdamped Langevin systems. When in contact with multiple heat baths at different temperatures, the system generally becomes nonconservative. This is because, even if $\boldsymbol{F}_t(\boldsymbol{z})$ is given by a potential force $-\nabla U_t(\boldsymbol{z})$, the thermodynamic force may not be expressed as $\boldsymbol{f}_t(\boldsymbol{z})=-\nabla \phi_t(\boldsymbol{z})$, which means that the system can be regarded as a nonconservative system. In this case, the housekeeping information flow does not vanish. The simplest example of this occurs when systems $\rm X$ and $\rm Y$ are in contact with heat baths at different temperatures. Behavior resembling that of the Feynman ratchet~\cite{parrondo1996criticism} or Maxwell's demon can then be observed. It would be interesting to consider the emergence of an excess demon and a housekeeping demon under such conditions. However, when considering multiple heat baths, the correspondence with optimal transport metrics such as the generalized $2$-Wasserstein distance becomes more complex. This is why we focus on the case of a uniform temperature.

This result is specific to the overdamped Langevin equation and does not directly extend to underdamped Langevin systems. This is because the 2-Wasserstein distance arising from optimal transport does not, in general, provide the lower bound on dissipation in underdamped Langevin systems. However, in information thermodynamics, underdamped Langevin systems exhibit phenomena such as feedback cooling~\cite{kim2004entropy,kim2007fluctuation,munakata2013feedback} that can be viewed as Maxwell’s demon~\cite{ito2011effects,horowitz2014second,rosinberg2016continuous,dechant2025precision}. It is therefore a highly intriguing direction to investigate whether similar housekeeping and excess information flows can be introduced in these underdamped Langevin systems. In this regard, methodologies developed in papers that discuss the optimal transport problem in underdamped Langevin systems~\cite{dechant2019thermodynamic,sabbagh2024wasserstein}, as well as approaches that introduce geometric decomposition in Markov jump systems~\cite{kolchinsky2024generalized}, fluid systems~\cite{yoshimura2024two}, and open quantum systems~\cite{yoshimura2025force} with odd degrees of freedom or reversible flux, may provide useful starting points.

Another important direction is to extend the present framework beyond bipartite systems. The bipartite assumption plays an essential role throughout the present work because it allows the thermodynamic force to be separated into contributions associated with the two subsystems, thereby enabling the definition of subsystem-wise excess and housekeeping information flows. In general nonbipartite systems, where the subsystem dynamics are directly coupled through the noise or mobility matrix, this separation is no longer straightforward, and the present framework cannot be directly extended. Nevertheless, some parts of the theory are expected to remain applicable. For example, although the second law of information thermodynamics can be formulated even for general nonbipartite systems~\cite{ito2013,ito2016backward,chetrite2019information}, the partial entropy production rates can no longer be obtained as a decomposition of the total entropy production rate~\cite{ito2020unified}. Consequently, extending the present definition of the partial excess and housekeeping entropy production rates appears to be nontrivial. By contrast, the local excess entropy production rate may be formulated solely from the coarse-grained dynamics of each subsystem and may therefore admit a natural extension beyond the bipartite setting. This suggests that the optimal-transport formulation for subsystem dynamics may naturally extend to more general nonbipartite systems. Clarifying which aspects of the present geometric decomposition survive in genuinely nonbipartite dynamics, and whether an alternative decomposition of information flow exists in such systems, remains an interesting direction for future work.

Just as discussions on the generative accuracy of diffusion models~\cite{sohl2015deep,song2020score,lipman2022flow} have been aided by thermodynamics based on optimal transport~\cite{ikeda2025}, combining optimal transport with information thermodynamics could improve our understanding of generative models. For example, when considering diffusion models operating within limited spaces such as latent spaces~\cite{rombach2022high} rather than the true data space, information thermodynamic methods could be applied by treating such models as diffusion processes within a subsystem. Indeed, it has been suggested that machine learning on neural networks could be approached from the perspective of information thermodynamics~\cite{Goldt2017}. Therefore, exploring whether such a framework can be applied to various generative AI methods involving optimal transport is a promising area for future research.

\begin{acknowledgments}
S.I. thanks Daiki Sekizawa and Masafumi Oizumi for their discussions on the Koopman mode decomposition. The authors thank Artemy Kolchinsky for his discussions on the geometric decomposition and optimal transport.
S.I.\ is supported by JSPS KAKENHI Grants No.~22H01141, No.~23H00467, and No.~24H00834, and UTEC-UTokyo FSI Research Grant Program, and JST ERATO Grant Number JPMJER2302.
R.N. is supported by JSR Fellowship, the University of Tokyo. A.D. is supported by JSPS KAKENHI Grants No. 24H00833 and 25K00926.
K.Y.\ is supported by the Special Postdoctoral Researchers Program at RIKEN, and JSPS KAKENHI Grants No.~22H01141.

\end{acknowledgments}
\appendix
\section{Variational formulas}
\label{appendix:variationalformula}
Here we prove variational formulas for the excess and housekeeping entropy production rates [Eqs.~\eqref{BBformula}, ~\eqref{TUR1}, ~\eqref{variational formulas2} and ~\eqref{TUR2}] and variational formulas for the local excess and local housekeeping entropy production rates [Eqs.~\eqref{BBformula2}, ~\eqref{TUR3}, ~\eqref{variationallocal}, ~\eqref{TUR4} and~\eqref{BBformula3}]. We first prove Eqs.~\eqref{BBformula}, ~\eqref{TUR1}, ~\eqref{variational formulas2} and ~\eqref{TUR2}. 

To prove Eq.~\eqref{BBformula}, we consider the quantity $\boldsymbol{f}'_t(\boldsymbol{z})$ that satisfies $\partial_t p_t (\boldsymbol{z}) = -\nabla \cdot [p_t(\boldsymbol{z})\mathsf{D}\boldsymbol{f}'_t(\boldsymbol{z})] $. Because $ \langle -\nabla \phi_t , \nabla  \phi_t+  \boldsymbol{f}'_t \rangle_{p_t \mathsf{D}} = \int d\boldsymbol{z} \phi_t(\boldsymbol{z}) \nabla \cdot [ p_t (\boldsymbol{z}) \mathsf{D}(\nabla  \phi_t(\boldsymbol{z})+  \boldsymbol{f}'_t (\boldsymbol{z}))]=0$, we obtain
\begin{align}
&\langle \boldsymbol{f}'_t, \boldsymbol{f}'_t\rangle_{p_t \mathsf{D}} \nonumber\\
&= \langle -\nabla \phi_t +\nabla  \phi_t+  \boldsymbol{f}'_t, -\nabla \phi_t +\nabla  \phi_t+  \boldsymbol{f}'_t \rangle_{p_t \mathsf{D}} \nonumber \\
&= \langle -\nabla \phi_t , -\nabla \phi_t \rangle_{p_t \mathsf{D}} +\langle  \nabla  \phi_t+  \boldsymbol{f}'_t,\nabla  \phi_t+  \boldsymbol{f}'_t \rangle_{p_t \mathsf{D}} \nonumber \\
&\geq \langle -\nabla \phi_t , -\nabla \phi_t \rangle_{p_t \mathsf{D}} = \dot{\Sigma}^{\rm ex}_t.
\end{align}
Here, $-\nabla \phi_t (\boldsymbol{z})$ satisfies  $\partial_t p_t (\boldsymbol{z}) = -\nabla \cdot [p_t(\boldsymbol{z})\mathsf{D} (-\nabla \phi_t (\boldsymbol{z}))] $. The minimum value $\dot{\Sigma}^{\rm ex}_t$ can  be achieved when $ \boldsymbol{f}'_t(\boldsymbol{z}) = -\nabla  \phi_t(\boldsymbol{z})$, and Eq.~\eqref{BBformula} \begin{align}
 \dot{\Sigma}^{\rm ex}_t &= \inf_{\boldsymbol{f}'_t(\boldsymbol{z})|\partial_t p_t = -\nabla \cdot ( p_t\mathsf{D}\boldsymbol{f}'_t)} \langle \boldsymbol{f}'_t, \boldsymbol{f}'_t\rangle_{p_t \mathsf{D}},
\end{align}
is verified.

To prove Eq.~\eqref{TUR1}, we consider the Cauchy-Schwarz inequality 
\begin{align}
 \dot{\Sigma}^{\rm ex}_t &= \langle -\nabla \phi_t, -\nabla \phi_t \rangle_{p_t \mathsf{D}} \nonumber \\
 &\geq \frac{(\langle -\nabla \psi, -\nabla \phi_t\rangle_{p_t \mathsf{D}})^2}{\langle -\nabla \psi, -\nabla \psi\rangle_{p_t \mathsf{D}}}\nonumber\\
 &= \frac{(\langle -\nabla \psi, \boldsymbol{f}_t\rangle_{p_t \mathsf{D}})^2}{\langle -\nabla \psi, -\nabla \psi\rangle_{p_t \mathsf{D}}},
\end{align}
where we used $\langle -\nabla \psi, \nabla \phi_t + \boldsymbol{f}_t\rangle_{p_t \mathsf{D}}= \int d\boldsymbol{z} \psi(\boldsymbol{z}) \nabla \cdot [ p_t (\boldsymbol{z}) \mathsf{D}(\nabla  \phi_t(\boldsymbol{z})+  \boldsymbol{f}_t (\boldsymbol{z}))]=0$. The maximum value can be achieved when $\nabla \psi (\boldsymbol{z}) \propto \nabla \phi_t (\boldsymbol{z})$, and Eq.~\eqref{TUR1},
\begin{align}
 \dot{\Sigma}^{\rm ex}_t 
 &= \sup_{\psi(\boldsymbol{z})} \frac{(\langle -\nabla \psi, \boldsymbol{f}_t\rangle_{p_t \mathsf{D}})^2}{\langle -\nabla \psi, -\nabla \psi\rangle_{p_t \mathsf{D}}},
\end{align}
is verified.

To prove Eq.~\eqref{variational formulas2}, we consider
\begin{align}
 &\langle \boldsymbol{f}_t +\nabla \psi, \boldsymbol{f}_t + \nabla \psi \rangle_{p_t \mathsf{D}} \nonumber\\
 &= \langle \boldsymbol{f}_t +\nabla (\phi_t -\phi_t + \psi), \boldsymbol{f}_t +\nabla (\phi_t - \phi_t +  \psi) \rangle_{p_t \mathsf{D}} \nonumber\\
 &=\langle \boldsymbol{f}_t +\nabla \phi_t , \boldsymbol{f}_t +\nabla \phi_t \rangle_{p_t \mathsf{D}} +\langle \nabla (\psi-\phi_t ), \nabla (\psi-\phi_t )\rangle_{p_t \mathsf{D}}\nonumber\\
 &\geq \langle \boldsymbol{f}_t +\nabla \phi_t , \boldsymbol{f}_t +\nabla \phi_t \rangle_{p_t \mathsf{D}} = \dot{\Sigma}^{\rm hk}_t,
 \end{align}
 where we used $\langle \boldsymbol{f}_t +\nabla \phi_t ,\nabla (\psi-\phi_t) \rangle_{p_t \mathsf{D}}= -\int d\boldsymbol{z} (\psi(\boldsymbol{z}) -\phi_t(\boldsymbol{z})) \nabla \cdot [ p_t (\boldsymbol{z}) \mathsf{D}(\boldsymbol{f}_t (\boldsymbol{z})+\nabla  \phi_t(\boldsymbol{z}))]=0$. The minimum value can be achieved when $\nabla \psi(\boldsymbol{z})= \nabla \phi_t(\boldsymbol{z})$, and Eq.~\eqref{variational formulas2},
 \begin{align}
 \dot{\Sigma}^{\rm hk}_t &= \inf_{\psi(\boldsymbol{z})} \langle \boldsymbol{f}_t +\nabla \psi, \boldsymbol{f}_t + \nabla \psi \rangle_{p_t \mathsf{D}},
\end{align}
is verified.

To prove Eq.~\eqref{TUR2}, we consider the quantity $\boldsymbol{f}'_t(\boldsymbol{z})$ that satisfies $0 = -\nabla \cdot ( p_t(\boldsymbol{z})\mathsf{D}\boldsymbol{f}'_t(\boldsymbol{z})) $. We obtain the Cauchy-Schwarz inequality 
\begin{align}
 \dot{\Sigma}^{\rm hk}_t &= \langle \boldsymbol{f}_t +\nabla \phi_t , \boldsymbol{f}_t +\nabla \phi_t \rangle_{p_t \mathsf{D}} \nonumber \\
 &\geq \frac{(\langle  \boldsymbol{f}_t',  \boldsymbol{f}_t +\nabla \phi_t\rangle_{p_t \mathsf{D}})^2}{\langle\boldsymbol{f}_t', \boldsymbol{f}_t' \rangle_{p_t \mathsf{D}}}\nonumber\\
 &= \frac{(\langle \boldsymbol{f}_t', \boldsymbol{f}_t\rangle_{p_t \mathsf{D}})^2}{\langle \boldsymbol{f}_t', \boldsymbol{f}_t' \rangle_{p_t \mathsf{D}}},
\end{align}
where we used $\langle \boldsymbol{f}_t', \nabla \phi_t \rangle_{p_t \mathsf{D}}= -\int d\boldsymbol{z} \phi_t (\boldsymbol{z}) \nabla \cdot [ p_t (\boldsymbol{z}) \mathsf{D}  \boldsymbol{f}'_t (\boldsymbol{z})]=0$. The maximum value can be achieved when $\boldsymbol{f}'_t (\boldsymbol{z}) \propto \boldsymbol{f}_t (\boldsymbol{z})+\nabla \phi_t (\boldsymbol{z})$, and Eq.~\eqref{TUR2},
\begin{align}
 \dot{\Sigma}^{\rm hk}_t &= \sup_{\boldsymbol{f}'_t (\boldsymbol{z})|-\nabla \cdot ( p_t\mathsf{D}\boldsymbol{f}'_t) =0} \frac{(\langle \boldsymbol{f}_t', \boldsymbol{f}_t\rangle_{p_t \mathsf{D}})^2}{\langle \boldsymbol{f}_t', \boldsymbol{f}_t' \rangle_{p_t \mathsf{D}}}, 
\end{align}
is verified.

We next prove Eqs.~\eqref{BBformula2}, ~\eqref{TUR3}, ~\eqref{variationallocal} and ~\eqref{TUR4} in parallel with the proofs of Eqs.~\eqref{BBformula}, ~\eqref{TUR1}, ~\eqref{variational formulas2} and ~\eqref{TUR2}. 

To prove Eq.~\eqref{BBformula2}, we consider the quantity ${\boldsymbol{f}^{\rm X}_t}'(\boldsymbol{x},\boldsymbol{y})$ that satisfies $\partial_t p^{\rm X}_t (\boldsymbol{x}) = - \int d\boldsymbol{y} \nabla_{\boldsymbol{x}} \cdot [ p_t(\boldsymbol{x},\boldsymbol{y})\mathsf{D}^{\rm X} {\boldsymbol{f}^{\rm X}_t}'(\boldsymbol{x}, \boldsymbol{y})] $. Because $ \langle -\nabla_{\boldsymbol{x}} \phi^{\rm X}_t , \nabla_{\boldsymbol{x}}  \phi^{\rm X}_t+  {\boldsymbol{f}^{\rm X}_t}' \rangle_{p_t \mathsf{D}^{\rm X}} = \int d\boldsymbol{x} \int d\boldsymbol{y} \phi^{\rm X}_t(\boldsymbol{x}) \nabla_{\boldsymbol{x}} \cdot [ p_t (\boldsymbol{x}, \boldsymbol{y}) \mathsf{D}^{\rm X}(\nabla_{\boldsymbol{x}}  \phi^{\rm X}_t(\boldsymbol{x})+  {\boldsymbol{f}^{\rm X}_t}' (\boldsymbol{x},\boldsymbol{y}))]=0$, we obtain
\begin{align}
&\langle {\boldsymbol{f}^{\rm X}_t}', {\boldsymbol{f}^{\rm X}_t}' \rangle_{p_t \mathsf{D}^{\rm X}} \nonumber\\
&= \dot{\Sigma}^{\rm localex;X}_t +\langle  \nabla_{\boldsymbol{x}}  \phi^{\rm X}_t+  {\boldsymbol{f}^{\rm X}_t}',\nabla_{\boldsymbol{x}}  \phi^{\rm X}_t+  {\boldsymbol{f}^{\rm X}_t}' \rangle_{p_t \mathsf{D}^{\rm X}} \nonumber \\
&\geq \dot{\Sigma}^{\rm localex;X}_t.
\end{align}
Here, $-\nabla_{\boldsymbol{x}} \phi^{\rm X}_t (\boldsymbol{x})$ satisfies  $\partial_t p^{\rm X}_t (\boldsymbol{x}) = -\nabla_{\boldsymbol{x}} \cdot [p^{\rm X}_t(\boldsymbol{x})\mathsf{D}^{\rm X} (-\nabla_{\boldsymbol{x}} \phi^{\rm X}_t (\boldsymbol{x}))] $. The minimum value $\dot{\Sigma}^{\rm localex;X}_t$ can  be achieved when ${\boldsymbol{f}^{\rm X}_t}'(\boldsymbol{x}, \boldsymbol{y})  = - \nabla_{\boldsymbol{x}}  \phi^{\rm X}_t(\boldsymbol{x})$, and Eq.~\eqref{BBformula2} is verified.

To prove Eq.~\eqref{TUR3}, we consider the Cauchy-Schwarz inequality 
\begin{align}
 \dot{\Sigma}^{\rm localex;X}_t 
 &\geq \frac{(\langle -\nabla_{\boldsymbol{x}} \psi^{\rm X}, -\nabla_{\boldsymbol{x}} \phi^{\rm X}_t\rangle_{p_t \mathsf{D}^{\rm X}})^2}{\langle -\nabla_{\boldsymbol{x}} \psi^{\rm X}, -\nabla_{\boldsymbol{x}} \psi^{\rm X} \rangle_{p_t \mathsf{D}^{\rm X}}}\nonumber\\
 &= \frac{(\langle -\nabla_{\boldsymbol{x}} \psi^{\rm X}, \boldsymbol{f}^{\rm X}_t\rangle_{p_t \mathsf{D}^{\rm X}})^2}{\langle -\nabla_{\boldsymbol{x}} \psi^{\rm X}, -\nabla_{\boldsymbol{x}} \psi^{\rm X}\rangle_{p_t \mathsf{D}^{\rm X}}},
\end{align}
where we used $\langle -\nabla_{\boldsymbol{x}} \psi^{\rm X}, \nabla_{\boldsymbol{x}} \phi^{\rm X}_t + \boldsymbol{f}^{\rm X}_t\rangle_{p_t \mathsf{D}^{\rm X}}= \int d\boldsymbol{x} \psi^{\rm X}(\boldsymbol{x}) \nabla_{\boldsymbol{x}} \cdot [ p^{\rm X}_t (\boldsymbol{x}) \mathsf{D}^{\rm X}(\nabla_{\boldsymbol{x}}  \phi^{\rm X}_t(\boldsymbol{x})+  \bar{\boldsymbol{f}}^{\rm X}_t (\boldsymbol{x}))]=0$. The maximum value can be achieved when $\nabla_{\boldsymbol{x}} \psi^{\rm X} (\boldsymbol{x}) \propto \nabla_{\boldsymbol{x}} \phi^{\rm X}_t (\boldsymbol{x})$, and Eq.~\eqref{TUR3} is verified.

To prove Eq.~\eqref{variationallocal}, we consider
\begin{align}
 &\langle \boldsymbol{f}^{\rm X}_t +\nabla_{\boldsymbol{x}} \psi^{\rm X}, \boldsymbol{f}^{\rm X}_t + \nabla_{\boldsymbol{x}} \psi^{\rm X} \rangle_{p_t \mathsf{D}^{\rm X}} \nonumber\\
 &=\dot{\Sigma}^{\rm localhk;X}_t +\langle \nabla_{\boldsymbol{x}}  (\psi^{\rm X}-\phi_t^{\rm X} ), \nabla_{\boldsymbol{x}}  (\psi^{\rm X}-\phi_t^{\rm X} )\rangle_{p_t \mathsf{D}^{\rm X}}\nonumber\\
 &\geq \dot{\Sigma}^{\rm localhk;X}_t,
 \end{align}
 where we used $\langle \boldsymbol{f}^{\rm X}_t +\nabla_{\boldsymbol{x}}  \phi^{\rm X}_t ,\nabla_{\boldsymbol{x}}  (\psi^{\rm X}-\phi^{\rm X}_t) \rangle_{p_t \mathsf{D}^{\rm X}}= \int d\boldsymbol{x} (\psi^{\rm X}(\boldsymbol{x}) -\phi^{\rm X}_t(\boldsymbol{x})) \! \nabla_{\boldsymbol{x}} \cdot [ p^{\rm X}_t (\boldsymbol{x}) \mathsf{D}^{\rm X}(\bar{\boldsymbol{f}}^{\rm X}_t (\boldsymbol{x})+\nabla_{\boldsymbol{x}} \phi^{\rm X}_t(\boldsymbol{x}))]=0$. The minimum value can be achieved when $\nabla_{\boldsymbol{x}} \psi^{\rm X}(\boldsymbol{x})= \nabla_{\boldsymbol{x}} \phi^{\rm X}_t(\boldsymbol{x})$, and Eq.~\eqref{variationallocal} is verified.

 To prove Eq.~\eqref{TUR4}, we consider the quantity ${\boldsymbol{f}^{\rm X}_t}'(\boldsymbol{x},\boldsymbol{y})$ that satisfies $0= - \int d\boldsymbol{y} \nabla_{\boldsymbol{x}} \cdot ( p_t(\boldsymbol{x},\boldsymbol{y})\mathsf{D}^{\rm X}{\boldsymbol{f}^{\rm X}_t}'(\boldsymbol{x},\boldsymbol{y} )) $. We obtain the Cauchy-Schwarz inequality 
\begin{align}
 \dot{\Sigma}^{\rm localhk;X}_t  &\geq \frac{(\langle  {\boldsymbol{f}^{\rm X}_t}',  \boldsymbol{f}^{\rm X}_t +\nabla_{\boldsymbol{x}} \phi^{\rm X}_t\rangle_{p_t \mathsf{D}^{\rm X}})^2}{\langle {\boldsymbol{f}^{\rm X}_t}', {\boldsymbol{f}^{\rm X}_t}' \rangle_{p_t \mathsf{D}^{\rm X}}}\nonumber\\
 &= \frac{(\langle {\boldsymbol{f}^{\rm X}_t}', \boldsymbol{f}^{\rm X}_t\rangle_{p_t \mathsf{D}^{\rm X}})^2}{\langle {\boldsymbol{f}^{\rm X}_t}', {\boldsymbol{f}^{\rm X}_t}' \rangle_{p_t \mathsf{D}^{\rm X}}},
\end{align}
where we used $\langle {\boldsymbol{f}^{\rm X}_t}', \nabla_{\boldsymbol{x}} \phi^{\rm X}_t \rangle_{p_t \mathsf{D}^{\rm X}}= -\int d\boldsymbol{x} \int d\boldsymbol{y} \phi^{\rm X}_t (\boldsymbol{x}) \nabla_{\boldsymbol{x}} \cdot [ p_t (\boldsymbol{x},\boldsymbol{y}) \mathsf{D}^{\rm X}  {\boldsymbol{f}^{\rm X}_t }'(\boldsymbol{x},\boldsymbol{y})]=0$. The maximum value can be achieved when ${\boldsymbol{f}^{\rm X}_t}' (\boldsymbol{x},\boldsymbol{y}) \propto \boldsymbol{f}^{\rm X}_t (\boldsymbol{x},\boldsymbol{y})+\nabla_{\boldsymbol{x}} \phi^{\rm X}_t (\boldsymbol{x})$, and Eq.~\eqref{TUR4} is verified.

We finally prove Eq.~\eqref{BBformula3}. We consider the quantity $\psi^{\rm X}(\boldsymbol{x})$ that satisfies 
$\partial_t p^{\rm X}_t(\boldsymbol{x}) = -\nabla_{\boldsymbol{x}} \cdot [p_t^{\rm X} (\boldsymbol{x}) \mathsf{D}^{\rm X} (-\nabla_{\boldsymbol{x}} \psi^{\rm X} (\boldsymbol{x}) )]$. Because $ \langle -\nabla_{\boldsymbol{x}} \phi^{\rm X}_t , -\nabla_{\boldsymbol{x}}  ( \psi^{\rm X}-\phi^{\rm X}_t) \rangle_{p^{\rm X}_t \mathsf{D}^{\rm X}} = \int d\boldsymbol{x} \int d\boldsymbol{y} \phi^{\rm X}_t(\boldsymbol{x}) \nabla_{\boldsymbol{x}} \cdot [ p_t (\boldsymbol{x}, \boldsymbol{y}) \mathsf{D}^{\rm X}\nabla_{\boldsymbol{x}} ( \phi^{\rm X}_t(\boldsymbol{x})-  \psi^{\rm X}(\boldsymbol{x}))]=0$, we obtain
\begin{align}
&\langle -\nabla_{\boldsymbol{x}}\psi^{\rm X}, -\nabla_{\boldsymbol{x}}\psi^{\rm X} \rangle_{p^{\rm X}_t \mathsf{D}^{\rm X}} \nonumber\\
=& \langle -\nabla_{\boldsymbol{x}}\phi^{\rm X}_t, -\nabla_{\boldsymbol{x}}\phi^{\rm X}_t \rangle_{p^{\rm X}_t \mathsf{D}^{\rm X}} \nonumber\\
&+\langle -\nabla_{\boldsymbol{x}} (\psi^{\rm X} - \phi^{\rm X}_t), -\nabla_{\boldsymbol{x}} (\psi^{\rm X} - \phi^{\rm X}_t) \rangle_{p^{\rm X}_t \mathsf{D}^{\rm X}} \nonumber \\
\geq&  \langle -\nabla_{\boldsymbol{x}}\phi^{\rm X}_t, -\nabla_{\boldsymbol{x}}\phi^{\rm X}_t \rangle_{p^{\rm X}_t \mathsf{D}^{\rm X}} \nonumber\\ 
=&\langle -\nabla_{\boldsymbol{x}}\phi^{\rm X}_t, -\nabla_{\boldsymbol{x}}\phi^{\rm X}_t \rangle_{p_t \mathsf{D}^{\rm X}} = \dot{\Sigma}_t^{\rm localex;X}.
\end{align}
The minimum value $\dot{\Sigma}_t^{\rm localex;X}$ can be achieved when $\nabla_{\boldsymbol{x}} \psi^{\rm X} (\boldsymbol{x})= \nabla_{\boldsymbol{x}} \phi^{\rm X}_t (\boldsymbol{x})$, and Eq.~\eqref{BBformula3} is verified.

\section{Generalized 2-Wasserstein distance and coordinate transformation}
\label{appendix:generalized2-Wasserstein}
We show that the generalized $2$-Wasserstein distance can be understood as the $2$-Wasserstein distance under the coordinate transformation. Using the notation $\boldsymbol{u}_t(\boldsymbol{z}) = \mathsf{D} \boldsymbol{f}'_t (\boldsymbol{z})$, the definition of the generalized $2$-Wasserstein distance [Eq.~\eqref{generalBBformula}] can be rewritten as
\begin{align}
 &\tilde{\mathcal{W}}^{\mathsf{D}^{-1}}_2 (p^{\rm ini}, p^{\rm fin}) \nonumber\\
 &= \sqrt{ \inf_{(\boldsymbol{u}_{s}(\boldsymbol{z}), \rho_{s}(\boldsymbol{z}))_{t\leq s \leq t+\Delta t}}  
 (\Delta t)\int_t^{t+\Delta t} ds \langle \boldsymbol{u}_{s}, \boldsymbol{u}_{s}\rangle_{\rho_{s} \mathsf{D}^{-1} } }\nonumber\\
 &{\rm s.t.} \: \: \partial_{s} \rho_{s} (\boldsymbol{z})= -\nabla \cdot ( \rho_{s} (\boldsymbol{z})\boldsymbol{u}_{s}(\boldsymbol{z})),\nonumber \\
 &\:\:\:\:\:\:\:\:\rho_t(\boldsymbol{z})=p^{\rm ini}(\boldsymbol{z}),  \: \rho_{t+\Delta t}(\boldsymbol{z})=p^{\rm fin}(\boldsymbol{z}).
 \label{generalBBformula2}
\end{align}
If we consider the coordinate transformation $\tilde{\boldsymbol{z}}=\mathsf{D}^{-1/2} \boldsymbol{z}$, we can introduce the new distributions $\tilde{\rho}_t (\tilde{\boldsymbol{z}})= \rho_t(\mathsf{D}^{1/2}  \tilde{\boldsymbol{z}})|{\rm det}(D^{1/2}) |$, $\tilde{p}^{\rm ini}(\tilde{\boldsymbol{z}})=p^{\rm ini}(\mathsf{D}^{1/2}  \tilde{\boldsymbol{z}})|{\rm det}(D^{1/2}) |$ and $\tilde{p}^{\rm fin}(\tilde{\boldsymbol{z}})= p^{\rm fin}(\mathsf{D}^{1/2}  \tilde{\boldsymbol{z}})|{\rm det}(D^{1/2}) |$, where $\mathsf{D}^{1/2}$ denotes the unique symmetric positive-definite principal square root of $\mathsf{D}$ that satisfies $\mathsf{D}=\mathsf{D}^{1/2} \mathsf{D}^{1/2}$ and $\mathsf{D}^{1/2}=( \mathsf{D}^{1/2})^{\top}$. We also have $\tilde{\rho}_t (\tilde{\boldsymbol{z}}) d\tilde{\boldsymbol{z}}= \rho_t(\boldsymbol{z})d\boldsymbol{z}$.
Under the coordinate transformation, the vector field is transformed from $\boldsymbol{u}_t (\boldsymbol{z})$ to $\tilde{\boldsymbol{u}}_t (\tilde{\boldsymbol{z}})=\mathsf{D}^{-1/2} \boldsymbol{u}_t (\mathsf{D}^{1/2} \tilde{\boldsymbol{z}})$. Therefore, the optimization problem [Eq.~\eqref{generalBBformula2}] can also be rewritten as 
\begin{align}
 &\tilde{\mathcal{W}}^{\mathsf{D}^{-1}}_2 (p^{\rm ini}, p^{\rm fin}) \nonumber\\
 &= \sqrt{ \inf_{(\tilde{\boldsymbol{u}}_{s}(\tilde{\boldsymbol{z}}), \tilde{\rho}_{s}(\tilde{\boldsymbol{z}}) )_{t\leq s \leq t+\Delta t}}  
 (\Delta t)\int_t^{t+\Delta t} d{s} \langle \tilde{\boldsymbol{u}}_{s}, \tilde{\boldsymbol{u}}_{s}\rangle_{\tilde{\rho}_{s} \mathsf{I}} }\nonumber\\
 &{\rm s.t.} \: \: \partial_{s} \tilde{\rho}_{s} (\tilde{\boldsymbol{z}})= -\nabla_{\tilde{\boldsymbol{z}}} \cdot ( \tilde{\rho}_{s} (\tilde{\boldsymbol{z}}) \tilde{\boldsymbol{u}}_{s}(\tilde{\boldsymbol{z}})), \nonumber\\  &\:\:\:\:\:\:\:\:\tilde{\rho}_t(\tilde{\boldsymbol{z}})=\tilde{p}^{\rm ini}(\tilde{\boldsymbol{z}}),  \: \tilde{\rho}_{t+\Delta t}(\tilde{\boldsymbol{z}})=\tilde{p}^{\rm fin}(\tilde{\boldsymbol{z}}),
\end{align}
which implies $\tilde{\mathcal{W}}^{\mathsf{D}^{-1}}_2 (p^{\rm ini}, p^{\rm fin})  =\mathcal{W}_2 (\tilde{p}^{\rm ini}, \tilde{p}^{\rm fin})$. 

Because $\mathcal{W}_2 (\tilde{p}^{\rm ini}, \tilde{p}^{\rm fin})$ satisfies the axioms of the metric, $\tilde{\mathcal{W}}^{\mathsf{D}^{-1}}_2 (p^{\rm ini}, p^{\rm fin})$ also satisfies the axioms of the metric, i.e., (i) $\tilde{\mathcal{W}}^{\mathsf{D}^{-1}}_2 (p^{\rm ini}, p^{\rm fin}) \geq 0$, (ii) $\tilde{\mathcal{W}}^{\mathsf{D}^{-1}}_2 (p^{\rm ini}, p^{\rm fin}) = 0 \Leftrightarrow p^{\rm ini}=p^{\rm fin}$, (iii) $\tilde{\mathcal{W}}^{\mathsf{D}^{-1}}_2 (p^{\rm ini}, p^{\rm fin}) =\tilde{\mathcal{W}}^{\mathsf{D}^{-1}}_2 ( p^{\rm fin}, p^{\rm ini})$ and (iv) $\tilde{\mathcal{W}}^{\mathsf{D}^{-1}}_2 (p^{\rm ini}, p^{\rm fin}) \leq \tilde{\mathcal{W}}^{\mathsf{D}^{-1}}_2 ( p^{\rm ini}, p^{\rm mid})+\tilde{\mathcal{W}}^{\mathsf{D}^{-1}}_2 ( p^{\rm mid}, p^{\rm fin})$. These axioms of the metric are immediately evident from the fact that $\tilde{\mathcal{W}}^{\mathsf{D}^{-1}}_2 (p^{\rm ini}, p^{\rm fin})  =\mathcal{W}_2 (\tilde{p}^{\rm ini}, \tilde{p}^{\rm fin})$, $p^{\rm ini}( \boldsymbol{z})= p^{\rm fin} ( \boldsymbol{z})\Leftrightarrow \tilde{p}^{\rm ini}(\tilde{\boldsymbol{z}})= \tilde{p}^{\rm fin}(\tilde{\boldsymbol{z}})$ and $\tilde{p}^{\rm mid} (\tilde{\boldsymbol{z}}) := p^{\rm mid}(\mathsf{D}^{1/2}  \tilde{\boldsymbol{z}})|{\rm det}(D^{1/2}) |$ satisfying $\mathcal{W}_2 (\tilde{p}^{\rm ini}, \tilde{p}^{\rm fin}) \leq \mathcal{W}_2 (\tilde{p}^{\rm ini}, \tilde{p}^{\rm mid})+\mathcal{W}_2 (\tilde{p}^{\rm mid}, \tilde{p}^{\rm fin})$
can be introduced.

We note that the noise in the Langevin equation [Eq.~\eqref{Langevineq}] becomes standard white Gaussian noise with independent components under this coordinate transformation as follows,
\begin{align}
   \dot{\tilde{\boldsymbol{z}}}(t) = \mathsf{D}^{-1/2}\mathsf{\mu} \boldsymbol{F}_t (\mathsf{D}^{1/2}\tilde{\boldsymbol{z}}(t)) + \sqrt{2 } \tilde{\boldsymbol{\xi}}_t,
\end{align}
where $\tilde{\boldsymbol{\xi}}_t :=\sqrt{T} \mathsf{D}^{-1/2}\mathsf{B} \boldsymbol{\xi}_t$ is the standard white Gaussian noise because $(\sqrt{T} \mathsf{D}^{-1/2}\mathsf{B}) (\sqrt{T} \mathsf{D}^{-1/2}\mathsf{B})^{\top} = \mathsf{I}$ and thus $\mathbb{E} [\tilde{\boldsymbol{\xi}}_t ]= \boldsymbol{0}$, $\mathbb{E}[ \tilde{\boldsymbol{\xi}}_t \tilde{\boldsymbol{\xi}}_{t'}^{\top} ]= \delta(t-t') \mathsf{I}$.
Therefore, the relation $\tilde{\mathcal{W}}^{\mathsf{D}^{-1}}_2 (p^{\rm ini}, p^{\rm fin}) =\mathcal{W}_2 (\tilde{p}^{\rm ini}, \tilde{p}^{\rm fin})$ can also be considered to be related to the expression of the excess entropy production rate in this coordinate-transformed Langevin system.

We discuss an analytical expression of the generalized $2$-Wasserstein distance in the Gaussian case. When the distribution $p_t(\boldsymbol{z})$ is Gaussian $p_t (\boldsymbol{z})\sim \mathcal{N}(\boldsymbol{m}_t, \mathsf{V}_t)$, the new distribution under the coordinate transformation $\tilde{p}_t (\tilde{\boldsymbol{z}}):= p_t (\mathsf{D}^{1/2}\tilde{\boldsymbol{z}})|\rm det(\mathsf{D}^{1/2})|$ is also Gaussian $\tilde{p}_t (\tilde{\boldsymbol{z}}) \sim \mathcal{N}(\tilde{\boldsymbol{m}}_t, \tilde{\mathsf{V}}_t)$, where $\tilde{\boldsymbol{m}}_t:= \mathsf{D}^{-1/2} \boldsymbol{m}_t$ and $\tilde{\mathsf{V}}_t:= \mathsf{D}^{-1/2} {\mathsf{V}}_t (\mathsf{D}^{-1/2})^{\top}$. From the formula for the $2$-Wasserstein distance under the Gaussian distribution, the generalized $2$-Wasserstein distance between $p_t(\boldsymbol{z})$ and $p_s(\boldsymbol{z})$ is calculated as
\begin{align}
&\tilde{\mathcal{W}}^{\mathsf{D}^{-1}}_2 (p_t, p_s) \nonumber\\ 
=&\sqrt{ \| \tilde{\boldsymbol{m}}_t -\tilde{\boldsymbol{m}}_s\|^2 + {\rm tr}[\tilde{\mathsf{V}}_t+\tilde{\mathsf{V}}_s -2(\tilde{\mathsf{V}}_t^{1/2}\tilde{\mathsf{V}}_s \tilde{\mathsf{V}}_t^{1/2})^{1/2}]}.
\end{align}
Similarly, the generalized $2$-Wasserstein distance between marginalized distributions $p^{\rm X}_t (\boldsymbol{x})\sim \mathcal{N}(\boldsymbol{m}^{\rm X}_t, \mathsf{V}^{\rm XX}_t)$ and $p^{\rm X}_s (\boldsymbol{x})\sim \mathcal{N}(\boldsymbol{m}^{\rm X}_s, \mathsf{V}^{\rm XX}_s)$ is also calculated as
\begin{align}
&\tilde{\mathcal{W}}^{(\mathsf{D}^{\rm X})^{-1}}_2 (p^{\rm X}_t, p^{\rm X}_s) \nonumber\\ 
=& \Big(\| \tilde{\boldsymbol{m}}^{\rm X}_t -\tilde{\boldsymbol{m}}^{\rm X}_s\|^2 \nonumber\\
&+ {\rm tr}[\tilde{\mathsf{V}}^{\rm XX}_t+\tilde{\mathsf{V}}^{\rm XX}_s -2((\tilde{\mathsf{V}}^{\rm XX}_t)^{1/2}\tilde{\mathsf{V}}^{\rm XX}_s (\tilde{\mathsf{V}}^{\rm XX}_t)^{1/2})^{1/2}] \Big)^{1/2},
\end{align}
where $\tilde{\boldsymbol{m}}^{\rm X}_t:= (\mathsf{D}^{\rm X})^{-1/2} \boldsymbol{m}^{\rm X}_t$ and $\tilde{\mathsf{V}}^{\rm XX}_t:= (\mathsf{D}^{\rm X})^{-1/2} {\mathsf{V}}^{\rm XX}_t ((\mathsf{D}^{\rm X})^{-1/2})^{\top}$. Here,  $(\mathsf{D}^{\rm X})^{1/2}$ is defined as the unique symmetric positive-definite principal square root of $\mathsf{D}^{\rm X}$ that satisfies $\mathsf{D}^{\rm X}=(\mathsf{D}^{\rm X})^{1/2} (\mathsf{D}^{\rm X})^{1/2}$ and $(\mathsf{D}^{\rm X})^{1/2} =[(\mathsf{D}^{\rm X})^{1/2} ]^{\top}$.

\bibliography{biblio.bib}

@article{maekawa2025geometric,
  title={Geometric decomposition of information flow: New insights into information thermodynamics},
  author={Maekawa, Yoh and Nagayama, Ryuna and Yoshimura, Kohei and Ito, Sosuke},
  journal={Physical Review Research},
  volume={8},
  number={2},
  pages={023292},
  year={2026},
  publisher={APS},
  url={https://doi.org/10.1103/trrq-tkw5}
}

@article{mandal2012work,
  title={Work and information processing in a solvable model of Maxwell’s demon},
  author={Mandal, Dibyendu and Jarzynski, Christopher},
  journal={Proceedings of the National Academy of Sciences},
  volume={109},
  number={29},
  pages={11641--11645},
  year={2012},
  publisher={National Academy of Sciences}
}

@article{barato2014stochastic,
  title={Stochastic thermodynamics with information reservoirs},
  author={Barato, Andre C and Seifert, Udo},
  journal={Physical Review E},
  volume={90},
  number={4},
  pages={042150},
  year={2014},
  publisher={APS}
}

@article{fujimoto2024game,
  title = {Game-theoretical approach to minimum entropy productions in information thermodynamics},
  author = {Fujimoto, Yuma and Ito, Sosuke},
  journal = {Phys. Rev. Res.},
  volume = {6},
  issue = {1},
  pages = {013023},
  numpages = {20},
  year = {2024},
  month = {Jan},
  publisher = {American Physical Society},
  doi = {10.1103/PhysRevResearch.6.013023},
  url = {https://link.aps.org/doi/10.1103/PhysRevResearch.6.013023}
}

@article{otsubo2020estimating,
    title = {Estimating entropy production by machine learning of short-time fluctuating currents},
  author = {Otsubo, Shun and Ito, Sosuke and Dechant, Andreas and Sagawa, Takahiro},
  journal = {Phys. Rev. E},
  volume = {101},
  issue = {6},
  pages = {062106},
  numpages = {19},
  year = {2020},
  month = {Jun},
  publisher = {American Physical Society},
  doi = {10.1103/PhysRevE.101.062106},
  url = {https://link.aps.org/doi/10.1103/PhysRevE.101.062106}
}

@book{seifert2025stochastic,
  title={Stochastic thermodynamics},
  author={Seifert, Udo},
  year={2025},
  publisher={CAMBRIDGE University Press}
}

@article{ito2020unified,
  title={Unified framework for the entropy production and the stochastic interaction based on information geometry},
  author={Ito, Sosuke and Oizumi, Masafumi and Amari, Shun-ichi},
  journal={Physical Review Research},
  volume={2},
  number={3},
  pages={033048},
  year={2020},
  publisher={APS}
}

@article{chetrite2019information,
  title={Information thermodynamics for interacting stochastic systems without bipartite structure},
  author={Chetrite, Raphael and Rosinberg, ML and Sagawa, T and Tarjus, G},
  journal={Journal of Statistical Mechanics: Theory and Experiment},
  volume={2019},
  number={11},
  pages={114002},
  year={2019},
  publisher={IOP Publishing and SISSA}
}

@book{de2013non,
  title={Non-equilibrium thermodynamics},
  author={De Groot, Sybren Ruurds and Mazur, Peter},
  year={2013},
  publisher={Courier Corporation}
}

@article{schnakenberg1976,
  title = {Network theory of microscopic and macroscopic behavior of master equation systems},
  author = {Schnakenberg, J.},
  journal = {Rev. Mod. Phys.},
  volume = {48},
  issue = {4},
  pages = {571--585},
  numpages = {0},
  year = {1976},
  month = {Oct},
  publisher = {American Physical Society},
  doi = {10.1103/RevModPhys.48.571},
  url = {https://link.aps.org/doi/10.1103/RevModPhys.48.571}
}

@article{casimir1945,
  title = {On Onsager's Principle of Microscopic Reversibility},
  author = {Casimir, H. B. G.},
  journal = {Rev. Mod. Phys.},
  volume = {17},
  issue = {2-3},
  pages = {343--350},
  numpages = {0},
  year = {1945},
  month = {Apr},
  publisher = {American Physical Society},
  doi = {10.1103/RevModPhys.17.343},
  url = {https://link.aps.org/doi/10.1103/RevModPhys.17.343}
}

@article{ohga2023,
  title = {Thermodynamic Bound on the Asymmetry of Cross-Correlations},
  author = {Ohga, Naruo and Ito, Sosuke and Kolchinsky, Artemy},
  journal = {Phys. Rev. Lett.},
  volume = {131},
  issue = {7},
  pages = {077101},
  numpages = {7},
  year = {2023},
  month = {Aug},
  publisher = {American Physical Society},
  doi = {10.1103/PhysRevLett.131.077101},
  url = {https://link.aps.org/doi/10.1103/PhysRevLett.131.077101}
}

@article{Onsager1931,
  title = {Reciprocal Relations in Irreversible Processes. I.},
  author = {Onsager, Lars},
  journal = {Phys. Rev.},
  volume = {37},
  issue = {4},
  pages = {405--426},
  numpages = {0},
  year = {1931},
  month = {Feb},
  publisher = {American Physical Society},
  doi = {10.1103/PhysRev.37.405},
  url = {https://link.aps.org/doi/10.1103/PhysRev.37.405}
}

@article{Onsager1931-2,
  title = {Reciprocal Relations in Irreversible Processes. II.},
  author = {Onsager, Lars},
  journal = {Phys. Rev.},
  volume = {38},
  issue = {12},
  pages = {2265--2279},
  numpages = {0},
  year = {1931},
  month = {Dec},
  publisher = {American Physical Society},
  doi = {10.1103/PhysRev.38.2265},
  url = {https://link.aps.org/doi/10.1103/PhysRev.38.2265}
}

@article{parrondo2015thermodynamics,
  title={Thermodynamics of information},
  author={Parrondo, Juan MR and Horowitz, Jordan M and Sagawa, Takahiro},
  journal={Nature physics},
  volume={11},
  number={2},
  pages={131--139},
  year={2015},
  publisher={Nature Publishing Group UK London},
  url={https://doi.org/10.1038/nphys3230}
}

@article{Schreiber2000,
  title = {Measuring Information Transfer},
  author = {Schreiber, Thomas},
  journal = {Phys. Rev. Lett.},
  volume = {85},
  issue = {2},
  pages = {461--464},
  numpages = {0},
  year = {2000},
  month = {Jul},
  publisher = {American Physical Society},
  doi = {10.1103/PhysRevLett.85.461},
  url = {https://link.aps.org/doi/10.1103/PhysRevLett.85.461}
}

@article{Touchette2000,
  title = {Information-Theoretic Limits of Control},
  author = {Touchette, Hugo and Lloyd, Seth},
  journal = {Phys. Rev. Lett.},
  volume = {84},
  issue = {6},
  pages = {1156--1159},
  numpages = {0},
  year = {2000},
  month = {Feb},
  publisher = {American Physical Society},
  doi = {10.1103/PhysRevLett.84.1156},
  url = {https://link.aps.org/doi/10.1103/PhysRevLett.84.1156}
}

@article{sagawa2008,
  title = {Second Law of Thermodynamics with Discrete Quantum Feedback Control},
  author = {Sagawa, Takahiro and Ueda, Masahito},
  journal = {Phys. Rev. Lett.},
  volume = {100},
  issue = {8},
  pages = {080403},
  numpages = {4},
  year = {2008},
  month = {Feb},
  publisher = {American Physical Society},
  doi = {10.1103/PhysRevLett.100.080403},
  url = {https://link.aps.org/doi/10.1103/PhysRevLett.100.080403}
}

@article{sagawa2012,
  title = {Fluctuation Theorem with Information Exchange: Role of Correlations in Stochastic Thermodynamics},
  author = {Sagawa, Takahiro and Ueda, Masahito},
  journal = {Phys. Rev. Lett.},
  volume = {109},
  issue = {18},
  pages = {180602},
  numpages = {5},
  year = {2012},
  month = {Nov},
  publisher = {American Physical Society},
  doi = {10.1103/PhysRevLett.109.180602},
  url = {https://link.aps.org/doi/10.1103/PhysRevLett.109.180602}
}

@article{Allahverdyan_2009,
doi = {10.1088/1742-5468/2009/09/P09011},
url = {https://doi.org/10.1088/1742-5468/2009/09/P09011},
year = {2009},
month = {sep},
publisher = {},
volume = {2009},
number = {09},
pages = {P09011},
author = {Allahverdyan, Armen E and Janzing, Dominik and Mahler, Guenter},
title = {Thermodynamic efficiency of information and heat flow},
journal = {Journal of Statistical Mechanics: Theory and Experiment},
}

@article{horowitz2010,
  title = {Nonequilibrium detailed fluctuation theorem for repeated discrete feedback},
  author = {Horowitz, Jordan M. and Vaikuntanathan, Suriyanarayanan},
  journal = {Phys. Rev. E},
  volume = {82},
  issue = {6},
  pages = {061120},
  numpages = {6},
  year = {2010},
  month = {Dec},
  publisher = {American Physical Society},
  doi = {10.1103/PhysRevE.82.061120},
  url = {https://link.aps.org/doi/10.1103/PhysRevE.82.061120}
}

@article{still2012,
  title = {Thermodynamics of Prediction},
  author = {Still, Susanne and Sivak, David A. and Bell, Anthony J. and Crooks, Gavin E.},
  journal = {Phys. Rev. Lett.},
  volume = {109},
  issue = {12},
  pages = {120604},
  numpages = {5},
  year = {2012},
  month = {Sep},
  publisher = {American Physical Society},
  doi = {10.1103/PhysRevLett.109.120604},
  url = {https://link.aps.org/doi/10.1103/PhysRevLett.109.120604}
}

@article{sagawapre2012,
  title = {Nonequilibrium thermodynamics of feedback control},
  author = {Sagawa, Takahiro and Ueda, Masahito},
  journal = {Phys. Rev. E},
  volume = {85},
  issue = {2},
  pages = {021104},
  numpages = {16},
  year = {2012},
  month = {Feb},
  publisher = {American Physical Society},
  doi = {10.1103/PhysRevE.85.021104},
  url = {https://link.aps.org/doi/10.1103/PhysRevE.85.021104}
}

@article{ito2013,
  title = {Information Thermodynamics on Causal Networks},
  author = {Ito, Sosuke and Sagawa, Takahiro},
  journal = {Phys. Rev. Lett.},
  volume = {111},
  issue = {18},
  pages = {180603},
  numpages = {6},
  year = {2013},
  month = {Oct},
  publisher = {American Physical Society},
  doi = {10.1103/PhysRevLett.111.180603},
  url = {https://link.aps.org/doi/10.1103/PhysRevLett.111.180603}
}

@article{Hartich_2014,
doi = {10.1088/1742-5468/2014/02/P02016},
url = {https://doi.org/10.1088/1742-5468/2014/02/P02016},
year = {2014},
month = {feb},
publisher = {IOP Publishing and SISSA},
volume = {2014},
number = {2},
pages = {P02016},
author = {Hartich, D and Barato, A C and Seifert, U},
title = {Stochastic thermodynamics of bipartite systems: transfer entropy inequalities and a Maxwell’s demon interpretation},
journal = {Journal of Statistical Mechanics: Theory and Experiment}
}

@article{horowitzesposito2014,
  title = {Thermodynamics with Continuous Information Flow},
  author = {Horowitz, Jordan M. and Esposito, Massimiliano},
  journal = {Phys. Rev. X},
  volume = {4},
  issue = {3},
  pages = {031015},
  numpages = {11},
  year = {2014},
  month = {Jul},
  publisher = {American Physical Society},
  doi = {10.1103/PhysRevX.4.031015},
  url = {https://link.aps.org/doi/10.1103/PhysRevX.4.031015}
}

@article{Owen2020,
  title = {Universal Thermodynamic Bounds on Nonequilibrium Response with Biochemical Applications},
  author = {Owen, Jeremy A. and Gingrich, Todd R. and Horowitz, Jordan M.},
  journal = {Phys. Rev. X},
  volume = {10},
  issue = {1},
  pages = {011066},
  numpages = {21},
  year = {2020},
  month = {Mar},
  publisher = {American Physical Society},
  doi = {10.1103/PhysRevX.10.011066},
  url = {https://link.aps.org/doi/10.1103/PhysRevX.10.011066}
}

@article{ito2016backward,
  title={Backward transfer entropy: Informational measure for detecting hidden Markov models and its interpretations in thermodynamics, gambling and causality},
  author={Ito, Sosuke},
  journal={Scientific reports},
  volume={6},
  number={1},
  pages={36831},
  year={2016},
  publisher={Nature Publishing Group UK London},
  url={https://doi.org/10.1038/srep36831}
}

@article{spinney2016,
  title = {Transfer entropy in physical systems and the arrow of time},
  author = {Spinney, Richard E. and Lizier, Joseph T. and Prokopenko, Mikhail},
  journal = {Phys. Rev. E},
  volume = {94},
  issue = {2},
  pages = {022135},
  numpages = {15},
  year = {2016},
  month = {Aug},
  publisher = {American Physical Society},
  doi = {10.1103/PhysRevE.94.022135},
  url = {https://link.aps.org/doi/10.1103/PhysRevE.94.022135}
}

@article{Crooks_2019,
doi = {10.1209/0295-5075/125/40005},
url = {https://doi.org/10.1209/0295-5075/125/40005},
year = {2019},
month = {mar},
publisher = {EDP Sciences, IOP Publishing and Società Italiana di Fisica},
volume = {125},
number = {4},
pages = {40005},
author = {Crooks, Gavin E. and Still, Susanne},
title = {Marginal and conditional second laws of thermodynamics},
journal = {Europhysics Letters}
}

@article{sagawa2010,
  title = {Generalized Jarzynski Equality under Nonequilibrium Feedback Control},
  author = {Sagawa, Takahiro and Ueda, Masahito},
  journal = {Phys. Rev. Lett.},
  volume = {104},
  issue = {9},
  pages = {090602},
  numpages = {4},
  year = {2010},
  month = {Mar},
  publisher = {American Physical Society},
  doi = {10.1103/PhysRevLett.104.090602},
  url = {https://link.aps.org/doi/10.1103/PhysRevLett.104.090602}
}

@article{auconi2019information,
  title={Information thermodynamics for time series of signal-response models},
  author={Auconi, Andrea and Giansanti, Andrea and Klipp, Edda},
  journal={Entropy},
  volume={21},
  number={2},
  pages={177},
  year={2019},
  publisher={MDPI},
  url={https://doi.org/10.3390/e21020177}
}

@article{ito2016information,
  title={Information flow and entropy production on Bayesian networks},
  author={Ito, Sosuke and Sagawa, Takahiro},
  journal={Mathematical Foundations and Applications of Graph Entropy},
  volume={6},
  pages={63--99},
  year={2016},
  publisher={Wiley Online Library},
  url={https://doi.org/10.1002/9783527693245.ch3}
}

@article{Ryota2022,
author = {Ryota Takaki  and Mauro L. Mugnai  and D. Thirumalai },
title = {Information flow, gating, and energetics in dimeric molecular motors},
journal = {Proceedings of the National Academy of Sciences},
volume = {119},
number = {46},
pages = {e2208083119},
year = {2022},
doi = {10.1073/pnas.2208083119},
URL = {https://www.pnas.org/doi/abs/10.1073/pnas.2208083119},
eprint = {https://www.pnas.org/doi/pdf/10.1073/pnas.2208083119}
}

@article{szilard1964decrease,
  title={On the decrease of entropy in a thermodynamic system by the intervention of intelligent beings},
  author={Szilard, Leo},
  journal={Behavioral Science},
  volume={9},
  number={4},
  pages={301--310},
  year={1964},
  publisher={Wiley Online Library}
}

@article{nagayama2025infinite,
  title = {Infinite variety of thermodynamic speed limits with general activities},
  author = {Nagayama, Ryuna and Yoshimura, Kohei and Ito, Sosuke},
  journal = {Phys. Rev. Res.},
  volume = {7},
  issue = {1},
  pages = {013307},
  numpages = {23},
  year = {2025},
  month = {Mar},
  publisher = {American Physical Society},
  doi = {10.1103/PhysRevResearch.7.013307},
  url = {https://link.aps.org/doi/10.1103/PhysRevResearch.7.013307}
}

@article{baratotur2015,
  title = {Thermodynamic Uncertainty Relation for Biomolecular Processes},
  author = {Barato, Andre C. and Seifert, Udo},
  journal = {Phys. Rev. Lett.},
  volume = {114},
  issue = {15},
  pages = {158101},
  numpages = {5},
  year = {2015},
  month = {Apr},
  publisher = {American Physical Society},
  doi = {10.1103/PhysRevLett.114.158101},
  url = {https://link.aps.org/doi/10.1103/PhysRevLett.114.158101}
}

@article{horowitz2020thermodynamic,
  title={Thermodynamic uncertainty relations constrain non-equilibrium fluctuations},
  author={Horowitz, Jordan M and Gingrich, Todd R},
  journal={Nature Physics},
  volume={16},
  number={1},
  pages={15--20},
  year={2020},
  publisher={Nature Publishing Group UK London},
  url={https://doi.org/10.1038/s41567-019-0702-6}
}

@article{oikawa2025experimentally,
  title={Experimentally achieving minimal dissipation via thermodynamically optimal transport},
  author={Oikawa, Shingo and Nakayama, Yohei and Ito, Sosuke and Sagawa, Takahiro and Toyabe, Shoichi},
  journal={Nature Communications},
  volume={16},
  number={1},
  pages={10424},
  year={2025},
  publisher={Nature Publishing Group UK London}
}

@ARTICLE{chen2020,
  author={Chen, Yongxin and Georgiou, Tryphon T. and Tannenbaum, Allen},
  journal={IEEE Transactions on Automatic Control}, 
  title={Stochastic Control and Nonequilibrium Thermodynamics: Fundamental Limits}, 
  year={2020},
  volume={65},
  number={7},
  pages={2979-2991},
  doi={10.1109/TAC.2019.2939625}}

@article{Dechantwasser_2022,
doi = {10.1088/1751-8121/ac4ac0},
url = {https://doi.org/10.1088/1751-8121/ac4ac0},
year = {2022},
month = {feb},
publisher = {IOP Publishing},
volume = {55},
number = {9},
pages = {094001},
author = {Dechant, Andreas},
title = {Minimum entropy production, detailed balance and Wasserstein distance for continuous-time Markov processes},
journal = {Journal of Physics A: Mathematical and Theoretical},
}

@article{vu2023,
  title = {Thermodynamic Unification of Optimal Transport: Thermodynamic Uncertainty Relation, Minimum Dissipation, and Thermodynamic Speed Limits},
  author = {Van Vu, Tan and Saito, Keiji},
  journal = {Phys. Rev. X},
  volume = {13},
  issue = {1},
  pages = {011013},
  numpages = {45},
  year = {2023},
  month = {Feb},
  publisher = {American Physical Society},
  doi = {10.1103/PhysRevX.13.011013},
  url = {https://link.aps.org/doi/10.1103/PhysRevX.13.011013}
}

@article{strasberg2013,
  title = {Thermodynamics of a Physical Model Implementing a Maxwell Demon},
  author = {Strasberg, Philipp and Schaller, Gernot and Brandes, Tobias and Esposito, Massimiliano},
  journal = {Phys. Rev. Lett.},
  volume = {110},
  issue = {4},
  pages = {040601},
  numpages = {5},
  year = {2013},
  month = {Jan},
  publisher = {American Physical Society},
  doi = {10.1103/PhysRevLett.110.040601},
  url = {https://link.aps.org/doi/10.1103/PhysRevLett.110.040601}
}

@article{toyabe2010experimental,
  title={Experimental demonstration of information-to-energy conversion and validation of the generalized Jarzynski equality},
  author={Toyabe, Shoichi and Sagawa, Takahiro and Ueda, Masahito and Muneyuki, Eiro and Sano, Masaki},
  journal={Nature physics},
  volume={6},
  number={12},
  pages={988--992},
  year={2010},
  publisher={Nature Publishing Group UK London},
  url={https://doi.org/10.1038/nphys1821}
}

@article{Leignton2024,
  title = {Information Arbitrage in Bipartite Heat Engines},
  author = {Leighton, Matthew P. and Ehrich, Jannik and Sivak, David A.},
  journal = {Phys. Rev. X},
  volume = {14},
  issue = {4},
  pages = {041038},
  numpages = {25},
  year = {2024},
  month = {Nov},
  publisher = {American Physical Society},
  doi = {10.1103/PhysRevX.14.041038},
  url = {https://link.aps.org/doi/10.1103/PhysRevX.14.041038}
}

@article{yada2022,
  title = {Quantum Fluctuation Theorem under Quantum Jumps with Continuous Measurement and Feedback},
  author = {Yada, Toshihiro and Yoshioka, Nobuyuki and Sagawa, Takahiro},
  journal = {Phys. Rev. Lett.},
  volume = {128},
  issue = {17},
  pages = {170601},
  numpages = {6},
  year = {2022},
  month = {Apr},
  publisher = {American Physical Society},
  doi = {10.1103/PhysRevLett.128.170601},
  url = {https://link.aps.org/doi/10.1103/PhysRevLett.128.170601}
}

@article{amano2022insights,
  title={Insights from an information thermodynamics analysis of a synthetic molecular motor},
  author={Amano, Shuntaro and Esposito, Massimiliano and Kreidt, Elisabeth and Leigh, David A and Penocchio, Emanuele and Roberts, Benjamin MW},
  journal={Nature Chemistry},
  volume={14},
  number={5},
  pages={530--537},
  year={2022},
  publisher={Nature Publishing Group UK London},
  url={https://doi.org/10.1038/s41557-022-00899-z}
}

@article{ikeda2025,
  title = {Speed-Accuracy Relations for Diffusion Models: Wisdom from Nonequilibrium Thermodynamics and Optimal Transport},
  author = {Ikeda, Kotaro and Uda, Tomoya and Okanohara, Daisuke and Ito, Sosuke},
  journal = {Phys. Rev. X},
  volume = {15},
  issue = {3},
  pages = {031031},
  numpages = {40},
  year = {2025},
  month = {Jul},
  publisher = {American Physical Society},
  doi = {10.1103/x5vj-8jq9},
  url = {https://link.aps.org/doi/10.1103/x5vj-8jq9}
}

@article{tasnim2021thermodynamic,
  title={Thermodynamic speed limits for co-evolving systems},
  author={Tasnim, Farita and Wolpert, David H},
  journal={arXiv preprint arXiv:2107.12471},
  year={2021},
  url={
https://doi.org/10.48550/arXiv.2107.12471
}
}

@article{dechant2025precision,
  title={Precision and cost of feedback cooling},
  author={Dechant, Andreas and H{\"u}pfl, Jakob and Kobayashi, Shuta and Ito, Sosuke and Rotter, Stefan},
  journal={arXiv preprint arXiv:2508.12875},
  year={2025},
  url={https://doi.org/10.48550/arXiv.2508.12875}
}

@article{ito2011effects,
    title = {Effects of error on fluctuations under feedback control},
  author = {Ito, Sosuke and Sano, Masaki},
  journal = {Phys. Rev. E},
  volume = {84},
  issue = {2},
  pages = {021123},
  numpages = {7},
  year = {2011},
  month = {Aug},
  publisher = {American Physical Society},
  doi = {10.1103/PhysRevE.84.021123},
  url = {https://link.aps.org/doi/10.1103/PhysRevE.84.021123}
}

@article{horowitz2014second,
  doi = {10.1088/1367-2630/16/12/125007},
url = {https://doi.org/10.1088/1367-2630/16/12/125007},
year = {2014},
month = {dec},
publisher = {IOP Publishing},
volume = {16},
number = {12},
pages = {125007},
author = {Horowitz, Jordan M and Sandberg, Henrik},
title = {Second-law-like inequalities with information and their interpretations},
journal = {New Journal of Physics}
}

@article{kim2004entropy,
   title = {Entropy Production of Brownian Macromolecules with Inertia},
  author = {Kim, Kyung Hyuk and Qian, Hong},
  journal = {Phys. Rev. Lett.},
  volume = {93},
  issue = {12},
  pages = {120602},
  numpages = {4},
  year = {2004},
  month = {Sep},
  publisher = {American Physical Society},
  doi = {10.1103/PhysRevLett.93.120602},
  url = {https://link.aps.org/doi/10.1103/PhysRevLett.93.120602}
}

@article{kim2007fluctuation,
   title = {Fluctuation theorems for a molecular refrigerator},
  author = {Kim, Kyung Hyuk and Qian, Hong},
  journal = {Phys. Rev. E},
  volume = {75},
  issue = {2},
  pages = {022102},
  numpages = {4},
  year = {2007},
  month = {Feb},
  publisher = {American Physical Society},
  doi = {10.1103/PhysRevE.75.022102},
  url = {https://link.aps.org/doi/10.1103/PhysRevE.75.022102}
}

@article{munakata2013feedback,
  doi = {10.1088/1742-5468/2013/06/P06014},
url = {https://doi.org/10.1088/1742-5468/2013/06/P06014},
year = {2013},
month = {jun},
publisher = {IOP Publishing and SISSA},
volume = {2013},
number = {06},
pages = {P06014},
author = {Munakata, T and Rosinberg, M L},
title = {Feedback cooling, measurement errors, and entropy production},
journal = {Journal of Statistical Mechanics: Theory and Experiment}
}

@article{kolchinsky2024generalized,
  title={Generalized free energy and excess/housekeeping decomposition in nonequilibrium systems: From large deviations to thermodynamic speed limits},
  author={Kolchinsky, Artemy and Dechant, Andreas and Yoshimura, Kohei and Ito, Sosuke},
  journal={Physical Review Research},
  volume={8},
  number={2},
  pages={023025},
  year={2026},
  publisher={American Physical Society},
  url={https://doi.org/10.1103/r48t-dghl}
}

@article{lan2012energy,
  title={The energy--speed--accuracy trade-off in sensory adaptation},
  author={Lan, Ganhui and Sartori, Pablo and Neumann, Silke and Sourjik, Victor and Tu, Yuhai},
  journal={Nature physics},
  volume={8},
  number={5},
  pages={422--428},
  year={2012},
  publisher={Nature Publishing Group UK London},
  url={https://doi.org/10.1038/nphys2276}
}

@article{ito2015maxwell,
  title={Maxwell’s demon in biochemical signal transduction with feedback loop},
  author={Ito, Sosuke and Sagawa, Takahiro},
  journal={Nature communications},
  volume={6},
  number={1},
  pages={7498},
  year={2015},
  publisher={Nature Publishing Group UK London},
  url={https://doi.org/10.1038/ncomms8498}
}

@article{hartich2016sensory,
  title = {Sensory capacity: An information theoretical measure of the performance of a sensor},
  author = {Hartich, David and Barato, Andre C. and Seifert, Udo},
  journal = {Phys. Rev. E},
  volume = {93},
  issue = {2},
  pages = {022116},
  numpages = {14},
  year = {2016},
  month = {Feb},
  publisher = {American Physical Society},
  doi = {10.1103/PhysRevE.93.022116},
  url = {https://link.aps.org/doi/10.1103/PhysRevE.93.022116}
}

@article{dechant2019thermodynamic,
  title={Thermodynamic interpretation of Wasserstein distance},
  author={Dechant, Andreas and Sakurai, Yohei},
  journal={arXiv preprint arXiv:1912.08405},
  url={https://doi.org/10.48550/arXiv.1912.08405},
  year={2019}
}

@article{yoshimura2024two,
  title = {Two applications of stochastic thermodynamics to hydrodynamics},
  author = {Yoshimura, Kohei and Ito, Sosuke},
  journal = {Phys. Rev. Res.},
  volume = {6},
  issue = {2},
  pages = {L022057},
  numpages = {5},
  year = {2024},
  month = {Jun},
  publisher = {American Physical Society},
  doi = {10.1103/PhysRevResearch.6.L022057},
  url = {https://link.aps.org/doi/10.1103/PhysRevResearch.6.L022057}
}

@article{yoshimura2025force,
   title = {Force-current structure in Markovian open quantum systems and its applications: Geometric housekeeping-excess decomposition and thermodynamic trade-off relations},
  author = {Yoshimura, Kohei and Maekawa, Yoh and Nagayama, Ryuna and Ito, Sosuke},
  journal = {Phys. Rev. Res.},
  volume = {7},
  issue = {1},
  pages = {013244},
  numpages = {24},
  year = {2025},
  month = {Mar},
  publisher = {American Physical Society},
  doi = {10.1103/PhysRevResearch.7.013244},
  url = {https://link.aps.org/doi/10.1103/PhysRevResearch.7.013244}
}

@article{jordan1998,
author = {Jordan, Richard and Kinderlehrer, David and Otto, Felix},
title = {The Variational Formulation of the Fokker--Planck Equation},
journal = {SIAM Journal on Mathematical Analysis},
volume = {29},
number = {1},
pages = {1-17},
year = {1998},
doi = {10.1137/S0036141096303359},

URL = { 
    
        https://doi.org/10.1137/S0036141096303359
    
    

},
eprint = { 
    
        https://doi.org/10.1137/S0036141096303359
    
    

}

}

@article{chennakesavalu2023,
  title = {Unified, Geometric Framework for Nonequilibrium Protocol Optimization},
  author = {Chennakesavalu, Shriram and Rotskoff, Grant M.},
  journal = {Phys. Rev. Lett.},
  volume = {130},
  issue = {10},
  pages = {107101},
  numpages = {6},
  year = {2023},
  month = {Mar},
  publisher = {American Physical Society},
  doi = {10.1103/PhysRevLett.130.107101},
  url = {https://link.aps.org/doi/10.1103/PhysRevLett.130.107101}
}

@article{zhong2024,
  title = {Beyond Linear Response: Equivalence between Thermodynamic Geometry and Optimal Transport},
  author = {Zhong, Adrianne and DeWeese, Michael R.},
  journal = {Phys. Rev. Lett.},
  volume = {133},
  issue = {5},
  pages = {057102},
  numpages = {9},
  year = {2024},
  month = {Jul},
  publisher = {American Physical Society},
  doi = {10.1103/PhysRevLett.133.057102},
  url = {https://link.aps.org/doi/10.1103/PhysRevLett.133.057102}
}

@article{kwon2024,
  title = {Unified hierarchical relationship between thermodynamic tradeoff relations},
  author = {Kwon, Euijoon and Park, Jong-Min and Lee, Jae Sung and Baek, Yongjoo},
  journal = {Phys. Rev. E},
  volume = {110},
  issue = {4},
  pages = {044131},
  numpages = {20},
  year = {2024},
  month = {Oct},
  publisher = {American Physical Society},
  doi = {10.1103/PhysRevE.110.044131},
  url = {https://link.aps.org/doi/10.1103/PhysRevE.110.044131}
}

@article{Delvenne2024,
  title = {Thermokinetic relations},
  author = {Delvenne, Jean-Charles and Falasco, Gianmaria},
  journal = {Phys. Rev. E},
  volume = {109},
  issue = {1},
  pages = {014109},
  numpages = {16},
  year = {2024},
  month = {Jan},
  publisher = {American Physical Society},
  doi = {10.1103/PhysRevE.109.014109},
  url = {https://link.aps.org/doi/10.1103/PhysRevE.109.014109}
}

@article{sabbagh2024wasserstein,
   title = {Wasserstein speed limits for Langevin systems},
  author = {Sabbagh, Ralph and Movilla Miangolarra, Olga and Georgiou, Tryphon T.},
  journal = {Phys. Rev. Res.},
  volume = {6},
  issue = {3},
  pages = {033308},
  numpages = {9},
  year = {2024},
  month = {Sep},
  publisher = {American Physical Society},
  doi = {10.1103/PhysRevResearch.6.033308},
  url = {https://link.aps.org/doi/10.1103/PhysRevResearch.6.033308}
}

@article{rosinberg2016continuous,
  doi = {10.1209/0295-5075/116/10007},
url = {https://doi.org/10.1209/0295-5075/116/10007},
year = {2016},
month = {nov},
publisher = {EDP Sciences, IOP Publishing and Società Italiana di Fisica},
volume = {116},
number = {1},
pages = {10007},
author = {Rosinberg, Martin Luc and Horowitz, Jordan M.},
title = {Continuous information flow fluctuations},
journal = {Europhysics Letters},
}

@article{tanogami2023universal,
 title={Universal bounds on the performance of information-thermodynamic engine},
  author={Tanogami, Tomohiro and Van Vu, Tan and Saito, Keiji},
  journal={Physical Review Research},
  volume={5},
  number={4},
  pages={043280},
  year={2023},
  publisher={APS}
}

@article{wolpert2020uncertainty,
  title = {Uncertainty Relations and Fluctuation Theorems for Bayes Nets},
  author = {Wolpert, David H.},
  journal = {Phys. Rev. Lett.},
  volume = {125},
  issue = {20},
  pages = {200602},
  numpages = {6},
  year = {2020},
  month = {Nov},
  publisher = {American Physical Society},
  doi = {10.1103/PhysRevLett.125.200602},
  url = {https://link.aps.org/doi/10.1103/PhysRevLett.125.200602}
}

@article{aurell2011optimal,
   title = {Optimal Protocols and Optimal Transport in Stochastic Thermodynamics},
  author = {Aurell, Erik and Mej\'{\i}a-Monasterio, Carlos and Muratore-Ginanneschi, Paolo},
  journal = {Phys. Rev. Lett.},
  volume = {106},
  issue = {25},
  pages = {250601},
  numpages = {4},
  year = {2011},
  month = {Jun},
  publisher = {American Physical Society},
  doi = {10.1103/PhysRevLett.106.250601},
  url = {https://link.aps.org/doi/10.1103/PhysRevLett.106.250601}
}

@article{Lee2022,
  title = {Speed Limit for a Highly Irreversible Process and Tight Finite-Time Landauer's Bound},
  author = {Lee, Jae Sung and Lee, Sangyun and Kwon, Hyukjoon and Park, Hyunggyu},
  journal = {Phys. Rev. Lett.},
  volume = {129},
  issue = {12},
  pages = {120603},
  numpages = {7},
  year = {2022},
  month = {Sep},
  publisher = {American Physical Society},
  doi = {10.1103/PhysRevLett.129.120603},
  url = {https://link.aps.org/doi/10.1103/PhysRevLett.129.120603}
}

@article{Zhen2021,
  title = {Universal Bound on Energy Cost of Bit Reset in Finite Time},
  author = {Zhen, Yi-Zheng and Egloff, Dario and Modi, Kavan and Dahlsten, Oscar},
  journal = {Phys. Rev. Lett.},
  volume = {127},
  issue = {19},
  pages = {190602},
  numpages = {7},
  year = {2021},
  month = {Nov},
  publisher = {American Physical Society},
  doi = {10.1103/PhysRevLett.127.190602},
  url = {https://link.aps.org/doi/10.1103/PhysRevLett.127.190602}
}

@article{proesmans2020,
  title = {Finite-Time Landauer Principle},
  author = {Proesmans, Karel and Ehrich, Jannik and Bechhoefer, John},
  journal = {Phys. Rev. Lett.},
  volume = {125},
  issue = {10},
  pages = {100602},
  numpages = {6},
  year = {2020},
  month = {Sep},
  publisher = {American Physical Society},
  doi = {10.1103/PhysRevLett.125.100602},
  url = {https://link.aps.org/doi/10.1103/PhysRevLett.125.100602}
}

@article{nagase2025thermodynamic,
  title = {Thermodynamic optimization of finite-time feedback protocols for Markov jump systems},
  author = {Nagase, Rihito and Sagawa, Takahiro},
  journal = {Phys. Rev. E},
  volume = {112},
  issue = {2},
  pages = {024118},
  numpages = {12},
  year = {2025},
  month = {Aug},
  publisher = {American Physical Society},
  doi = {10.1103/td2s-819q},
  url = {https://link.aps.org/doi/10.1103/td2s-819q}
}

@article{nagase2024thermodynamically,
  title = {Thermodynamically optimal information gain in finite-time measurement},
  author = {Nagase, Rihito and Sagawa, Takahiro},
  journal = {Phys. Rev. Res.},
  volume = {6},
  issue = {3},
  pages = {033239},
  numpages = {16},
  year = {2024},
  month = {Sep},
  publisher = {American Physical Society},
  doi = {10.1103/PhysRevResearch.6.033239},
  url = {https://link.aps.org/doi/10.1103/PhysRevResearch.6.033239}
}

@article{kamijima2025finite,
  title = {Finite-time thermodynamic bounds and trade-off relations for information processing},
  author = {Kamijima, Takuya and Funo, Ken and Sagawa, Takahiro},
  journal = {Phys. Rev. Res.},
  volume = {7},
  issue = {1},
  pages = {013329},
  numpages = {22},
  year = {2025},
  month = {Mar},
  publisher = {American Physical Society},
  doi = {10.1103/PhysRevResearch.7.013329},
  url = {https://link.aps.org/doi/10.1103/PhysRevResearch.7.013329}
}

@article{kamijima2025optimal,
   title = {Optimal finite-time Maxwell's demons in Langevin systems},
  author = {Kamijima, Takuya and Takatsu, Asuka and Funo, Ken and Sagawa, Takahiro},
  journal = {Phys. Rev. Res.},
  volume = {7},
  issue = {2},
  pages = {023159},
  numpages = {9},
  year = {2025},
  month = {May},
  publisher = {American Physical Society},
  doi = {10.1103/PhysRevResearch.7.023159},
  url = {https://link.aps.org/doi/10.1103/PhysRevResearch.7.023159}
}

@inproceedings{sohl2015deep,
  title={Deep unsupervised learning using nonequilibrium thermodynamics},
  author={Sohl-Dickstein, Jascha and Weiss, Eric and Maheswaranathan, Niru and Ganguli, Surya},
  booktitle={International conference on machine learning},
  pages={2256--2265},
  year={2015},
  organization={pmlr}
}

@article{Goldt2017,
  title = {Stochastic Thermodynamics of Learning},
  author = {Goldt, Sebastian and Seifert, Udo},
  journal = {Phys. Rev. Lett.},
  volume = {118},
  issue = {1},
  pages = {010601},
  numpages = {5},
  year = {2017},
  month = {Jan},
  publisher = {American Physical Society},
  doi = {10.1103/PhysRevLett.118.010601},
  url = {https://link.aps.org/doi/10.1103/PhysRevLett.118.010601}
}

@inproceedings{rombach2022high,
  title={High-resolution image synthesis with latent diffusion models},
  author={Rombach, Robin and Blattmann, Andreas and Lorenz, Dominik and Esser, Patrick and Ommer, Bj{\"o}rn},
  booktitle={Proceedings of the IEEE/CVF conference on computer vision and pattern recognition},
  pages={10684--10695},
  year={2022}
}

@article{lipman2022flow,
  title={Flow matching for generative modeling},
  author={Lipman, Yaron and Chen, Ricky TQ and Ben-Hamu, Heli and Nickel, Maximilian and Le, Matt},
  journal={The Eleventh International Conference on Learning Representations（ICLR 2023）},
  year={2023}
}

@article{song2020score,
  title={Score-based generative modeling through stochastic differential equations},
  author={Song, Yang and Sohl-Dickstein, Jascha and Kingma, Diederik P and Kumar, Abhishek and Ermon, Stefano and Poole, Ben},
  journal={in International Conference on Learning Representations (ICLR) },
  year={2021}
}

@article{Vanden2010,
  title = {Three faces of the second law. II. Fokker-Planck formulation},
  author = {Van den Broeck, Christian and Esposito, Massimiliano},
  journal = {Phys. Rev. E},
  volume = {82},
  issue = {1},
  pages = {011144},
  numpages = {7},
  year = {2010},
  month = {Jul},
  publisher = {American Physical Society},
  doi = {10.1103/PhysRevE.82.011144},
  url = {https://link.aps.org/doi/10.1103/PhysRevE.82.011144}
}

@misc{sekimoto2010stochastic,
  title={Stochastic energetics},
  author={Sekimoto, Ken},
  year={2010},
  publisher={Springer Berlin}
}

@article{aurell2012refined,
  title={Refined second law of thermodynamics for fast random processes},
  author={Aurell, Erik and Gawedzki, Krzysztof and Mej{\'\i}a-Monasterio, Carlos and Mohayaee, Roya and Muratore-Ginanneschi, Paolo},
  journal={Journal of statistical physics},
  volume={147},
  number={3},
  pages={487--505},
  year={2012},
  publisher={Springer},
  url={https://doi.org/10.1007/s10955-012-0478-x}
}

@book{villani2008optimal,
  title={Optimal transport: old and new},
  author={Villani, C{\'e}dric and others},
  volume={338},
  year={2008},
  publisher={Springer}
}

@article{maas2011gradient,
  title = {Gradient flows of the entropy for finite Markov chains},
journal = {Journal of Functional Analysis},
volume = {261},
number = {8},
pages = {2250-2292},
year = {2011},
issn = {0022-1236},
doi = {https://doi.org/10.1016/j.jfa.2011.06.009},
url = {https://www.sciencedirect.com/science/article/pii/S0022123611002278},
author = {Jan Maas}
}

@article{maes2014nonequilibrium,
  title={A nonequilibrium extension of the Clausius heat theorem},
  author={Maes, Christian and Neto{\v{c}}n{\`y}, Karel},
  journal={Journal of Statistical Physics},
  volume={154},
  number={1},
  pages={188--203},
  year={2014},
  publisher={Springer},
  url={https://doi.org/10.1007/s10955-013-0822-9}
}

@article{dechant2022geometric,
  title = {Geometric decomposition of entropy production in out-of-equilibrium systems},
  author = {Dechant, Andreas and Sasa, Shin-ichi and Ito, Sosuke},
  journal = {Phys. Rev. Res.},
  volume = {4},
  issue = {1},
  pages = {L012034},
  numpages = {6},
  year = {2022},
  month = {Mar},
  publisher = {American Physical Society},
  doi = {10.1103/PhysRevResearch.4.L012034},
  url = {https://link.aps.org/doi/10.1103/PhysRevResearch.4.L012034}
}

@article{benamou2000computational,
  title={A computational fluid mechanics solution to the Monge-Kantorovich mass transfer problem},
  author={Benamou, Jean-David and Brenier, Yann},
  journal={Numerische Mathematik},
  volume={84},
  number={3},
  pages={375--393},
  year={2000},
  publisher={Springer-Verlag Berlin/Heidelberg}
}

@article{dechant2022geometric2,
   title = {Geometric decomposition of entropy production into excess, housekeeping, and coupling parts},
  author = {Dechant, Andreas and Sasa, Shin-ichi and Ito, Sosuke},
  journal = {Phys. Rev. E},
  volume = {106},
  issue = {2},
  pages = {024125},
  numpages = {22},
  year = {2022},
  month = {Aug},
  publisher = {American Physical Society},
  doi = {10.1103/PhysRevE.106.024125},
  url = {https://link.aps.org/doi/10.1103/PhysRevE.106.024125}
}

@article{sekizawa2025koopman,
author = {Daiki Sekizawa  and Sosuke Ito  and Masafumi Oizumi },
title = {Koopman mode decomposition of thermodynamic dissipation in nonlinear Langevin dynamics},
journal = {Proceedings of the National Academy of Sciences},
volume = {123},
number = {25},
pages = {e2530617123},
year = {2026},
URL = {https://www.pnas.org/doi/abs/10.1073/pnas.2530617123},
}

@article{sekizawa2024decomposing,
  title = {Decomposing Thermodynamic Dissipation of Linear Langevin Systems via Oscillatory Modes and Its Application to Neural Dynamics},
  author = {Sekizawa, Daiki and Ito, Sosuke and Oizumi, Masafumi},
  journal = {Phys. Rev. X},
  volume = {14},
  issue = {4},
  pages = {041003},
  numpages = {29},
  year = {2024},
  month = {Oct},
  publisher = {American Physical Society},
  doi = {10.1103/PhysRevX.14.041003},
  url = {https://link.aps.org/doi/10.1103/PhysRevX.14.041003}
}

@article{ito2024geometric,
  title={Geometric thermodynamics for the Fokker--Planck equation: stochastic thermodynamic links between information geometry and optimal transport},
  author={Ito, Sosuke},
  journal={Information geometry},
  volume={7},
  number={Suppl 1},
  pages={441--483},
  year={2024},
  publisher={Springer},
  url={https://doi.org/10.1007/s41884-023-00102-3}
}

@article{nagayama2025geometric,
   title = {Geometric thermodynamics of reaction-diffusion systems: Thermodynamic trade-off relations and optimal transport for pattern formation},
  author = {Nagayama, Ryuna and Yoshimura, Kohei and Kolchinsky, Artemy and Ito, Sosuke},
  journal = {Phys. Rev. Res.},
  volume = {7},
  issue = {3},
  pages = {033011},
  numpages = {46},
  year = {2025},
  month = {Jul},
  publisher = {American Physical Society},
  doi = {10.1103/PhysRevResearch.7.033011},
  url = {https://link.aps.org/doi/10.1103/PhysRevResearch.7.033011}
}

@article{yoshimura2023housekeeping,
  title = {Housekeeping and excess entropy production for general nonlinear dynamics},
  author = {Yoshimura, Kohei and Kolchinsky, Artemy and Dechant, Andreas and Ito, Sosuke},
  journal = {Phys. Rev. Res.},
  volume = {5},
  issue = {1},
  pages = {013017},
  numpages = {24},
  year = {2023},
  month = {Jan},
  publisher = {American Physical Society},
  doi = {10.1103/PhysRevResearch.5.013017},
  url = {https://link.aps.org/doi/10.1103/PhysRevResearch.5.013017}
}

@article{hatano2001steady,
  title = {Steady-State Thermodynamics of Langevin Systems},
  author = {Hatano, Takahiro and Sasa, Shin-ichi},
  journal = {Phys. Rev. Lett.},
  volume = {86},
  issue = {16},
  pages = {3463--3466},
  numpages = {0},
  year = {2001},
  month = {Apr},
  publisher = {American Physical Society},
  doi = {10.1103/PhysRevLett.86.3463},
  url = {https://link.aps.org/doi/10.1103/PhysRevLett.86.3463}
}

@article{nakazato2021geometrical,
  title = {Geometrical aspects of entropy production in stochastic thermodynamics based on Wasserstein distance},
  author = {Nakazato, Muka and Ito, Sosuke},
  journal = {Phys. Rev. Res.},
  volume = {3},
  issue = {4},
  pages = {043093},
  numpages = {16},
  year = {2021},
  month = {Nov},
  publisher = {American Physical Society},
  doi = {10.1103/PhysRevResearch.3.043093},
  url = {https://link.aps.org/doi/10.1103/PhysRevResearch.3.043093}
}

@inproceedings{massey1990causality,
  title={Causality, feedback and directed information},
  author={Massey, James L2},
  booktitle={Proc. Int. Symp. Inf. Theory Applic.(ISITA-90)},
  volume={2},
  pages={1},
  year={1990}
}

@article{parrondo1996criticism,
  title={Criticism of Feynman's analysis of the ratchet as an engine},
  author={Parrondo, Juan MR and Espa{\~n}ol, Pep},
  journal={American Journal of Physics},
  volume={64},
  number={9},
  pages={1125--1129},
  year={1996},
  publisher={[Woodbury, NY, etc. Published for the American Association of Physics~…}
}

@book{cover1999elements,
  title={Elements of information theory},
  author={Cover, Thomas M},
  year={1999},
  publisher={John Wiley \& Sons}
}

@ARTICLE{olga2022,
  author={Movilla Miangolarra, Olga and Taghvaei, Amirhossein and Chen, Yongxin and Georgiou, Tryphon T.},
  journal={IEEE Control Systems Letters}, 
  title={Geometry of Finite-Time Thermodynamic Cycles With Anisotropic Thermal Fluctuations}, 
  year={2022},
  volume={6},
  number={},
  pages={3409-3414},
  doi={10.1109/LCSYS.2022.3184912}}

@article{li2023wasserstein,
  title={Wasserstein information matrix},
  author={Li, Wuchen and Zhao, Jiaxi},
  journal={Information Geometry},
  volume={6},
  number={1},
  pages={203--255},
  year={2023},
  publisher={Springer},
  url={https://doi.org/10.1007/s41884-023-00099-9}
}

@article{matsumoto2025learning,
doi = {10.1088/1742-5468/adf4be},
url = {https://doi.org/10.1088/1742-5468/adf4be},
year = {2025},
journal = {Journal of Statistical Mechanics: Theory and Experiment},
month = {sep},
publisher = {IOP Publishing},
volume = {2025},
number = {9},
pages = {093202},
author = {Matsumoto, Kenshin and Sasa, Shin-ichi and Dechant, Andreas},
title = {Learning rate matrix and information-thermodynamic trade-off relation}
}

\end{document}